\documentclass[12pt]{article}%
\pdfoutput=1

\usepackage{rotating}
\usepackage{cite}
\usepackage[all]{xy}
\usepackage{amsmath,amssymb,amsthm,amsfonts,mathrsfs}
\theoremstyle{definition}

\usepackage{mathtools}
\usepackage{amsfonts}
\usepackage{amssymb}
\usepackage{authblk}
\usepackage{bbm}
\usepackage{bm}
\usepackage{caption}
\usepackage{epsfig}
\usepackage{epstopdf}
\usepackage{heck}
\usepackage[margin=1in]{geometry}
\usepackage{tabularx}
\usepackage[usenames,dvipsnames]{xcolor}
\usepackage{graphicx}
\numberwithin{equation}{section}
\usepackage{tikz}
\usetikzlibrary{decorations.pathreplacing}
\usetikzlibrary{calc}

\usepackage{verbatim}

\usepackage{hyperref}
\usepackage[nameinlink,capitalize]{cleveref}

\makeatletter
\newcommand*{\diff}{\@ifnextchar^{\DIFF}{\DIFF^{}}}
\def\DIFF^#1{\mathop{\mathrm{\mathstrut d}}\nolimits^{#1}\gobblespace}
\newcommand*{\bigdiff}{\@ifnextchar^{\BIGDIFF}{\BIGDIFF^{}}}
\def\BIGDIFF^#1{\mathop{\mathrm{\mathstrut \mathcal{D}}}\nolimits^{#1}\gobblespace}
\def\gobblespace{\futurelet\diffarg\opspace}
\def\opspace{%
	\let\DiffSpace\!%
	\ifx\diffarg(%
		\let\DiffSpace\relax
	\else
		\ifx\diffarg[%
			\let\DiffSpace\relax
		\else
			\ifx\diffarg\{%
				\let\DiffSpace\relax
			\fi
		\fi
	\fi
	\DiffSpace
}
\makeatother

\usepackage{tikz}
\usetikzlibrary{arrows}
\usetikzlibrary{arrows.meta}
\usetikzlibrary{shapes.geometric,calc,arrows, positioning,shapes.misc,decorations.markings}
\tikzset{
  big arrow/.style={
    decoration={markings,mark=at position 1 with {\arrow[scale=2,#1]{>}}},
    postaction={decorate},
    shorten >=0.4pt},
  big arrow/.default=black}

\pgfdeclarelayer{edgelayer}
\pgfdeclarelayer{nodelayer}
\pgfsetlayers{edgelayer,nodelayer,main}
\tikzstyle{none}=[inner sep=0pt]

\tikzstyle{NodeCross}=[draw, shape=circle, cross out, inner sep=0pt, minimum size=6pt,line width=0.25mm]
\tikzstyle{Circle}=[draw, shape=circle, black, fill=black, inner sep=0pt, minimum size=6pt]
\tikzstyle{Star}=[draw, shape=star, fill=black, star points=6, inner sep=0pt, minimum size=8pt]

\tikzstyle{DashedLine}=[-, densely dashed, line width=0.25mm]
\tikzstyle{DottedLine}=[-, dotted, line width=0.25mm]
\tikzstyle{ThickLine}=[-, line width=0.25mm]
\tikzstyle{ArrowLineRight}=[-, -{Stealth[scale=1.75]}, line width=0.15mm, scale=5]
\tikzstyle{RedLine}=[-, draw={rgb,255: red,191; green,0; blue,0}, fill=none, line width=0.25mm]
\tikzstyle{DashedLineThin}=[-, densely dashed, line width=0.125mm, fill=none, draw=black]
\tikzstyle{ArrowLineRed}=[-, draw={rgb,255: red,191; green,0; blue,0}, -{Stealth[scale=1.75]}, line width=0.15mm, scale=5]

\newcommand{\bea}{\begin{eqnarray}}
\newcommand{\eea}{\end{eqnarray}}
\newcommand{\be}{\begin{equation}}
\newcommand{\ee}{\end{equation}}
\newcommand{\ba}{\begin{aligned}}
\newcommand{\ea}{\end{aligned}}
\newcommand{\bit}{\begin{itemize}}
\newcommand{\eit}{\end{itemize}}
\newcommand{\ben}{\begin{enumerate}}
\newcommand{\een}{\end{enumerate}}

\renewcommand{\ni}{\noindent}

\newcommand{\lb}{\left(}
\newcommand{\rb}{\right)}

\newcommand{\lbbb}{\left\{}
\newcommand{\rbbb}{\right\}}
\newcommand{\tn}[1]{\textnormal{#1}}

\newcommand{\Z}{{\mathbb Z}}
\newcommand{\R}{{\mathbb R}}
\newcommand{\C}{{\mathbb C}}

\renewcommand{\P}{{\mathbb P}}

\begin{document}

\date{March 2022}

\title{0-Form, 1-Form and 2-Group Symmetries via \\[4mm] Cutting and Gluing of Orbifolds}

\institution{PENN}{\centerline{$^{1}$Department of Physics and Astronomy, University of Pennsylvania, Philadelphia, PA 19104, USA}}
\institution{PENNmath}{\centerline{$^{2}$Department of Mathematics, University of Pennsylvania, Philadelphia, PA 19104, USA}}
\institution{MARIBOR}{\centerline{${}^{3}$Center for Applied Mathematics and Theoretical Physics, University of Maribor, Maribor, Slovenia}}
\institution{CERN}{\centerline{${}^{4}$CERN Theory Department, CH-1211 Geneva, Switzerland}}

\authors{
Mirjam Cveti\v{c}\worksat{\PENN,\PENNmath,\MARIBOR,\CERN}\footnote{e-mail: \texttt{cvetic@physics.upenn.edu}},
Jonathan J. Heckman\worksat{\PENN,\PENNmath}\footnote{e-mail: \texttt{jheckman@sas.upenn.edu}},\\[4mm]
Max H\"ubner\worksat{\PENN}\footnote{e-mail: \texttt{hmax@sas.upenn.edu}}, and
Ethan Torres\worksat{\PENN}\footnote{e-mail: \texttt{emtorres@sas.upenn.edu}}
}

\abstract{Orbifold singularities of M-theory constitute the building blocks of a broad class of supersymmetric quantum field theories (SQFTs). In this paper we show how the local data of these geometries determines
global data on the resulting higher symmetries of these systems. In particular, via a process of cutting and gluing,
we show how local orbifold singularities encode the 0-form, 1-form and 2-group symmetries of
the resulting SQFTs. Geometrically, this is obtained
from the possible singularities which extend to the boundary of the non-compact geometry.
The resulting category of boundary conditions then captures these symmetries, and
is equivalently specified by the orbifold homology of the boundary geometry.
We illustrate these general points in the context of a number of examples, including 5D superconformal field theories
engineered via orbifold singularities, 5D gauge theories engineered via singular elliptically
fibered Calabi-Yau threefolds, as well as 4D SQCD-like theories engineered via M-theory on non-compact $G_2$ spaces.}

{\small \texttt{\hfill UPR-1317-T}}

{\small \texttt{\hfill CERN-TH-2022-053}}

\maketitle

\setcounter{tocdepth}{2}

\tableofcontents

\newpage

\section{Introduction}

Compactification provides a bridge connecting higher-dimensional quantum gravity to lowe-dimensional quantum field theories.
In the context of M- and F-theory compactification, the general procedure for obtaining interacting
quantum field theories necessarily involves the study of localized singularities and branes. The degrees of freedom of the resulting supersymmetric quantum field theory (SQFT) are localized in a small neighborhood, and can be decoupled from bulk gravitational degrees of freedom.

In the limit where lower-dimensional gravity decouples, global symmetries can emerge, and these serve to constrain the dynamics of the resulting theories. Recently it has been appreciated that in addition to the standard 0-form symmetries which act on local operators, higher-form symmetries acting on extended objects often provide additional important topological data \cite{Gaiotto:2014kfa} (see also \cite{Gaiotto:2010be, Aharony:2013hda, DelZotto:2015isa}). Especially in $D > 4$ spacetime dimensions, geometry / brane constructions are the only tool available for directly constructing the resulting SQFTs, and in $D \leq 4$ spacetime dimensions, the resulting string constructions also provide
invaluable tools in studying non-perturbative phenomena. Given this, a natural question is whether the geometry of the string compactification can be used to extract this important global data of the resulting SQFTs.

Recently much progress has been made in understanding the spectrum of extended objects which behave as non-dynamical
defects in the resulting SQFT. For example, in the case of 6D superconformal field theories (SCFTs) realized via F-theory on an elliptically fiber Calabi-Yau threefold $X \rightarrow B$, the defect group associated with non-dynamical strings results from D3-branes stretched on non-compact 2-cycles in the base $B$ \cite{DelZotto:2015isa}.\footnote{See e.g. \cite{Heckman:2018jxk, Argyres:2022mnu}  for recent reviews of 6D SCFTs.} Similar considerations hold for a wide class of M-theory compactifications, where stretched M2-branes and M5-branes can result in a rich spectrum of extended objects (see e.g. \cite{Morrison:2020ool, Albertini:2020mdx,Bhardwaj:2020phs}). In all of these cases, the general idea is that dynamical states obtained from branes wrapped on compact cycles can partially screen the non-dynamical objects. The resulting ``defect group'' is then obtained from the non-dynamical defects modulo such dynamical degrees of freedom.

In the context of M-theory, the SQFT limit necessarily involves dealing with a non-compact geometry $X$ which will contain singularities in the internal geometry. These singularities can often be quite intricate, and can also involve non-isolated components which extend out to the boundary $\partial X$. In such situations, reading off the structure of higher symmetries is necessarily more delicate. One approach to these issues is to explicitly resolve the singular geometry $X$. This was used in \cite{Morrison:2020ool,Albertini:2020mdx,Tian:2021cif} to explicitly construct the resulting (relative) homology cycles, though it rapidly becomes quite complicated to explicitly track all of this data. In \cite{DelZotto:2022fnw} an alternative approach was developed for 5D SCFTs obtained from the orbifold singularities $\mathbb{C}^3 / \Gamma$ for $\Gamma$ a finite subgroup of $SU(3)$. Instead of explicitly performing any resolutions, the data of 1-form symmetries was extracted from the adjacency matrix of the corresponding 5D BPS quiver (see \cite{Closset:2019juk}), as well as from the fundamental group for the (possibly singular) boundary geometry $S^5 / \Gamma$. It was also conjectured in \cite{DelZotto:2022fnw} that the structure of candidate 2-group symmetries is closely correlated with the abelianization of $\Gamma$ itself.

\begin{figure}[t!]
\centering
\includegraphics[scale=0.45, trim = {0cm 3.0cm 0cm 3.0cm}]{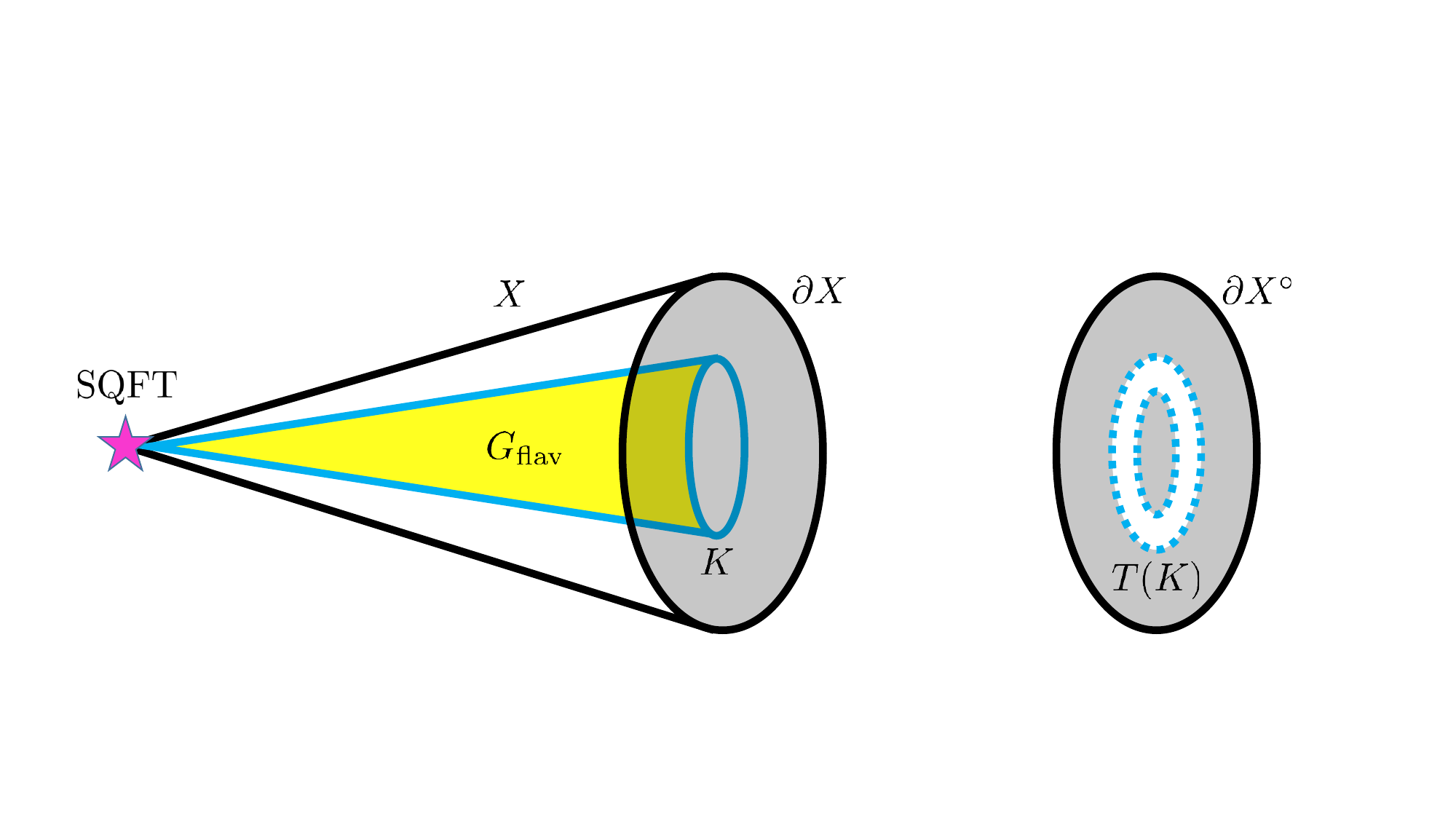}
\caption{Depiction of an SQFT realized at a localized region of a non-compact geometry $X$, with boundary $\partial X$. Flavor branes can extend out to infinity and intersect the boundary along a subspace $K$. Our procedure for extracting the global form of the flavor symmetry, 1-form symmetry and possible 2-group structures involves working with $\partial X^{\circ} = \partial X \backslash T(K)$, where $T(K)$ is a tubular neighborhood of $K$. The Mayer-Vietoris exact sequence then yields the relevant physical structures directly from geometry.}
\label{fig:setup}
\end{figure}

Broadly speaking, orbifold singularities comprise a large class of geometries and also serve as the building blocks of more general
geometries with singularities. As such, they are an excellent theoretical laboratory for the study of higher symmetries.
In this context it is worth noting that in local M-theory models with a weakly coupled Lagrangian description,
the topology of the resulting geometries can always be viewed as glued together from various orbifold geometries. For example,
in the realization of 4D SQFTs via M-theory on a local $G_2$ space, a three-manifold of ADE singularities provides the gauge group and further enhancement in the ADE type at points of the geometry yields localized chiral matter (see e.g. \cite{Acharya:2001gy, Pantev:2009de, Braun:2018vhk, Barbosa:2019bgh}). In many cases, the study of higher symmetries reduces to understanding the analogous question for orbifolds and their subsequent recombination into more general geometries.

Our aim in this paper will be to study the structure of 0-form, 1-form and 2-group symmetries in systems with localized orbifold singularities.
More precisely, we focus on geometries where the flavor group originates from ``flavor branes,''
as generated by local singularities of the form $\mathbb{C}^2 / \Gamma_{ADE}$ for $\Gamma_{ADE} \subset SU(2)$ a finite subgroup of $SU(2)$.
Our analysis will center on the non-abelian contribution to the flavor symmetry. We neglect other possible contributions to the 0-form symmetry such as those coming from the R-symmetry as well as possible $U(1)$ global symmetries, and other discrete symmetry factors.\footnote{There can in principle also be contributions to the 0-form symmetry from isometries and ``accidental'' enhancements at strong coupling, but in what follows we focus purely on localized contributions. When such phenomena occur, additional restrictions apply.} By itself, such an orbifold singularity will realize a 7D super Yang-Mills theory with gauge algebra of ADE type (in the simply laced case).\footnote{We can also treat non-simply laced gauge groups and matter enhancements by including suitable automorphism twists \cite{Bershadsky:1996nh} and / or frozen fluxes \cite{Witten:1997bs, deBoer:2001wca, Tachikawa:2015wka}, but in what follows we mainly focus on the simply laced case unless otherwise stated.} The global structure of the 7D gauge group depends on the specification of a fractional 4-form flux, as captured in the boundary $\partial \mathbb{C}^2 / \Gamma_{ADE} \cong S^3 / \Gamma_{ADE}$ \cite{Albertini:2020mdx,Cvetic:2021sxm}. One of our tasks will be to consider more general configurations with multiple flavor brane factors, and through a process of cutting and gluing, extract the resulting global structure of the 0-form and 1-form symmetries directly from the boundary geometry. This process of cutting and gluing also furnishes a precise prediction for possible 2-group symmetries, as captured by a non-trivial action of the 0-form symmetry on the 1-form symmetries.\footnote{For recent work on the physics of 2-groups, see e.g. \cite{Baez:2005sn, Sati:2008eg, Sati:2009ic, Fiorenza:2010mh, Fiorenza:2012tb, Kapustin:2013uxa,Benini:2018reh, Cordova:2018cvg,Cordova:2020tij,Apruzzi:2021vcu, Bhardwaj:2021wif, Apruzzi:2021mlh, Genolini:2022mpi, DelZotto:2022fnw}.} See figure \ref{fig:setup} for a depiction of the general geometric setup.

Our strategy for reading off this data is as follows. For a general M-theory model on $X$ with flavor branes, even the boundary is singular. The task of understanding higher symmetries then amounts to understanding the category of possible boundary conditions on $\partial X$. Now, in the boundary geometry $\partial X$, the flavor branes are singularities localized on (possibly topologically trivial) sub-manifolds $K \subset \partial X$. Given multiple flavor branes wrapping the subspaces $K_1,...,K_n$, each one will contribute to the corresponding 0-form symmetry. The geometry provides important information on the center of this candidate 0-form symmetry, and we compute it by first deleting the flavor loci, namely by considering the non-compact geometry $\partial X^{\circ} = \partial X \backslash K$ with $K = K_1 \cup ... \cup K_n$. Each local contribution has the form of an ADE singularity fibered over $K_i$, and in this patch we can independently specify a choice of fractional $G_4$-flux which in turn specifies the local structure of the 0-form symmetry. Gluing these local contributions back together via the Mayer-Vietoris sequence then specifies the global 0-form symmetry, which we denote as:
\begin{equation}
G \equiv \widetilde{G} / \mathcal{C},
\end{equation}
with $\widetilde{G}$ the ``naive'' flavor symmetry group which acts on both genuine local operators as well as non-genuine local operators, the latter of which are only defined as the endpoints of line operators.\footnote{Note that all local operators that transform projectively are non-genuine, since otherwise the $G$ should be extended to act faithfully on them, but the converse is not necessarily true. For example, given QCD with $N_f$ fundamental quarks, a quark is not a genuine local operator since it is not gauge invariant but still transforms faithfully under $SU(N_f)$. Other than forming composites, one may attach a quark to a fundamental color Wilson line to give a gauge-invariant configuration which displays what we mean when we say that non-genuine operators only exist as ends of line operators.}

Here, $\mathcal{C}$ is a subgroup of the center of $\widetilde{G}$. In many cases, $\widetilde{G} = G_1 \times ... \times G_n$, with all $G_i$ simply connected, though in some cases, the answer from geometry already anticipates a less naive result.

By a similar token, the 1-form symmetries $\mathcal{A}$ of an orbifold geometry can be read off from a suitably defined notion of $H_2(X,\partial X) / H_{2}(X) \cong \mathrm{Ab}[\pi_{1}(\partial X)]$, even when the boundary geometry has orbifold fixed points. More generally, we can again use Mayer-Vietoris to determine the resulting 1-form symmetry when these orbifold singularities are glued together to produce a more general boundary topology. The specific technique for accomplishing this in the special case of 5D orbifold SCFTs in terms of data associated purely with the orbifold group action was introduced in \cite{DelZotto:2022fnw}.

Putting these pieces together, it is natural to ask whether there is a 2-group structure, as obtained from a non-trivial action of the 0-form symmetry on the spectrum of lines (which are acted on by the 1-form symmetry). In general, this is a challenging question, but for geometries which can be decomposed into local orbifold singularities, we propose a general prescription which passes several non-trivial checks,
at least in those cases where the geometry faithfully captures the 0-form symmetry (no accidental enhancements). As explained in \cite{Bhardwaj:2021wif, Lee:2021crt}, the 1-form symmetry group can be extended by considering additional lines which become equivalent upon quotienting by $\mathcal{C}$. This yields a pair of short exact sequences:
\begin{align}
0 & \rightarrow \mathcal{C} \rightarrow Z_{\widetilde{G}} \rightarrow Z_G \rightarrow 0 \label{absoexacto} \\
0 & \rightarrow \mathcal{A} \rightarrow \widetilde{\mathcal{A}} \rightarrow \mathcal{C} \rightarrow 0 \label{absoexacto2}
\end{align}
where $Z_G$ and $Z_{\widetilde{G}}$ denote the respective centers of the groups $G$ and $\widetilde{G}$.
The first short exact sequence tells us the precise quotient from the ``naive''
simply connected 0-form flavor group $\widetilde{G}$ and its
quotient by a subgroup $\mathcal{C}$ of the center $Z_{\widetilde{G}} \subset \widetilde{G}$, namely we can lift it to a short exact sequence on the groups: $1 \rightarrow \mathcal{C} \rightarrow \widetilde{G} \rightarrow G \rightarrow 1$.
The second short exact sequence tells us the structure of the 1-form symmetries: $\widetilde{\mathcal{A}}$ denotes the
``naive'' 1-form symmetry where we neglect the possibility of lines with endpoints charged under the 0-form symmetry,
and $\mathcal{A}$ is the true 1-form symmetry obtained by corrections associated with the group $\mathcal{C}$.
Applying the Bockstein homomorphism $\beta: H^{2}(B G, \mathcal{C}) \rightarrow H^{3}(B G, \mathcal{A})$ then yields a corresponding Postnikov class. When the image is non-trivial, this detects the presence of a non-trivial 2-group structure.
A sufficient condition for this to occur is that the short exact sequence involving the 1-form symmetries
does not split, namely $\widetilde{\mathcal{A}} \neq \mathcal{A} \oplus \mathcal{C}$.\footnote{In some cases it can happen that $Z_{G}$ is trivial even though $G$ is non-trivial. The important point is that the 2-group structure really makes reference to the Lie group rather than center, namely the Postnikov class detects a non-split 2-group as captured by an element of $H^{3}(B G , \mathcal{A})$. This is still detected by the geometry precisely when the short exact sequence for the true and naive 1-form symmetries of line (\ref{absoexacto2}) is non-split.}

From this perspective, identifying a 2-group symmetry amounts to identifying the geometric origin of each of the terms appearing in lines (\ref{absoexacto}) and (\ref{absoexacto2}). Using the Mayer-Vietoris exact sequence for the different contributions to the singular cohomology, we provide a geometric interpretation for all the terms of this pair of short exact sequences. This same structure can also be directly extracted from orbifold (co)homology. For example, the 1-form symmetries are obtained via the orbifold homology short exact sequence:
\begin{equation}
0 \rightarrow H_{1}^{\mathrm{twist}}(\partial X) \rightarrow H_{1}^{\mathrm{orb}}(\partial X) \rightarrow H_{1}(\partial X) \rightarrow 0,
\end{equation}
where each term is to be identified with the terms in the Pontryagin dual exact sequence:
\begin{equation}\label{eq:SESintro}
0 \rightarrow \mathcal{C}^{\vee} \rightarrow \widetilde{\mathcal{A}}^{\vee} \rightarrow \mathcal{A}^{\vee} \rightarrow 0,
\end{equation}
with $H^{\vee} = \mathrm{Hom}(H,U(1))$ the Pontryagin dual of a group $H$.
Another benefit of working directly in terms of the orbifold (co)homology of the boundary geometry is that it provides an efficient
means of extracting the relevant higher symmetries even when the associated orbifold singularities result from non-abelian group actions.

To test this general proposal, we present a number of examples. As a first case, we return to the 5D orbifold SCFTs
recently considered in \cite{DelZotto:2022fnw}. In this case, we are dealing with orbifold singularities of the form
$\mathbb{C}^3 / \Gamma$ with $\Gamma$ a finite subgroup of $SU(3)$. In many cases, the boundary geometry
has non-trivial flavor brane loci, each of which is locally defined by an ADE singularity. Using our general prescription,
we determine the resulting global 0-form symmetry. Since the 1-form symmetry can also be read
off directly from the data of the orbifold group action \cite{DelZotto:2022fnw}, we can read off all of this data,
including the 2-group symmetry directly from the geometry. We primarily focus on the case of $\Gamma$ an abelian group,
but we also show that the orbifold (co)homology of the boundary geometry $S^5 / \Gamma$ naturally extends to the case of
$\Gamma$ non-abelian as well.

As another class of examples, we consider 5D SQFTs obtained from circle compactification of the partial tensor branch of some
6D SCFTs. In these cases, the flavor brane locus involves a corresponding singular elliptic fibration which amounts to an affine extension of the intersecting homology spheres for a flavor brane. We show that this has no material effect on the resulting higher symmetries and present explicit results in a number of cases. These include the case of $\mathfrak{so}$ gauge theories with an $\mathfrak{sp}$ flavor symmetry algebra and the case of conformal matter on a partial tensor branch. In all these cases, the entire geometry can be decomposed into a collection of orbifold singularities each locally of the form $\mathbb{C}^3 / \mathbb{Z}_k$ for appropriate $k$.

As a more involved example, we also consider the case of SQCD-like theories engineered via M-theory on a local $G_2$ space.
Although an explicit construction of the corresponding special holonomy space remains an outstanding open question, the \textit{topological} data associated with the boundary geometry can be extracted by using the standard type IIA to M-theory lift in which $SU(N)$ D6-branes wrapped on three-manifolds are replaced by three-manifolds of A-type singularities. Similar considerations apply for $SO/Sp$ gauge theories. Using some well-known realizations of such field theories in type IIA brane systems \cite{Ooguri:1997ih, Feng:2005gw}, we determine candidate 0-form, 1-form and 2-group symmetries for these systems.

The rest of this paper is organized as follows. In section \ref{sec:PRESCRIPTION} we present our general strategy for extracting 0-form and higher symmetries from orbifold geometries. In particular, we show that the higher symmetry structure is encoded in the relative homology of the boundary space itself. In section \ref{sec:5Dorb} we apply these general considerations in the case of 5D orbifold SCFTs $\mathbb{C}^3 / \Gamma$. Section \ref{sec:ELLIPTIC} considers a class of 5D SQFTs obtained from elliptically fibered Calabi-Yau threefolds with a singular elliptic fiber. Section \ref{sec:G2} provides a similar analysis in the case of SQCD-like theories engineered from M-theory on local $G_2$ spaces. We present our conclusions and potential areas for future investigation in section \ref{sec:CONC}. Appendix \ref{app:G2LIFT} presents some additional details on the topological data of SQCD-like models realized via $G_2$ spaces. Appendix \ref{app:AppendixSU3} presents some additional aspects of group actions by finite abelian subgroups of $SU(3)$ on $\mathbb{C}^3$.

\section{Symmetries via Cutting and Gluing of Orbifolds} \label{sec:PRESCRIPTION}

In this section we present our general prescription for reading off symmetries of SQFTs engineered from the gluing of
orbifold singularities. We primarily consider M-theory
compactified on a space $X$ which contains various orbifold singularity loci which can potentially extend out to the boundary
$\partial X$. We assume that these are Kleinian singularities $\mathbb{C}^2 / \Gamma_{i}$, with $\Gamma_{i} \subset SU(2)$ a finite subgroup.
In each local patch, the resulting flavor 6-brane specifies a simply connected simple Lie group of ADE type $G_i$. For now, we ignore the possibility of non-simply laced Lie groups, as can result from an outer automorphism twist \cite{Bershadsky:1996nh} and / or frozen fluxes \cite{Witten:1997bs, deBoer:2001wca, Tachikawa:2015wka}, but we return to this issue in sections \ref{sec:ELLIPTIC} and \ref{sec:G2}.

Reading off the global 0-form symmetry for M-theory compactified on $X$ amounts to determining the appropriate way to piece together this local data across all of $\partial X$. Gluing together these building blocks follows directly from an application of the Mayer-Vietoris long exact sequence in singular homology \cite{Mayer, Vietoris}. The main idea is to work in the boundary $\partial X$, excise the flavor brane loci, and then determine possible identifications across multiple flavor factors after taking account of such gluings. More precisely, the Mayer-Vietoris sequence detects the appropriate way to glue together the \textit{centers} of the various $G_i$, as captured by the abelianization $\mathrm{Ab}[\pi_{1}(S^3 / \Gamma_{i})]$, and this is all we really need to extract the global 0-form symmetry. Building on the results of \cite{DelZotto:2022fnw}, this also allows us to read off the global 1-form symmetry, as well as candidate 2-group structures.

The rest of this section is organized as follows. We begin by briefly reviewing the interplay between defects and higher symmetries.
After this, we show how to compute this data using singular (co)homology and the Mayer-Vietoris sequence. We next observe that
precisely the same data can also be read off from the local orbifold (co)homology of the geometry.

\subsection{Defects, Symmetries and 2-Groups}\label{ssec:Defects}

To frame the analysis to follow, in this section we present a brief review of defects and the action of various higher symmetries on such structures. Recall that the defects of a quantum field theory involve heavy non-dynamical objects which extend in some number of directions of the spacetime. For each such extended object, there is a corresponding $p$-form potential which couples to its worldvolume. Some of this charge can be screened by dynamical states of the theory, but importantly, this can leave behind an unscreened remnant.
The defect group
\be
\mathbb{D}=\underset{m}{\bigoplus} \, \mathbb{D}^{(m)}
\ee
consists of equivalence classes $\mathbb{D}^{(m)}$ of mutually non-local $m$-dimensional defects. Defects contributing to such classes are topological and invertible with the latter implying an abelian group structure for their fusion algebra. The equivalence relation declares pairs of $m$-dimensional defects equivalent whenever there exists an $(m-1)$-dimensional defect living at their junction, screening one into the other. Defect groups were introduced in \cite{DelZotto:2015isa} within the context of 6D SCFTs and have been studied from the viewpoint of geometric engineering in \cite{Heckman:2017uxe, GarciaEtxebarria:2019caf, Dierigl:2020myk,Morrison:2020ool,Albertini:2020mdx,Apruzzi:2020zot, Bhardwaj:2020ruf, Bhardwaj:2020avz, DelZotto:2020sop,Gukov:2020btk,Cvetic:2021sxm,Braun:2021sex,Cvetic:2021maf,Debray:2021vob,Apruzzi:2021mlh,Closset:2021lwy,
Apruzzi:2021nmk,Bhardwaj:2021mzl, Tian:2021cif, DelZotto:2022fnw} and references therein.

Turning to the specific case of M-theory on a supersymmetry preserving background geometry $X$, we engineer a corresponding SQFT $\mathcal{T}_{X}$. The defects of $\mathcal{T}_X$ are constructed by wrapping M2- and M5-branes on non-compact cycles of the internal space $X$, and we can accordingly distinguish two contributions to the group of $m$-dimensional defects
\be
\mathbb{D}^{(m)} = \mathbb{D}^{(m)}_{\textnormal{M2}}\oplus \mathbb{D}^{(m)}_{\textnormal{M5}} \,.
\ee
Further, both these subgroups are equally characterized by the non-compact cycles the respective branes wrap and therefore determined by singular homology groups of $X$ \cite{Morrison:2020ool,Albertini:2020mdx}
\be\ba\label{eq:CyclesDefects}
\mathbb{D}^{(m)}_{\textnormal{M2}}&= \frac{H_{3-m}(X,\partial X)}{H_{3-m}(X)} \cong  H_{3-m-1}(\partial X)\big|_\textnormal{trivial} \\
\mathbb{D}^{(m)}_{\textnormal{M5}}&= \frac{H_{6-m}(X,\partial X)}{H_{6-m}(X)} \cong  H_{6-m-1}(\partial X)\big|_\textnormal{trivial}\,.
\ea\ee
Here $|_\textnormal{trivial}$ restricts the group of torsion cycles to the subgroup trivializing under the inclusion $\partial X \hookrightarrow X$ with the isomorphism taken from the long exact sequence in relative homology for the pair $(X,\partial X)$ \cite{Morrison:2020ool,Albertini:2020mdx}. We remark that in all cases we consider, the group is always a finite order group, i.e. it is purely torsion.

Theories with non-trivial defect groups have phase ambiguities and a vector of partition functions. As such, they are more properly viewed as relative theories \cite{Witten:1998wy, Seiberg:2011dr, Freed:2012bs, Gaiotto:2014kfa, DelZotto:2015isa, Gukov:2020btk}.\footnote{One can view these theories as the edge modes of a bulk TQFT which has a Hilbert space of states which is non-trivial.}
These same phase ambiguities also appear in the braid relations between defect operators, as captured by the Dirac pairing:
\be
\langle\, \cdot \,, \cdot \, \rangle\,:\quad \mathbb{D}^{(m)} \times  \mathbb{D}^{(n-m-2)}~\rightarrow~ \mathbb{R}/\mathbb{Z}
\ee
where $n=\textnormal{dim}\,X$. Two defects are mutually local whenever they pair trivially. Well-defined or absolute theories are obtained by restricting the spectrum of defects to a maximal subset $\mathbb{P}\subset \mathbb{D}$ of mutually local defects.  Such maximal sets of mutually local defects are referred to as polarizations. Geometrically the Dirac pairing takes the form of the linking pairing on the boundary homology torsion groups of equation \eqref{eq:CyclesDefects}. The choice of polarization $\mathbb{P} \subset \mathbb{D}$ determines the global structures of a theory by fixing its spectrum of extended operators. The higher-form symmetry groups $\mathcal{A}_{\mathrm{higher}}$ of such absolute theories are then determined by the Pontryagin dual:
\be
\mathcal{A}_{\mathrm{higher}} = \mathbb{P}^{\vee}.
\ee
In what follows, we shall mainly focus on the case of a preferred electric polarization dictated by wrapped M2-branes, and so will often
keep the polarization data implicit.

This is to be contrasted with the case of 0-form global symmetries which act on the local operators of the theory.
If we do not distinguish between genuine local operators and those which simply specify the endpoints of line operators, we get a ``naive'' flavor symmetry group $\widetilde{G}$, with Lie algebra $\mathfrak{g}$. There can in principle be additional discrete factors for the global 0-form symmetry, but we neglect these as well as possible $U(1)$ symmetry factors. For the most part, we also neglect possible non-trivial mixing with the R-symmetry (when present) of an SQFT. Now, it can often happen that the actual flavor symmetry group is itself screened. This is to be expected due to the presence of extended objects such as line operators. Consequently, the actual flavor symmetry may be a quotient of $\widetilde{G}$ by a subgroup $\mathcal{C} \subset Z_{\widetilde{G}}$ of the center of $\widetilde{G}$. The flavor symmetry is then given by \cite{Bhardwaj:2021ojs, Apruzzi:2021vcu}:
\begin{equation}
G = \widetilde{G} / \mathcal{C}.
\end{equation}

Higher-form symmetries can intertwine to higher-group structures, and in this work we focus on 2-groups, as captured by a non-trivial interplay of between a 0-form symmetry and 1-form symmetry of the theory. The data of a 2-group is specified by (see e.g. \cite{Benini:2018reh}):\footnote{
For additional foundational work on 2-groups, see e.g. \cite{Pantev:2005zs, Pantev:2005rh, Pantev:2005wj, Baez:2005sn, Sati:2008eg, Sati:2009ic, Fiorenza:2010mh, Fiorenza:2012tb, Kapustin:2013uxa, Sharpe:2015mja}.}
\begin{itemize}
\item Flavor symmetry group $G$
\item 1-form symmetry group $\mathcal{A}$
\item Symmetry action $\rho: G \rightarrow \textnormal{Aut}\big(\mathcal{A}\big)$
\item Postnikov class $P \in H^3(BG,\mathcal{A})$.
\end{itemize}
The Postnikov class $P$ determines an obstruction to turning on backgrounds for the flavor symmetry independently from those of 1-form symmetries. Since the influence of $G$ on the 2-group structure is only through its discrete center, it is often enough to restrict attention to just the center $Z_G \subset G$. This is all to the good, because what we can actually detect in the geometry is precisely $Z_G$, with the rest of $G$ being obtained from physical (i.e. non-geometric) ingredients such as wrapped M2-branes.

This 2-group structure can be packaged in terms of various equivalence classes of line operators in $\mathcal{T}_X$. This is detailed in Appendix A of \cite{Lee:2021crt} (see also \cite{Bhardwaj:2021ojs}),
which we now briefly review, with a few specializations of importance for our geometric analysis.
In $\mathcal{T}_X$, let us define the following sets which are abelian groups under line operator fusion:
\begin{eqnarray*}
  \mathcal{A}^{\vee} & \equiv & \{\mathrm{line \; operators \; modulo \; local \; operator \; interfaces}\} \\
  \widetilde{\mathcal{A}}^{\vee} & \equiv &\{\mathrm{line \; operators \; modulo \; local \; operator \; interfaces \; with \; faithful \; action \; under} \; G\} \\
  \mathcal{C}^{\vee} & \equiv & \mathrm{ker}(\widetilde{\mathcal{A}}^{\vee} \rightarrow \mathcal{A}^{\vee}).
\end{eqnarray*}
By construction, we have the following short exact sequence for the Pontryagin dual objects:\footnote{Working in terms of the original groups, we also have the short exact sequence (all arrows reversed):
$0 \rightarrow \mathcal{A} \rightarrow \widetilde{\mathcal{A}} \rightarrow \mathcal{C} \rightarrow 0.$}
\be\label{eq:SESlines}
0 \rightarrow \mathcal{C}^{\vee} \rightarrow \widetilde{\mathcal{A}}^{\vee} \rightarrow \mathcal{A}^{\vee} \rightarrow 0.
\ee
Note that the class of line operators in $\mathcal{C}^\vee$ can be stated as:\footnote{Equivalently, $\mathcal{C}^\vee$ consists of those line operators which are screened by local operators in projective representations of $G$.}
\begin{equation}\label{eq:chat}
  \mathcal{C}^{\vee} = \{\mathrm{line \; ops \; which \; can \; end \; on \; local \; ops \; in \; a\;  projective \; reps \; of} \;  Z_G  \}
\end{equation}
where here $Z_G$ is the center of $G$ and projective simply  means ``not faithful''. One can then consider the smallest extension of $Z_G$ such that projective representations of local operators in $\mathcal{T}_X$ transform in a faithful representation, and this is simply $Z_{\widetilde{G}}$. This can be rephrased as the short exact sequence
\be\label{eq:CG}
0~\xrightarrow[]{} ~  \mathcal{C}  ~\xrightarrow[]{} ~ Z_{\widetilde{G}} ~\xrightarrow[]{} ~ Z_{G}  ~\rightarrow ~0\,.
\ee
This is how the geometry encodes the short exact sequence for the corresponding Lie groups:
\begin{equation}
1 ~\xrightarrow[]{} ~  \mathcal{C}  ~\xrightarrow[]{} ~ \widetilde{G} ~\xrightarrow[]{} ~ G  ~\rightarrow ~1\,,
\end{equation}
and we denote its extension class as $w_2\in H^2(B G, \mathcal{C})$.
From this, we we can form the long exact sequence:
\be\label{eq:longboi}
 0 ~\xrightarrow[]{}  ~ \mathcal{A} ~\xrightarrow[]{}  ~  \widetilde{\mathcal{A}} ~\xrightarrow[]{} ~ \widetilde{G} ~\xrightarrow[]{} ~ G  ~\rightarrow ~ 1\,,
\ee
where the analogous extension class is given by $\beta(w_2)\in H^3(B G, \mathcal{A})$ where:
\begin{equation}
\beta : H^{2}(B G , \mathcal{C}) \rightarrow H^{3}(B G, \mathcal{A})
\end{equation}
is the Bockstein homomorphism associated to the Pontryagin dual of line (\ref{eq:SESlines}). Observe that $\beta (w_2)$ is then the Postnikov class, $P$, mentioned earlier as a key defining feature of a 2-group structure. As a final comment, the astute reader will notice that in comparison with other discussions in the literature, the geometry really detects restrict attention to the center of the Lie groups $G$ and $\widetilde{G}$. The only subtlety here is that the 2-group really makes reference to the full $G$ and $\widetilde{G}$. The main point is that the geometry provides us with the correct answer for $G$, and we can still, via physical considerations thus determine the corresponding 2-group structure. Indeed, in such situations the prediction of a non-split 2-group follows from having a non-split short exact sequence for the 1-form symmetries, namely $0 \rightarrow \mathcal{A} \rightarrow \widetilde{\mathcal{A}} \rightarrow \mathcal{C} \rightarrow 0$.

\subsection{Flavor Symmetry and 2-Groups via Singular Homology}
\label{sec:SingHom}

Having reviewed the interplay between defects, higher symmetries and 2-groups, we now turn to the core task of
extracting this information from a given M-theory background $X$. Our aim will be to extract both the global 0-form symmetry, as well as the 1-form symmetry, and possible intertwining due to the 2-group. We confine our discussion to flavor symmetries localized on geometrized 6-branes, i.e. M-theory singularities which are locally of the form $\mathbb{C}^2 / \Gamma_{i}$ for $\Gamma_{i}$ a finite subgroup of $SU(2)$.

Our aim will be to determine these global structures directly from the singular homology of the asymptotic boundary $\partial X$, thereby complementing a similar analysis for the 1-form symmetry presented in \cite{Morrison:2020ool,Albertini:2020mdx}. When cutting out the orbifold loci,\footnote{We take $K$ to support an ADE singularity.} $K\subset \partial X$, there are several types of (relative) homology cycles that one may consider. The goal of this section is then to establish a dictionary between these various homology groups and the equivalence classes of $\mathcal{T}_X$-line operators by means of wrapped M2-branes. The 2-group structure and flavor symmetry\footnote{Strictly speaking, this is not expected to capture 0-form symmetry from isometries, nor from flavor enhancements.} then appear naturally in the geometric definitions. We leave implicit the extension to the case of wrapped M5-branes since it is quite similar to
the M2-brane analysis.

We begin by introducing notation. We denote the non-compact components of each flavor brane locus on $\partial X$ by $K_i$ and associate with each a flavor symmetry algebra $\mathfrak{g}_i$ of simple Lie algebra type. For ease of exposition, we focus on the case where this Lie algebra is of ADE type, but our method naturally extends to further twists by outer automorphisms, a point we return to in sections \ref{sec:ELLIPTIC} and \ref{sec:G2}. The boundary singularities $\partial X\cap K_i$ are assumed to be disjoint. We further require the first homology group of $X\cap K_i$ to be torsion-free. We define the smooth boundary to be
\be
\partial X^\circ= \partial X \setminus \cup_i K_i
\ee
and denote a tubular neighborhood of the boundary singularities by $T(K)=\cup_{i\,} T(K_i) \subset \partial X$.

We now discuss some immediate consequences of these restrictions. First, we note that $\partial X^\circ$ is connected. Next we consider the lift of the embedding $j: \partial X^\circ \hookrightarrow \partial X$ to homology and notice that the kernel of the map
\be
j_1\,:\quad H_1(\partial X^\circ)~\rightarrow ~ H_1(\partial X)
\ee
is a torsion subgroup of $\textnormal{Tor}\,H_1(\partial X^\circ)$. By assumption, the normal geometry of $K_i$ is locally of the form  $\mathbb{C}^2/\Gamma_{i}$ for $\Gamma_{i} \subset SU(2)$ a finite subgroup, and therefore we have $K_i$ linked by $S^3/\Gamma_{i}$.
The only 1-cycles created by the excision of $K_i$ are therefore torsional. Finally we note that the tubular neighborhood $T(K_i)$ deformation retracts to $K_i$ and therefore also the first homology group of $T(K_i)$ is torsion-free.

Together, these pieces can be packaged into the Mayer-Vietoris sequence:\footnote{Here we have defined the following inclusions: $\ell:T(K)\hookrightarrow \partial X$, $\iota_A: \partial X^\circ\cap T(K)\hookrightarrow \partial X^\circ$, and $\iota_B: \partial X^\circ\cap T(K)\hookrightarrow T(K)$. We further define $\iota=(\iota_A,\iota_B)$.}
\be\label{eq:MV1}
\dots~\xrightarrow[]{\,\partial_{k+1} \,} ~H_{k}\big(\partial X^\circ\cap T(K) \big)  ~\xrightarrow[]{\,\iota_{k} \,} H_k\big(\partial X^\circ\big)\oplus H_k\big(T(K)\big) ~\xrightarrow[]{j_k-\ell_k }  ~  H_k\big(\partial X\big) ~\xrightarrow[]{\,\partial_k\, }  ~\dots.
\ee
Our claim is that each term in this sequence admits a physical interpretation in terms of the short exact sequences of lines (\ref{eq:SESlines}) and (\ref{eq:CG}) and introduced in our review of defects, symmetries and 2-groups. Consequently, we can give a fully geometric interpretation of these structures. We now give our proposal and then establish how it descends to the 2-group structure of $\mathcal{T}_X$.

\begin{figure}[t!]
\centering
\includegraphics[width=0.8\textwidth, angle=0]{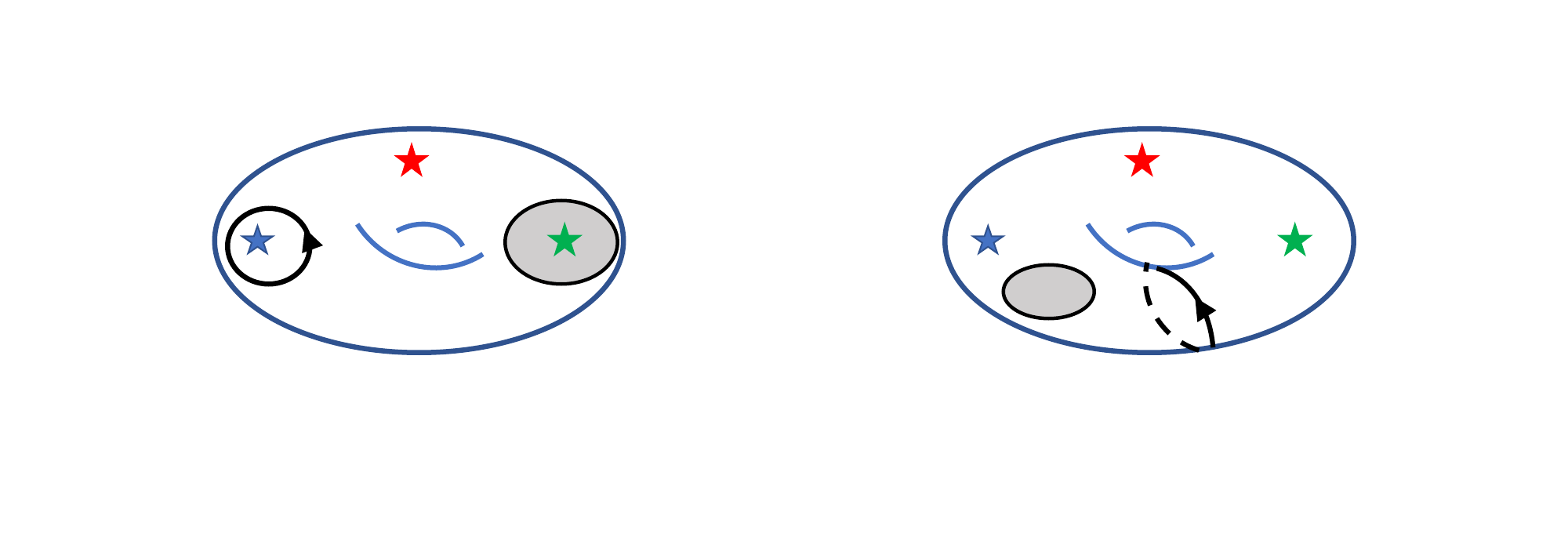}
\caption{On the left, we depict $\partial X$ with various orbifold singularities (stars) and an element in $\mathcal{C}^\vee$ (black loop). The gray disk bounds an element of $\mathcal{C}^\vee$, and shows that this element is topologically trivial under the inclusion of the orbifold loci back into $\partial X^\circ$. This motivates the identification $\textnormal{Tor\,ker}\,(j_1)=\mathcal{C}^\vee$. On the right, we depict an element in $\mathcal{A}$ with a black loop, and illustrate a gray disk to highlight that the topological equivalence relation inherent in $H_1(\partial X)$ is the same as that of $\mathcal{A}^\vee$ when wrapping $M2$-branes.}
\label{fig:M2lineoperators}
\end{figure}

\noindent {\bf Flavor Symmetry:} We begin by showing how the geometry encodes the global form of the flavor symmetry, namely
how the short exact sequence (\ref{eq:CG}) can be defined using the boundary geometry of $\partial X$.
The simplest of the three objects to identify in (\ref{eq:CG}) is the center $Z_{\widetilde G}$. For each component $K_i$
of the flavor symmetry locus, we have already seen that there is a linking with an $S^3 / \Gamma_i$. The naive center of each factor is then:
\begin{equation}
Z_{i} = \mathrm{Tor} H_{1}\big(\partial X^\circ\cap T(K_i) \big)^{\vee}.
\end{equation}
Indeed, it is well-known that the geometry detects the center of the corresponding simply-laced Lie group $G_i$ via the abelianization $\mathrm{Ab}[\mathrm{\pi}_1(S^3 / \Gamma_i)] = \mathrm{Ab}[\Gamma_{i}]$. In many cases, then, the naive flavor group will just be a product of these simply connected Lie group factors. We note that in some cases, the linking of flavor branes in the geometry already detects a ``slightly less naive'' answer than the one obtained from simply taking the product of simply connected flavor symmetry factors, a point we return to in sections \ref{sec:ELLIPTIC} and \ref{sec:G2}. In any case, the center of the naive flavor group is:
\begin{equation}
Z_{\widetilde{G}} = Z_{1} \oplus ... \oplus Z_{n},
\end{equation}
and we have the geometric identification:
\be
\boxed{
Z_{\widetilde{G}}^{\vee} = \bigoplus_i  \textnormal{Tor}\, H_{1}\big(\partial X^\circ\cap T(K_i) \big).}
\ee
How is this encoded in the M-theory degrees of freedom? Physically, an M2-brane wrapped on a cycle $\gamma_i \in \textnormal{Tor}\, H_{1}\big(\partial X^\circ\cap T(K_i) \big)$ for a given $i$, is part of the twisted sector of the ADE locus labelled by $i$. This implies that it is part of a representation of $Z_{i}$. If the worldvolume of the M2-brane is
\begin{align}\label{eq:m2worldvol}
  & \{\mathrm{Cone}(\gamma_i) \}\times L
\end{align}
where $L$ is a line in the spacetime of the SQFT, and $\mathrm{Cone}(\gamma_{i})$ amounts to extending the cycle $\gamma_{i}$ in $\partial X$ to the interior singularity where the SQFT is localized. These M2-branes are then (center) flavor Wilson lines with center representation specified by its geometric definition. Each of these flavor Wilson lines can be screened by a local operator in a representation of $Z_i$, which is clear geometrically by taking the worldvolume of the M2 to be
\begin{equation}\label{eq:M2voldisk}
  \textnormal{Cone}(D)\times [0,\infty)/\sim
\end{equation}
where $\sim$ identifies $\textnormal{Cone}(\partial D)$ as a fiber over $(0,\infty)$ and $\textnormal{Cone}(D\backslash \partial D)$ as a fiber over $\{0\}$. This is illustrated by the gray disk on the left-side of figure \ref{fig:M2lineoperators}. Considering the entirety of $\partial X$, there may be other disks with boundary $\gamma_i$ which are in trivial representations of $Z_i$, and possibly non-trivial representations of $\bigoplus_{i\neq j} Z_j$. This motivates calling $\bigoplus_i Z_{i}$ the center of the \textit{naive flavor group} because all non-genuine local operators built this way from M2 branes transform faithfully under $\bigoplus_i Z_{i}$ and we generally expect that a subgroup will transform projectively under a finite quotient $\bigoplus_i Z_{i}/\mathcal{C}$ which coincides with the center of the true flavor symmetry $Z_G$. Our goal then is to give a geometric interpretation to this quotient and understand how it connects with our discussion of different equivalence relations of line operators.

We claim that (the Pontryagin dual of) the subgroup $\mathcal{C} \subset Z_{\widetilde{G}}$ of the center of $\widetilde{G}$ is encoded in the geometry via:
\be\label{eq:chatgeo}
\boxed{\mathcal{C}^\vee = \textnormal{Tor}\,H_{1}\big(\partial X^\circ\cap T(K) \big)/ \textnormal{ker}(\iota_1)=\textnormal{Tor}\,\textnormal{ker} (j_1)\,}
\ee
where here, we have the ``gluing maps'' of the Mayer-Vietoris sequence $\iota_{1}: H_{1}(\partial X^\circ \cap T(K)) \rightarrow H_{1}(\partial X^{\circ})  $ and $j_{1}: H_{1}(\partial X^{\circ}) \rightarrow H_{1} (\partial X)$. Indeed the second equality follows from the general way in which a long exact sequence can be repackaged in terms of a collection of short exact sequences.\footnote{\label{foot:les} In a long exact sequence $\dots~\xrightarrow[]{\,f_1 \,} ~A_1  ~\xrightarrow[]{\,f_2 \,} A_2 ~\xrightarrow[]{ \,f_3 \,} ~\dots$, we can always write down an associated short exact sequence for each element. For example, for $A_1$ this is  $0~\xrightarrow[]{} ~K_1  ~\xrightarrow[]{} A_1 ~\xrightarrow[]{ } K_2 ~\xrightarrow[]{ }  ~0$ where $K_i=\textnormal{ker}(f_{i+1})=\textnormal{im}(f_i)=\textnormal{coker}(f_{i-1})$ (the last equality is only true for abelian groups).} To interpret this physically, we will study $\textnormal{Tor}\,\textnormal{ker} (j_1)$ which means torsion elements of $H_1(\partial X^\circ)$ which trivialize upon the inclusion $j:\partial X^\circ\hookrightarrow \partial X$. So similar to our discussion of the naive center flavor symmetry, given a class $\gamma\in \textnormal{Tor}\,\textnormal{ker} (j_1)$ we can wrap an M2-brane on $\{\mathrm{Cone}(\gamma) \}\times L$ to get a (true) flavor Wilson line operator in $\mathcal{T}_X$.

To establish equation \eqref{eq:chatgeo}, it then suffices to demonstrate that:
\begin{itemize}\label{it:claim}
  \item An M2-brane wrapped on a cycle in $\textnormal{Tor}\,\textnormal{ker} (j_1)$ should become, in $\mathcal{T}_X$, a line operator that cannot be screened by a $\oplus_i Z_{i}$ singlet local operator, but must be able to be screened by a local operator which is in a non-trivial representation of $\oplus_i Z_{i}$.
\end{itemize}
For elements in $\textnormal{Tor}\,\textnormal{ker} (j_1)$, the corresponding line operators cannot be screened by a naive flavor singlet local operators, but must be screened by a local operator which is in a non-trivial representation of $\oplus_i Z_{F_i}$. Such a local operator, $\mathcal{O}_{\mathrm{proj}}$, would then not transform faithfully under the true flavor symmetry, because if it did then we could multiply the endpoint of the line by some genuine local operator $\mathcal{O}_{gen.loc.}$ to obtain a singlet under $Z_{\widetilde{G}} = \oplus_i Z_{i}$ which contradicts the hypothesis. Pictorially, the gray disk on the left-side of figure \ref{fig:M2lineoperators} must always pass through the ADE singularity, i.e. the local operator screening the flavor Wilson line maintains its $Z_{\widetilde{G}} = \oplus_i Z_{i}$-charge. Since cycles in $\textnormal{Tor}\,\textnormal{ker} (j_1)$ are only of this form, we recover the claim above.


Moving on to our geometric proposal for the flavor symmetry, we can already see from the first equality in \eqref{eq:chatgeo} and the Pontryagin dual of the short exact sequence (\ref{eq:CG}) that the center of the flavor symmetry is:\footnote{Again, this neglects symmetry enhancements and isometries.}
\be\label{eq:flavsymdef}
\boxed{Z_{G}^\vee=\textnormal{Tor} \; \textnormal{ker}(\iota_1) \,},
\ee
that is, this is the center flavor symmetry which survives after quotienting by $\widetilde{G}$ by $\mathcal{C}^{\vee}$.

\medskip

\noindent {\bf 1-Form Symmetry and 2-Group:}
Let us now turn to the higher symmetries of the system. For ease of exposition we focus on the ``electric'' contribution to the 1-form symmetries, i.e. those coming from M2-branes wrapped on boundary 1-cycles.\footnote{This subtlety is not much of an issue in 5D SQFTs, but it does make an important appearance in $ D \leq 4 $ spacetime dimensions}. Now, from our previous discussion, we expect to get line operators in $\mathcal{T}_{X}$ from M2-branes wrapped on 1-cycles of the boundary geometry. The (Pontryagin dual of the) ``naive'' collection of line operators $\widetilde{\mathcal{A}}$ is then simply:
\begin{equation}\label{eq:hath1}
 \boxed{ \widetilde{\mathcal{A}}^{\vee} = \textnormal{Tor} \; H_1(\partial X^\circ).}
\end{equation}
Next we note that the number of connected components of $T(K)$ agrees with those of $\partial X^\circ \cap T(K)$. The Mayer-Vietoris sequence \eqref{eq:MV1} is therefore exact in degree zero. Using general properties of long exact sequences (see footnote \ref{foot:les}), one extracts two short exact sequences:
\be\label{eq:SES4}
0~\xrightarrow[]{} ~  \textnormal{ker}(\iota_1) ~\xrightarrow[]{} ~ H_{1}\big(\partial X^\circ\cap T(K) \big) ~\xrightarrow[]{} ~ \frac{H_{1}\big(\partial X^\circ\cap T(K) \big)}{\textnormal{ker}(\iota_1)}  ~\rightarrow ~0\,,
\ee
\be\label{eq:SES5}
0~\xrightarrow[]{} ~  \frac{H_{1}\big(\partial X^\circ\cap T(K) \big)}{\textnormal{ker}(\iota_1)} ~\xrightarrow[]{\,\iota_{1} \,} ~ H_1\big(\partial X^\circ\big)\oplus H_1(T(K)) ~\xrightarrow[]{j_1- \ell_1 } ~ H_1(\partial X)  ~\rightarrow ~0\,.
\ee
Our interest of course is in the restriction of the above short exact sequences to short exact sequences of their torsion subgroups, which is possible due to our assumption that $T(K)$ is torsion-free. Note that $\iota_1$ is a group homomorphism mapping torsion subgroups onto torsion subgroups. In the present context, the (electric) 1-form symmetry is \cite{Albertini:2020mdx, Morrison:2020ool}:
\be
\mathcal{A} \cong \bigg(\textnormal{Tor}\, H_1(\partial X)\bigg)^\vee .
\ee
Collapsing the two short exact sequences (\ref{eq:SES4}) and (\ref{eq:SES5}), we have the following long exact sequence
\be\label{eq:SES3}
0~\xrightarrow[]{} ~  \textnormal{ker}(\iota_1) ~\xrightarrow[]{} ~ H_{1}\big(\partial X^\circ\cap T(K) \big) ~\xrightarrow[]{} ~ H_1\big(\partial X^\circ\big)\oplus H_1(T(K)) ~\xrightarrow[]{j_1- \ell_1 } ~ H_1(\partial X) ~\rightarrow ~0\,.
\ee
which fully characterizes the 2-group geometrically. Indeed, the 2-group structure (including the global form of the flavor symmetry and 1-form symmetry) is given by the long exact sequence of line (\ref{eq:longboi}):
\be
 0 ~\xrightarrow[]{}  ~ \mathcal{A} ~\xrightarrow[]{}  ~  \widetilde{\mathcal{A}} ~\xrightarrow[]{} ~ \widetilde{G} ~\xrightarrow[]{} ~ G  ~\rightarrow ~ 1\,,
\ee
and the geometry detects the centers of the Lie groups $\widetilde{G}$ and $G$.
The extension class of this sequence is classified by the Postnikov class $\beta (w_2) \in H^3(BG,\mathcal{A})$, where $w_2$ is the extension class of the short exact sequence (\ref{eq:SES4}), and $\beta$ is the Bockstein homomorphism associated to line (\ref{eq:SES5}).

Summarizing, we have now given a geometric characterization of the 0-form, 1-form and 2-group symmetries.

\subsection{Comparison with Orbifold Homology}

In the previous subsection we gave a general prescription for how to read off the flavor group and higher symmetries of $\mathcal{T}_{X}$ directly from the singular homology of $\partial X$. Rather than directly performing such excisions in the boundary geometry it is natural to ask whether we can replace some of these structures by a suitable notion of an orbifold (co)homology theory. Our goal will be to show why this is to be expected on general grounds, as well highlight an example. We will present a more involved example which makes use of orbifold homology in section \ref{sec:5Dorb}. These examples will explicitly show that the first orbifold homology group carries important physical data, which in case of wrapping M2 branes gives the naive 1-form symmetry, and leave a more detailed study of the physics of the higher orbifold (co)homology groups for future work.

Recall that when $\partial X$ is smooth, the (electric) 1-form defect group is obtained from wrapped M2-branes:
\begin{equation}\label{eq:defectgrp1}
  \mathbb{D}^{(1)}_{\mathrm{elec}} = \textnormal{Tor}\,  H_{1}(\partial X) \vert_{\mathrm{trivial}}.
\end{equation}
The presence of orbifold singularities in $\partial X$ introduces an interplay between this group and the 0-form flavor symmetry $G$. A priori, when the $(n-1)$-manifold $\partial X$ has orbifold singularities, there are two natural modifications that one can consider. The first is to excise the singularity and assign appropriate boundary conditions to fields along the $(n-2)$-dimensional manifold which surrounds the singular locus. This was the guiding principle of the previous section. The second is to generalize $H_1$ to a suitable orbifold homology, $H_1^{\mathrm{orb}}$, which captures quotient data in addition to the standard topological 1-cycles. One could then ask what new information such an ``orbifold defect group'',
\begin{equation}\label{eq:}
  \mathbb{D}^{(1)}_{\mathrm{orb}}=\textnormal{Tor}\,  H^{\mathrm{orb}}_{1}(\partial X),
\end{equation}
would contain? In quite general terms, this latter approach would be assigning a geometric engineering Hilbert space to the $(n-1)$-dimensional orbifold boundary manifold, while the former approach (i.e. that of the previous subsection) would be assigning a geometric engineering Category of Boundary Conditions to the $(n-1)$-boundary manifold which itself has a $(n-2)$-boundary. One then gets a geometric engineering Hilbert space by choosing boundary conditions appropriate for a given orbifold.\footnote{In categorical language, this would be a Hom-vector space between two objects in a category labeled by $(n-2)$-manifolds with boundary conditions.} Therefore, seeing how these two approaches might complement each other is clearly of interest.

To understand what we mean by orbifold homology, let us first consider how to define it for the case when the orbifold is a global quotient, $X/H$, what Thurston refers to as a ``good'' orbifold \cite{Thurston}, where $X$ is a space acted on by a finite group $H$. We then have the following definition\footnote{To explain the notation, $EH$ is the universal principle $H$-bundle over $BH$. Since the action of $H$ on $X$ induces a map $X\rightarrow BH$, both $X$ and $EH$ are equipped with natural maps to $BH$ so $\times_H$ simply means their relative product with respect to these maps.}
\begin{equation}
  H_*^{\mathrm{equiv}}(X/H) \equiv H_*(EH\times_H X )
\end{equation}
although the left-hand side is often referred to as $H_*^{\mathrm{equiv}}(X)$, this notation makes the comparison to orbifold homology more clear. Now from the natural projection $EH\times_H X\rightarrow X/H$, we have a projection on first homology groups
\begin{equation}\label{eq:projorb1}
   p:H^{\mathrm{equiv}}_1(X/H)\rightarrow H_1(X/H)
\end{equation}
and dually an inclusion in first cohomology: $H^1_{\mathrm{equiv}}\hookrightarrow H^1(X/H)$. We define
\begin{equation}\label{eq:projorb3}
 H^{\mathrm{twist}}_1 \equiv \textnormal{ker}\; p
\end{equation}
as the twisted (i.e. fractional) cycles. One can then rephrase (\ref{eq:projorb1}) as the short exact sequence:
\begin{equation}\label{eq:projorb2}
 0\; \rightarrow \; \textnormal{ker}\; p \; \rightarrow  \; H^{\mathrm{equiv}}_1(X/H)\; \rightarrow \; H_1(X/H)\; \rightarrow \; 0.
\end{equation}
By inspection, this is quite similar to the exact sequence of line (\ref{eq:SES5}) in singular homology,
which we reproduce here for convenience of the reader:
\begin{equation}
0~\xrightarrow[]{} ~  \frac{H_{1}\big(\partial X^\circ\cap T(K) \big)}{\textnormal{ker}(\iota_1)} ~\xrightarrow[]{\,\iota_{1} \,} ~ H_1\big(\partial X^\circ\big)\oplus H_1(T(K)) ~\xrightarrow[]{j_1- \ell_1 } ~ H_1(\partial X)  ~\rightarrow ~0\,.
\end{equation}
In fact, these two sequences turn out to be equivalent thanks to unpublished work by Thurston which can be used to define an appropriate
orbifold homotopy group $\pi^{\mathrm{orb}}_1$ . Moreover, these definitions carry over to the case when the orbifold singularities are
not globally defined quotients, precisely the situation we need for the present work (see \cite{Thurston} as well as p.12 of \cite{Davis}).

A key result is that for an orbifold, $\partial X$, with orbifold loci $K$ that have codimension greater than 2, $\pi^{\mathrm{orb}}_1$ has the  presentation $\pi^{\mathrm{orb}}_1(\partial X)=\pi_1(\partial X^\circ)$ where $\partial X^\circ \equiv \partial X\backslash K$. Note that there is an orbifold version of Hurewicz theorem, $\textnormal{Ab}[\pi^{\mathrm{orb}}_1] = H^{\mathrm{orb}}_1$, which relates orbifold homology to singular homology (see \cite{moerdijk1997simplicial}):
\begin{equation}\label{eq:h1orb}
  H^{\mathrm{orb}}_1(\partial X)=H_1(\partial X^\circ).
\end{equation}
An important application of this relation is when $\partial X$ is a global quotient of a simply-connected space by some group $\Gamma$, where instead of understanding how to cut out singular loci and computing via the Mayer-Vietoris sequence, one can simply use the fact that $H^{\mathrm{orb}}_1(\partial X)=\textnormal{Ab}[\Gamma]$.

We finish with discussing a simple example. Consider an \textit{isolated} singularity of the form $\mathbb{C}^n/\Gamma$ where $\Gamma \in SU(n)$ ($n\geq 2$). From the relation to equivariant homology, $H^{\mathrm{orb}}_1(\mathbb{C}^n/\Gamma)=H_1(B\Gamma)=\textnormal{Ab}[\Gamma]$, where the last equality is a standard result which follows from the assumption that the group action of $\Gamma$ is fixed point free.
This agrees with $H_1\big((\mathbb{C}^n/\Gamma) \; \backslash \{ 0\}\big)$ since
\begin{equation}\label{eq:manipulation}
  (\mathbb{C}^n/\Gamma) \; \backslash \{ 0\}\cong \mathbb{R}\times S^{2n-1}/\Gamma
\end{equation}
and $H_1$ of the righthand side is $\textnormal{Ab}[\Gamma]$.

When $\Gamma$ has a fixed point locus, suitable modifications of these expressions are required, but they can again be handled using orbifold homology. We now turn to some examples of this sort in the context of 5D SCFTs engineered via orbifold singularities.

\section{5D SCFTs from $\mathbb{C}^3/\Gamma$} \label{sec:5Dorb}

Having presented a general prescription for reading off symmetries via cutting and gluing of orbifold singularities, we now turn to some explicit examples. In this section we consider 5D SCFTs $\mathcal{T}_X$ engineered in M-theory by the Calabi-Yau threefold $X=\mathbb{C}^3/\Gamma$ with finite $\Gamma\subset SU(3)$. Recently, the higher-form symmetries of such 5D orbifold SCFTs were studied in \cite{Tian:2021cif, DelZotto:2022fnw} (see also \cite{Albertini:2020mdx, Morrison:2020ool}). Our aim will be to show how the considerations of section \ref{sec:PRESCRIPTION} recover these structures and also enable us to extract the global form of the flavor symmetry localized on 6-branes and the intertwined 2-group structure. As a general comment, the orbifold $\mathbb{C}^3 / \Gamma$ may also include contributions to the flavor symmetry from isometries, as well as possible non-trivial mixing between flavor symmetries and the $SU(2)$ R-symmetry of the SCFT. Our analysis will not include such subtleties, but it would be interesting to study them in the present framework. For various aspects of flavor symmetries in 5D SCFTs, see e.g. \cite{DelZotto:2017pti, Jefferson:2017ahm, Jefferson:2018irk, Apruzzi:2018nre, Closset:2018bjz, Closset:2019juk, Apruzzi:2019vpe, Apruzzi:2019opn, Apruzzi:2019enx, Apruzzi:2019kgb, Bhardwaj:2019fzv, Bhardwaj:2020ruf, Bhardwaj:2020avz, Apruzzi:2021vcu, Tian:2021cif, DelZotto:2022fnw}.

To frame the discussion to follow, we first recall that the 1-form symmetry is captured by the singular homology group $H_{1}(\partial X)$. The group $H_1(\partial X)$ has already been computed in \cite{DelZotto:2022fnw} as the abelianization of $\pi_1(\partial X)$ which was in turn computed using a theorem by Armstrong \cite{armstrong_1968}:\smallskip

\noindent \textit{Let $\Gamma$ be a discontinuous group of homeomorphisms of a path connected, simply connected, locally compact metric space $Y$, and let $H$ be the normal subgroup of $\Gamma$ generated by those elements which have fixed points. Then the fundamental group of the orbit space $Y/\Gamma$ is isomorphic to the factor group $\Gamma/ H$. }\smallskip

Orbifold homology provides a streamlined way to access this, as well as the other contributions to the candidate 2-group structure. Indeed, as already noted in section \ref{sec:PRESCRIPTION}, the ``naive'' 1-form symmetry $\mathcal{A}$, the true 1-form symmetry, and the central quotienting subgroup $\mathcal{C}$ all descend from appropriate orbifold homology groups. In terms of the short exact sequence for 1-form symmetries:
\begin{equation}
0 \rightarrow \mathcal{C}^{\vee} \rightarrow \widetilde{\mathcal{A}}^{\vee} \rightarrow \mathcal{A}^{\vee} \rightarrow 0 \\
\end{equation}
each term is given by:
\begin{equation}\label{orbo}
0 \rightarrow H_{1}^{\mathrm{twist}}(S^5 / \Gamma) \rightarrow H_{1}^{\mathrm{orb}}(S^5 / \Gamma) \rightarrow H_{1}(S^5 / \Gamma) \rightarrow 0
\end{equation}
or, in terms of the data of the group, we have the identifications:
\begin{align}\label{eq:Solutions}
\widetilde{\mathcal{A}} & = \mathrm{Ab}[\Gamma]\\
\mathcal{A} & = \mathrm{Ab}[\Gamma / H]\\
\mathcal{C} & = \mathrm{Ab}[\Gamma] / \mathrm{Ab}[\Gamma / H].
\end{align}
Note in particular that the orbifold homology computation is directly sensitive to the fixed point locus specified by the group $H$, and this is precisely where the flavor 6-branes are localized in the boundary geometry. Now, precisely when the exact sequence of line (\ref{orbo}) does not split, we expect to get a non-trivial 2-group structure, precisely as conjectured in \cite{DelZotto:2022fnw}.

Of course, it is also important to directly verify this structure using our procedure of ``cutting and gluing''. Our aim in the remainder of this section will be to present a general analysis of this in the special case where $\Gamma$ is abelian. This provides a complementary way to isolate the individual contributions to the flavor symmetry, and also serves as a crosscheck on our orbifold homology calculation. While it would be interesting to also consider the same singular homology computation for non-abelian $\Gamma$, this is somewhat more involved and we defer this task to future work.

Restricting now to the special case of $\Gamma$ abelian, our aim will be to directly extract via singular homology
the geometric origin of each of the terms appearing in the pair of short exact sequences:
\begin{align}
0 & \rightarrow \mathcal{C} \rightarrow Z_{\widetilde{G}} \rightarrow Z_{G} \rightarrow 0 \\
0 & \rightarrow \mathcal{C}^{\vee} \rightarrow \widetilde{\mathcal{A}}^{\vee} \rightarrow \mathcal{A}^{\vee} \rightarrow 0.
\end{align}
As already stated, our analysis of the global form of the flavor symmetry will center on the piece coming from localized 6-brane contributions.

The rest of this section is organized as follows. We begin by specifying in more detail the orbifold singularities $\mathbb{C}^3 / \Gamma$. In this case, methods from toric geometry provide a convenient way to encode possible singular loci in the boundary geometry. With this in place, we then turn to the case of $\Gamma = \mathbb{Z}_n$, where we divide our analysis up according to the number of singular loci in the boundary geometry. We then turn to a similar analysis for $\Gamma = \mathbb{Z}_n \times \mathbb{Z}_m$. In this case, the structure of the flavor symmetry has a non-trivial dependence on $n$ and $m$, but the 1-form symmetry and 2-group structure is always trivial. We also present, when available, some examples which have a gauge theory phase, since one can in principle cross-check our geometric answer using such methods. In some cases, however, no known gauge theory phase is available, but the answer from geometry is unambiguous.

\subsection{Abelian $\Gamma\subset SU(3)$}

We now turn to the toric geometry of the Calabi-Yau orbifold singularities $X=\mathbb{C}^3/\Gamma$  with $\Gamma \subset{SU(3)}$ a finite abelian subgroup. Precisely because the group action embeds in the maximal torus of $SU(3)$, this group action is compatible with the torus action on $\mathbb{C}^3$ and as such all of these examples are toric manifolds. This was exploited in \cite{Albertini:2020mdx, Morrison:2020ool, Tian:2021cif} to perform explicit resolutions of the singular geometry, and thus determine the 1-form symmetry. Our goal here will be to avoid doing any blowups and instead directly obtain the symmetries from a suitable cutting and gluing of orbifold singularities.

There are two general choices for an abelian subgroup, as given by $\Gamma = \mathbb{Z}_n$ and $\Gamma=\mathbb{Z}_n \times \mathbb{Z}_m$ with $m$ dividing $n$, see Appendix \ref{app:AppendixSU3} for a detailed discussion of possible subgroups. The resulting group actions admit following parameterizations:
\begin{enumerate}
\item $\Gamma=\Z_n$\,: The action on $\mathbb{C}^3$ is $(z_1,z_2,z_3)  \sim (\omega^{k_1} z_1,\omega^{k_2}  z_2,\omega^{k_3} z_3)$ where $\omega$ is a primitive $n^{th}$ root of unity and the $k_i$ are positive integers satisfying $k_1+k_2+k_3=n$. Define $q_i=n/\textnormal{gcd}(n,k_i)$. We require the group action to be faithful which is the case precisely when $\textnormal{lcm}(q_1,q_2,q_3)=n$. We shall sometimes use the notation $\frac{1}{n}(k_1 , k_2, k_3)$ to indicate this group action.\smallskip

(1)' ~$\Gamma=\Z_N\times \Z_M=\Z_{NM}$\,: Subclass of actions with $\textnormal{gcd}(N,M)=1$ and  $ (z_1,z_2,z_3) \sim$
 ${}\qquad{}$ $(  \omega z_1,\eta  z_2,(\omega\eta)^{-1}z_3)$ with $\omega,\eta$ primitive $N^{th}$ and $M^{th}$ roots of unity and $n=NM$.

\item $\Gamma=\Z_n\times \Z_m$\,: The action on $\mathbb{C}^3$ is $(z_1,z_2,z_3) \sim (\omega^{k_1} z_1,\omega^{k_2}  z_2,\omega^{k_3} z_3) \sim (  z_1,\eta  z_2,\eta^{-1}z_3)$ with $\omega$ and $\eta$ primitive $n^{th}$ and $m^{th}$ roots of unity and integers $k_i$ constrained as above in case $(1)$ and $n=mm'$. We further require trivial intersection between $\Z_n$ and $\Z_m$ realized by restricting to actions with $\textnormal{gcd}(n,k_1)$ and $m$ co-prime. When $n=m$ we can chose generators as $ (z_1,z_2,z_3) \sim  (  \omega z_1,\eta  z_2,(\omega\eta)^{-1}z_3) $.

(2)' ~$\Gamma=\Z_N\times \Z_M$\,: Subclass of actions with $\textnormal{gcd}(N,M)=g\geq 2$ and  $ (z_1,z_2,z_3) \sim$
 ${}\qquad{}$ $(  \omega z_1,\eta  z_2,(\omega\eta)^{-1}z_3)$ with $\omega,\eta$ primitive $N^{th}$ and $M^{th}$ roots of unity. The integers ${}\qquad{}$  $n,m$ follow upon regrouping prime factors of $N,M$.
\end{enumerate}
These unitary group actions restrict to the asymptotic boundary $\partial X$ which is modelled on an $S^5$ with unit radius acted on by $\Gamma$. The fixed point loci of these two actions are non-compact and their intersection with the boundary, denoted $K$, admit the following characterization according to $|K|$, the number of connected components of the fixed locus.

\begin{enumerate}
\item $\Gamma=\Z_n$: The locus $K$ consists of a circle's worth of $A_{g_i-1}$ singularities located at $|z_i|=1$ where $g_i=\textnormal{gcd}(n,k_i)$. The $\Z_{g_i}$ subgroup folding the singularity is generated by $\omega^{n/g_i}$. We can have $|K|=0,1,2,3$ depending on the group action.

(1)' ~$\Gamma=\Z_N\times \Z_M=\Z_{NM}$\,: Subclass with $|K|=2$ and an $A_{M-1},A_{N-1}$ singularity ${}\qquad{}$ along circles $|z_1|,|z_2|=1$ respectively.

\item $\Gamma=\Z_n\times \Z_m$: The locus $K$ consists of three circle worths of $A_{g_i'-1}$ singularities located at $|z_i|=1$. Here $g_i'=m\;\! \textnormal{gcd}(m',k_i)$ where $n=mm'$. In all casses $g_i'\geq m$, we therefore have $|K|=3$ independent of the group action. When $n=m$ we have $g_i'=n$ and three circles worth of $A_{n-1}$ singularities.

(2)' ~$\Gamma=\Z_N\times \Z_M$\,: Subclass with an $A_{M-1},A_{N-1},A_{g-1}$ singularity along circles ${}\qquad{}$ ${}\qquad{}$\,$|z_1|,|z_2|,|z_3|=1$ respectively.
\end{enumerate}

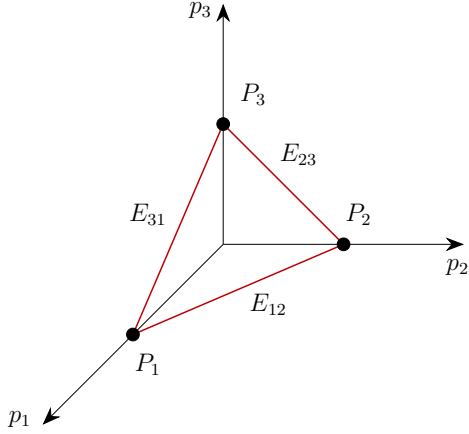
\begin{figure}[t!]
\center
\scalebox{0.8}{
\begin{tikzpicture}
	\begin{pgfonlayer}{nodelayer}
		\node [style=none] (0) at (0, 0) {};
		\node [style=none] (1) at (-3, -3) {};
		\node [style=none] (2) at (0, 4) {};
		\node [style=none] (3) at (4, 0) {};
		\node [style=none] (4) at (-1.5, -1.5) {};
		\node [style=none] (5) at (2, 0) {};
		\node [style=none] (6) at (0, 2) {};
		\node [style=none] (7) at (-0.375, 3.9) {$p_3$};
		\node [style=none] (8) at (-3.375, -2.9) {$p_1$};
		\node [style=none] (9) at (3.9, -0.375) {$p_2$};
		\node [style=none] (13) at (2.25, 0.5) {$P_2$};
		\node [style=none] (14) at (0.5, 2.5) {$P_3$};
		\node [style=none] (15) at (-1.25, -2) {$P_1$};
		\node [style=Circle] (16) at (0, 2) {};
		\node [style=Circle] (17) at (2, 0) {};
		\node [style=Circle] (18) at (-1.5, -1.5) {};
		\node [style=none] (19) at (0.75, -1) {$E_{12}$};
		\node [style=none] (20) at (1.25, 1.5) {$E_{23}$};
		\node [style=none] (21) at (-1.25, 0.5) {$E_{31}$};
	\end{pgfonlayer}
	\begin{pgfonlayer}{edgelayer}
		\draw [style=ArrowLineRight] (0.center) to (1.center);
		\draw [style=ArrowLineRight] (0.center) to (3.center);
		\draw [style=ArrowLineRight] (0.center) to (2.center);
		\draw [style=RedLine] (4.center) to (6.center);
		\draw [style=RedLine] (6.center) to (5.center);
		\draw [style=RedLine] (5.center) to (4.center);
	\end{pgfonlayer}
\end{tikzpicture}
}
\caption{Sketch of the base $\Delta$ of the torus fibration $\pi:\partial X\rightarrow \Delta$. }
\label{fig:BaseBoundary}
\end{figure}

\begin{figure}
\centering
\scalebox{0.8}{
\begin{tikzpicture}
	\begin{pgfonlayer}{nodelayer}
		\node [style=none] (0) at (-5, -1) {};
		\node [style=none] (1) at (-3, 2.5) {};
		\node [style=none] (2) at (-1, -1) {};
		\node [style=none] (3) at (0, -1) {};
		\node [style=none] (4) at (2, 2.5) {};
		\node [style=none] (5) at (4, -1) {};
		\node [style=none] (6) at (5, -1) {};
		\node [style=none] (7) at (7, 2.5) {};
		\node [style=none] (8) at (9, -1) {};
		\node [style=none] (9) at (2, -1) {};
		\node [style=none] (11) at (6, 0.75) {};
		\node [style=none] (12) at (8, 0.75) {};
		\node [style=none] (13) at (7, -1) {};
		\node [style=none] (14) at (7, 0) {};
		\node [style=none] (15) at (-3, 3) {$P_3$};
		\node [style=none] (16) at (-1, -1.5) {$P_2$};
		\node [style=none] (17) at (-5, -1.5) {$P_1$};
		\node [style=none] (18) at (2, 3) {$P_3$};
		\node [style=none] (19) at (4, -1.5) {$P_2$};
		\node [style=none] (20) at (0, -1.5) {$P_1$};
		\node [style=none] (21) at (7, 3) {$P_3$};
		\node [style=none] (22) at (9, -1.5) {$P_2$};
		\node [style=none] (23) at (5, -1.5) {$P_1$};
		\node [style=none] (24) at (-3, -2.5) {$|\pi(K)|=1$};
		\node [style=none] (25) at (2, -2.5) {$|\pi(K)|=2$};
		\node [style=none] (26) at (7, -2.5) {$|\pi(K)|=3$};
		\node [style=none] (29) at (-3, -0.5) {$E_{12}$};
		\node [style=none] (30) at (2.5, -0.5) {$I_{12}$};
		\node [style=none] (31) at (2.5, 0.5) {$I_3$};
		\node [style=none] (32) at (6.25, 0) {$I_{31}$};
		\node [style=none] (33) at (7.5, -0.5) {$I_{12}$};
		\node [style=none] (34) at (7.25, 0.75) {$I_{23}$};
		\node [style=Circle] (27) at (2, 0) {};
		\node [style=Circle] (28) at (7, 0) {};
	\end{pgfonlayer}
	\begin{pgfonlayer}{edgelayer}
		\draw [style=ThickLine] (0.center) to (1.center);
		\draw [style=ThickLine] (1.center) to (2.center);
		\draw [style=ThickLine] (3.center) to (4.center);
		\draw [style=ThickLine] (4.center) to (5.center);
		\draw [style=ThickLine] (5.center) to (3.center);
		\draw [style=ThickLine] (6.center) to (7.center);
		\draw [style=ThickLine] (7.center) to (8.center);
		\draw [style=ThickLine] (8.center) to (6.center);
		\draw [style=RedLine] (0.center) to (2.center);
		\draw [style=RedLine] (4.center) to (9.center);
		\draw [style=RedLine] (13.center) to (14.center);
		\draw [style=RedLine] (14.center) to (12.center);
		\draw [style=RedLine] (14.center) to (11.center);
	\end{pgfonlayer}
\end{tikzpicture}	}
\caption{Sketches of the deformation retracts for the base $\Delta$. In the first and second and third configuration the orbifold locus projects to $P_3$ and $P_1, P_2$ and $P_1, P_2 , P_3$ respectively. The $T^3/\:\!\Gamma$ fibration then deformation retracts to a fibration over the graphs $\mathfrak{G}$ marked red. We depict a decomposition of $\mathfrak{G}$ into intervals $I_*$ with one endpoint on the boundary of $\Delta$.}
\label{fig:bases}
\end{figure}
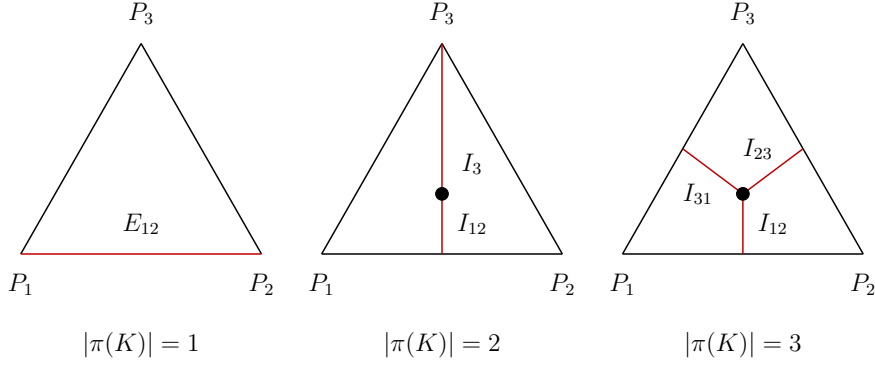

The components $K_i$ are always circles and located at the vanishing locus of two coordinates. They are therefore conveniently parameterized by standard toric coordinates. For $\mathbb{C}^3$ these read $p_i=|z_i|^2$ and $\theta_i=\textnormal{arg}\,z_i$ with three-torus fiber
\be\label{eq:T3Fibration}
T^3=\lbbb (\theta_1,\theta_2,\theta_3)\rbbb\,.
\ee
This fibration restricts to the boundary five-sphere of $\mathbb{C}^3$ with triangle base
\be
\Delta=\lbbb p_1+p_2+p_3=1 \rbbb
\ee
whose corners $P_i$ and edges $E_{jk}$ are labelled as shown in figure \ref{fig:BaseBoundary}. Along edges and at corners the three-torus fiber degenerates to a two-torus and circle respectively. The abelian actions preserve the torus fiber $T^3$ and both the quotient $X=\mathbb{C}^3/\Gamma$ and its boundary inherit this fibration which for the boundary reads
\be
T^3_\Gamma ~\hookrightarrow ~ \partial X  ~\xrightarrow[]{\,\pi \, } ~ \Delta\,.
\ee
Here $T^3_\Gamma=T^3/\;\!\Gamma$ are three-tori as $\Gamma$ is a subgroup of a continuous abelian action on $T^3$. The orbifold locus now clearly projects to the corners
\be
\pi(K)\subset\lbbb P_1,P_2,P_3 \rbbb \,.
\ee
The smooth boundary $\partial X^\circ$ is therefore fibered over $\Delta \setminus \pi (K)$ which deformation retracts onto a one-dimensional subspace $\mathfrak{G}\subset \Delta$. We denote the induced deformation retract of the total space by $\partial X^\circ_r$ and therefore
\be
H_*(\partial X^\circ)\cong H_*(\partial X^\circ_r)\,.
\ee
The subspace $\mathfrak{G}$ is an interval, when $|\pi(K)|=1,2$ and a $Y$-shaped graph when $|\pi(K)|=3$. See figure \ref{fig:bases}.

We now discuss the topology of $\partial X^\circ_r$ for these different cases. First, we treat the different cases associated with $\Gamma = \mathbb{Z}_n$ and then turn to the cases with $\Gamma = \mathbb{Z}_n \times \mathbb{Z}_m$.

\subsubsection{$\Gamma = \Z_n$ and $|K| = 0$}\label{sec:ZN0}

Consider first the case that there are no fixed loci in the boundary geometry, namely $|K| = 0$. The boundary geometry is a generalized lens space $S^5 / \Gamma$ with a fixed point free action on the $S^5$. This occurs whenever the $k_i$ are all relatively prime to $n$. For such orbifold group actions, there is no orbifold fixed point locus on $\partial X$ to speak of, since the orbifold group action on $S^5$ is, by definition, fixed point free.
In this case, $\mathcal{A}^{\vee} \cong \pi_{1}(S^5/ \mathbb{Z}_n) = \mathbb{Z}_n$,
and $\mathcal{C} = Z_{\widetilde{G}} = Z_G = 0$ in the pair of exact sequences:
\begin{align}
0 & \rightarrow \mathcal{C} \rightarrow Z_{\widetilde{G}} \rightarrow Z_{G} \rightarrow 0\\
0 & \rightarrow \mathcal{C}^{\vee} \rightarrow \widetilde{\mathcal{A}}^{\vee} \rightarrow \mathcal{A} \rightarrow 0.
\end{align}

As an example of this sort, consider $\Gamma = \mathbb{Z}_3$ with group action $\frac{1}{3}(1,1,1)$, in the obvious notation. This results in the celebrated $E_0$ Seiberg theory \cite{Seiberg:1996bd} as obtained from a collapsing $\mathbb{CP}^2$ in the local geometry $\mathcal{O}(-3) \rightarrow \mathbb{CP}^2$ (see also \cite{Morrison:1996xf, Douglas:1996xp}). Let us also comment that in this case, there is indeed an additional contribution to the 0-form symmetry since we can permute the three holomorphic coordinates. This generates a $\mathbb{Z}_3$ global symmetry. Thankfully, however, this decouples from the higher symmetries \cite{Albertini:2020mdx}.

\subsubsection{$\Gamma= \Z_n$ and $|K| = 1$}
\label{sec:Zn1}

Consider next the case of one fixed locus for the orbifold group action, namely $|K| = 1$. Now, in this case, we can always choose coordinates such that the fixed locus projects to the corner $\pi(K)=P_3$. The smooth boundary $\partial X^\circ$ then deformation retracts to a fibration over the edge $E_{12}$ resulting in a lens space
\be
 \partial X^\circ_{r}=S^3/\Gamma
 \ee
where $\Gamma$ acts on $S^3$ with order $n$ and without fixed points because $(z_1,z_2,0)\sim (z_1^{k_1},z_2^{k_2},0)$. Let $K$ support an $A_{g-1}$ singularity, then we find overall
\be\ba
\textnormal{Tor}\,H_2( \partial X)&=0\\
\textnormal{Tor}\,H_1( \partial X^\circ\cap T(K))&= \Z_g\\
\textnormal{Tor}\, H_1( \partial X^\circ)&= \Z_n\\
\textnormal{Tor}\,H_1( \partial X)&= \Z_{n/g},
\ea\ee
where the first line follows from results in \cite{Kawasaki:1973wl}. This determines the maps:
\begin{align}
\iota_1 &: ~\mathrm{Tor}\, H_1(\partial X^{\circ} \cap T(K)) \rightarrow \mathrm{Tor}\, H_1(\partial X^\circ) \\
j_1 &: ~\mathrm{Tor}\, H_1(\partial X^{\circ}) \rightarrow \mathrm{Tor}\, H_{1}(\partial X)
\end{align}
to be multiplication by $n/g$ (for $\iota_1$) and modding by $n/g$ (for $j_1$).
In particular we have $\textnormal{ker}(\iota_1)=0$, therefore, in the pair of exact sequences:
\begin{align}
0 & \rightarrow \mathcal{C} \rightarrow Z_{\widetilde{G}} \rightarrow Z_{G} \rightarrow 0 \\
0 & \rightarrow \mathcal{C}^{\vee} \rightarrow \widetilde{\mathcal{A}}^{\vee} \rightarrow \mathcal{A}^{\vee} \rightarrow 0,
\end{align}
we have:
\begin{align}
0 & \rightarrow \mathbb{Z}_g \rightarrow \mathbb{Z}_g \rightarrow 0 \rightarrow 0 \\
0 & \rightarrow \mathbb{Z}_g \rightarrow \mathbb{Z}_n \rightarrow \mathbb{Z}_{n/g} \rightarrow 0.
\end{align}
In particular, we can now extract the global 0-form symmetry and the 1-form symmetry:
 \be
G=SU(g)/\Z_g\,, \qquad \mathcal{A}=\Z_{n/g}\,.
\ee
The short exact sequence characterizing 2-groups becomes:
\be
0~\xrightarrow[]{} ~  \Z_g ~\xrightarrow[]{} ~  \Z_n ~\xrightarrow[]{ } ~  \Z_{n/g} ~\rightarrow ~0\,.
\ee
This sequence is non-split whenever $n/g$ is divisible by any prime factor of $g$ and in these cases we have a non-trivial 2-group.

As an important special case, consider $\Gamma = \mathbb{Z}_{2n}$ with group action
$\frac{1}{2n}(1,1,2n-2)$ (note that we have rescaled $n$ by a factor of $2$ to match to the presentation commonly found in the literature). In this case, we have a fixed point locus along $z_1 = z_2 = 0$, and a flavor 6-brane supporting an $A_1$ singularity. From our general considerations presented above, we have $\mathcal{A} = \mathbb{Z}_n$, and also find $\mathcal{C} = \mathbb{Z}_2$. So in these cases, we also expect a flavor group $G = SO(3) = SU(2) / \mathbb{Z}_2$. Moreover, there is a non-trivial 2-group when $n$ is even since in that case $\mathbb{Z}_{2n} \neq \mathbb{Z}_n \oplus \mathbb{Z}_2$. The case of $n$ even has a 5D description in terms of $SU(n)_n$ gauge theory, and the corresponding gauge theory analysis of \cite{Apruzzi:2021vcu} is in accord with our results.

\subsubsection{$\Gamma= \Z_n$ and $|K| = 2$}
\label{sec:Zn2}

Consider next the case with $\Gamma = \mathbb{Z}_n$ and $|K| = 2$, namely we have two separate flavor 6-branes extending out to the boundary. In this case, it is convenient to choose coordinates such that the orbifold locus projects to $\pi(K_i)=P_i$ and set $g_i=\textnormal{gcd}(n,k_i)$ for $i=1,2$. Observe that since we require the group action to be faithful we have $\mathrm{gcd}(g_1,g_2) = 1$, i.e. $g_1$ and $g_2$ are co-prime, as otherwise by the condition $k_1+k_2+k_3=n$ the integer $k_3$ would share divisors with $k_1,k_2$.

The smooth boundary deformation retracts to a fibration over the interval $I=I_{12}\cup I_3$. Here $I_{12}$ and $I_3$ are intervals ending on $E_{12}$ and at $P_3$ respectively, see figure \ref{fig:bases}. We therefore have the covering
\be\label{eq:MVCoveringFirst}
\partial X^\circ_{r} = \pi^{-1}(I_{12})\cup \pi^{-1}(I_3)
\ee
The intervals $I_{12}$ and $ I_3$ retract to a point on the edge $E_{12}$ and the corner $P_3$ respectively and lifting these retraction to the full space we find $\pi^{-1}(I_{12})$ and $\pi^{-1}(I_3)$ to retract to the fibers above these, denoted $T^2_{12}, S^1_3$ respectively. The intersection of $\pi^{-1}(I_{12})\cap \pi^{-1}(I_{3})=T^3_\Gamma$ is a copy of the three-torus fiber. We now apply Mayer-Vietoris sequence to the covering \eqref{eq:MVCoveringFirst}. The sequence is exact in degree zero and has no 2-cycles  \cite{Kawasaki:1973wl} and therefore we find the short exact sequence
\be
0 ~\rightarrow ~ H_1(T^3_\Gamma) ~\xrightarrow[]{}~ H_1(T^2_{12}) \oplus  H_1(S^1_{3}) ~\rightarrow~  H_1(X^\circ_{r}) ~\rightarrow~ 0
\ee
We now denote the maps into the central factors by
\be\ba\label{eq:iotas}
\iota_{11}\,&:~ H_1(T^3_\Gamma) ~\rightarrow ~H_1(T^2_{12}) \\
\iota_{12}\,&:~ H_1(T^3_\Gamma) ~\rightarrow ~H_1(S^1_{3})
\ea\ee
and now reparametrize the fiber $T^3_\Gamma$ following the coordinate change  $(z_1,z_2,z_3) \rightarrow (z_1',z_2',z_3') = (z_1,z_2,z_1z_2z_3)$ which splits the fiber as $T^3_\Gamma=T^2_\Gamma \times S^1$. Here $S^1$ is the diagonal circle which is not acted on by $\Gamma\subset SU(3)$. We see that $\iota_{11}$ is surjective while $\iota_{12}$ is multiplication by $n$.

With this we find overall
\be\ba
\textnormal{Tor}\,H_2( \partial X)&=0\\
\textnormal{Tor}\,H_1( \partial X^\circ\cap T(K))&=\Z_{g_1g_2} \cong \mathbb{Z}_{g_1} \times \mathbb{Z}_{g_2}\\
\textnormal{Tor}\, H_1( \partial X^\circ)&=\Z_n\\
\textnormal{Tor}\,H_1( \partial X)&=\Z_{n/g_1g_2}
\ea\ee
where in the second line we used the fact that $g_1$ and $g_2$ are co-prime.
This determines the maps:
\begin{align}
\iota_1 &:~ \mathrm{Tor}\, H_1(\partial X^{\circ} \cap T(K)) \rightarrow \mathrm{Tor}\, H_1(\partial X^\circ) \\
j_1 &:~ \mathrm{Tor}\, H_1(\partial X^{\circ}) \rightarrow \mathrm{Tor}\, H_{1}(\partial X),
\end{align}
where $j_1$ is therefore modding out by $n/g_1g_2$ and $\textnormal{ker}(\iota_1)=0$. In our pair of short exact sequences:
\begin{align}
0 & \rightarrow \mathcal{C} \rightarrow Z_{\widetilde{G}} \rightarrow Z_{G} \rightarrow 0 \\
0 & \rightarrow \mathcal{C}^{\vee} \rightarrow \widetilde{\mathcal{A}}^{\vee} \rightarrow \mathcal{A}^{\vee} \rightarrow 0,
\end{align}
we now have:
\begin{align}
0 & \rightarrow \mathbb{Z}_{g_1 g_2} \rightarrow \mathbb{Z}_{g_1 g_2} \rightarrow 0 \rightarrow 0 \\
0 & \rightarrow \mathbb{Z}_{g_1 g_2} \rightarrow \mathbb{Z}_n \rightarrow \mathbb{Z}_{n / g_1 g_2} \rightarrow 0. \label{secondo}
\end{align}
In particular, the global form of the flavor symmetry and the 1-form symmetry are:
 \be
G=SU(g_1)/\Z_{g_1}\times SU(g_2)/\Z_{g_2}\,, \qquad \mathcal{A}=\Z_{n/{g_1g_2}}\,,
\ee
where we have used the fact that $g_1$ and $g_2$ are co-prime.
Finally, the short exact sequence characterizing 2-groups is controlled by the second short exact sequence:
\be\label{splitto}
0~\xrightarrow[]{} ~  \Z_{g_1g_2} ~\xrightarrow[]{} ~  \Z_n ~\xrightarrow[]{} ~  \Z_{n/g_1g_2} ~\rightarrow ~0\,.
\ee
This sequence is non-split whenever $n/{g_1g_2}$ is divisible by any prime factor
of either $g_i$ and in these cases we have a non-trivial 2-group.

Let us now turn to a few examples. Consulting table 5 of reference \cite{Tian:2021cif}, we see that some such orbifold group actions also have a gauge theory phase, which can be used as a cross-check on our proposed higher symmetries. Consider $\Gamma = \mathbb{Z}_6$ with group action $\frac{1}{6}(1,2,3)$. This theory has a gauge theory phase consisting of an $SU(2)$ gauge group and two flavors in the fundamental representation, denoted as $SU(2) - 2F$. In particular it has naive flavor group $\widetilde{G} = SU(2) \times SU(3)$. From our analysis, we have $g_1 = 2$ and $g_2 = 3$ so we expect the global form of the flavor symmetry is $SU(2) / \mathbb{Z}_2 \times SU(3) / \mathbb{Z}_3$, and that it has trivial 1-form symmetry (and thus trivial 2-group as well).

We now give some examples of such 5D SCFTs with a gauge theory phase of the subclass type (1)'. Following table 6 of \cite{Tian:2021cif}, consider the case $\Gamma = \mathbb{Z}_5 \times \mathbb{Z}_2$ which is equivalent to $\Gamma=\Z_{10}$ generated by $\frac{1}{10}(4,5,1)$. The gauge theory phase of this case is given by $SU(2)_0 - SU(2) - 2F$. From our general considerations, the global form of the flavor symmetry is then given by $SU(5) / \mathbb{Z}_5 \times SU(2) / \mathbb{Z}_2$.

As another example, consider the case $\Gamma = \mathbb{Z}_{12}$ with group action $\frac{1}{12}(1,2,9)$. This has a gauge theory phase $SU(4)_{4} - SU(2)_0$, and $\widetilde{G} = SU(2) \times SU(3)$. From our analysis, we have $g_1 = 2$ and $g_2 = 3$. Here, we expect $G = SU(2) / \mathbb{Z}_2 \times SU(3) / \mathbb{Z}_3$ and a non-trivial 1-form symmetry $\mathcal{A} = \mathbb{Z}_{2}$. In this case, the sequence (\ref{splitto}) does not split because $12 / 6 = 2$ is divisible by $g_1 = 2$, so we also expect a non-trivial 2-group.

\subsubsection{$\Gamma= \Z_n$ and $|K| = 3$}
\label{sec:Zn3}

The last case with $\Gamma = \mathbb{Z}_n$ has $|K| = 3$, i.e. three boundary flavor components. This can occur when $n$ has at least three distinct prime factors and $\mathrm{gcd}(k_i , n) = g_i \geq 2$ and all $g_i$ co-prime. In this case, the orbifold locus projects to $\pi(K_i)=P_i$ and the smooth boundary deformation retracts to $I_{12}\cup I_{23} \cup I_{31}$. See figure \ref{fig:bases}.

Let us consider the $T^3$ fibration \eqref{eq:T3Fibration} prior to taking the quotient by $\Gamma=\Z_n$. It is straightforward to see that this fibration restricted to any interval $I=I_{12}\cup I_{23}$ is topologically $S^1\times S^3$ and that the factor of $S^1$ collapses along the edge $E_{13}$. We then glue in the fibers projecting to $I_{13}$ using the Mayer-Vietoris sequence and find a simply connected space with no fixed points under the $\Z_n$ action. Now we can apply Armstrong's theorem and find
\be
H_1(X^\circ_{r}) =\Z_n\,.
\ee
With this result we now identify the subspaces of $X^\circ_{r}$ projecting to pairs of intervals
\be\ba\label{eq:TorsionGroups}
  U_1 &= \pi^{-1}(I_{31}\cup I_{12})=S^1_1\times S^3/\Z_{g_1} \\
   U_2 &= \pi^{-1}(I_{12}\cup I_{23})=S^1_2\times S^3/\Z_{g_2}\\
    U_3 &=\pi^{-1}(I_{23}\cup I_{31})=S^1_3\times S^3/\Z_{g_3}\\
\ea\ee
where the rightmost equalities follow from the fact that $\pi^{-1}(I_{ij}\cup I_{jk})$ can be identified with the boundary of a local neighborhood of the ADE singularity projecting to $P_j$. Any pair of $U_i$ constitute a covering of $X^\circ_{r}$ and from the corresponding Mayer-Vietoris sequence it follows that the torsional 1-cycles in $H_1(U_i)$ embed non-trivially into $H_1(X^\circ_{r})=\Z_n$. This implies that the map
\be
\iota_1 : \mathrm{Tor}\, H_1(\partial X^{\circ} \cap T(K)) \rightarrow \mathrm{Tor}\, H_{1}(\partial X^{\circ})
\ee
has trivial kernel. With this we find overall
\be\ba
\textnormal{Tor}\,H_2( \partial X)&=0\\
\textnormal{Tor}\,H_1( \partial X^\circ\cap T(K))&=\Z_{g_1g_2g_3} \cong \mathbb{Z}_{g_1} \times \mathbb{Z}_{g_2} \times \mathbb{Z}_{g_3}\\
\textnormal{Tor}\, H_1( \partial X^\circ)&=\Z_n\\
\textnormal{Tor}\,H_1( \partial X)&=\Z_{n/g_1g_2g_3}
\ea\ee
where in the second line we used the fact that the $g_i$ are all co-prime.
This determines the map $\iota_1$ to be multiplication by $g_1g_2g_3$ and the map
\begin{align}
j_1 &:~ \mathrm{Tor}\, H_1(\partial X^{\circ}) \rightarrow \mathrm{Tor}\, H_{1}(\partial X),
\end{align}
where $j_1$ is therefore modding out by $n/g_1g_2g_3$. In our pair of short exact sequences:
\begin{align}
0 & \rightarrow \mathcal{C} \rightarrow Z_{\widetilde{G}} \rightarrow Z_{G} \rightarrow 0 \\
0 & \rightarrow \mathcal{C}^{\vee} \rightarrow \widetilde{\mathcal{A}}^{\vee} \rightarrow \mathcal{A}^{\vee} \rightarrow 0 ,
\end{align}
we now have:
\begin{align}
0 & \rightarrow \mathbb{Z}_{g_1 g_2 g_3} \rightarrow \mathbb{Z}_{g_1 g_2 g_3} \rightarrow 0 \rightarrow 0 \\
0 & \rightarrow \mathbb{Z}_{g_1 g_2 g_3} \rightarrow \mathbb{Z}_n \rightarrow \mathbb{Z}_{n / g_1 g_2 g_3} \rightarrow 0. \label{secondo}
\end{align}
In particular, the global form of the flavor symmetry and the 1-form symmetry are:
 \be
G=SU(g_1)/\Z_{g_1}\times SU(g_2)/\Z_{g_2} \times SU(g_3) / \mathbb{Z}_{g_3} \,, \qquad \mathcal{A}=\Z_{n/{g_1g_2g_3}}\,,
\ee
where we have used the fact that the $g_i$ are co-prime.
Finally, the short exact sequence characterizing 2-groups is controlled by the second short exact sequence:
\be\label{splitto}
0~\xrightarrow[]{} ~  \Z_{g_1g_2g_3} ~\xrightarrow[]{} ~  \Z_n ~\xrightarrow[]{} ~  \Z_{n/g_1g_2g_3} ~\rightarrow ~0\,.
\ee
This sequence is non-split whenever $n/{g_1g_2g_3}$ is divisible by any prime factor
of any of the $g_i$ and in these cases we have a non-trivial 2-group.

For this class of examples we are unaware of a known gauge theory phase which we can use to possibly cross-check our statements. Nevertheless, we can still specify example group actions which have $\Gamma = \mathbb{Z}_n$ and $|K| = 3$. To illustrate, we can take $\mathbb{Z}_{30}$ with group action $\frac{1}{30}(2,3,25)$. For this case the greatest common divisors $g_{i} = \mathrm{gcd}(k_i , n)$ are $g_1 = 2, g_2 = 3, g_3 = 5$ so we expect a flavor symmetry group $SU(2) / \mathbb{Z}_2 \times SU(3) / \mathbb{Z}_{3} \times SU(5) / \mathbb{Z}_5$, and trivial 1-form symmetry and 2-group.

Similar considerations hold for other choices, and one way to generate examples is simply to require $n$ to be divisible by three
distinct prime factors. To get a non-trivial 1-form symmetry, the multiplicity of one of these prime factors must be greater than one, and this needs to correlate with the choice of $k_i$. As an example which has a non-trivial 1-form symmetry, we can take $n = 60 = 2^2 3 5$ so that $\Gamma = \mathbb{Z}_{60}$. We specify the orbifold group action by $\frac{1}{60}(2,3,55)$. In this case, $G = SU(2) / \mathbb{Z}_2 \times SU(3) / \mathbb{Z}_3 \times SU(5) / \mathbb{Z}_5$ and the 1-form symmetry is $\mathcal{A} = \mathbb{Z}_2$. Since the short exact sequence for 1-form symmetries does not split, we also see that there is a 2-group present.

\subsubsection{$\Gamma= \Z_n\times \Z_m$ and $|K| = 3$}
\label{sec:ZnZm3}

The final case to consider is $\Gamma = \mathbb{Z}_n \times \mathbb{Z}_m$ with $|K| = 3$, namely three distinct components for the flavor locus. We begin by analyzing the subclass (2)' of these actions which are parametrized as $\Gamma = \mathbb{Z}_N \times \mathbb{Z}_M$ with $g=\textnormal{gcd}(N,M)\geq 2$. In this case the smooth boundary deformation retracts to a fibration over $I_{12}\cup I_{23} \cup I_{31}$. See figure \ref{fig:bases}. We consider the open sets
\be\ba
  U_1 &= \pi^{-1}(I_{31}\cup I_{12})=S^1_1\times S^3/\Z_M \\
   U_2 &= \pi^{-1}(I_{12}\cup I_{23})=S^1_2\times S^3/\Z_N\\
    U_3 &=\pi^{-1}(I_{23}\cup I_{31})=S^1_3\times S^3/\Z_g\\
\ea\ee
where the rightmost equalities follow from the fact that $\pi^{-1}(I_{ij}\cup I_{jk})$ can be identified with the boundary of a local neighborhood of the ADE singularity projecting to $P_j$. Applying the Mayer-Vietoris sequence to the cover $\partial X^\circ_{r}=U_1\cup U_2$ we find the sequence
\be
0 ~\rightarrow ~ H_1(T^2_{12}) ~\xrightarrow[]{\,\,}~ H_1(U_1) \oplus H_1(U_2)  ~\rightarrow~  H_1(\partial X^\circ_r) ~\rightarrow~ 0
\ee
The generators of $H_1(T^2_{12})$ map onto $S^1_1,S^1_2$ and the torsion factors in the lens spaces are inherited by $H_1(\partial X^\circ_r) $ so we find:
\be
H_1(\partial X^\circ) = \Z_N\oplus \Z_M.
\ee
Returning to the map $\iota_1 : \mathrm{Tor}\, H_1(\partial X^{\circ} \cap T(K)) \rightarrow \mathrm{Tor}\, H_{1}(\partial X^{\circ})$, we see that
$\textnormal{ker}\,\iota_1=\Z_g$, which sits diagonally in $Z_{\widetilde{G}}=\Z_N\times \Z_M\times \Z_g$. Since the 1-form symmetry for all these cases is trivial (see e.g. \cite{Albertini:2020mdx, Morrison:2020ool, Tian:2021cif, DelZotto:2022fnw}), it suffices to specify the global form of the flavor symmetry. Returning to our short exact sequence for the centers:
\begin{equation}
0 \rightarrow \mathcal{C} \rightarrow Z_{\widetilde{G}} \rightarrow Z_{G} \rightarrow 0,
\end{equation}
we have:
\begin{equation}
0 \rightarrow \frac{\mathbb{Z}_{N} \times \mathbb{Z}_{M} \times \mathbb{Z}_{g}}{\mathbb{Z}_{g}}  \rightarrow \mathbb{Z}_{N} \times \mathbb{Z}_{M} \times \mathbb{Z}_g \rightarrow \mathbb{Z}_{g} \rightarrow 0.
\end{equation}

As a consequence, the global flavor symmetry extracted from geometry is:
\begin{equation}\label{eq:FlavorGroup}
G = \frac{SU(N) \times SU(M) \times SU(g)}{\mathbb{Z}_{N} \times \mathbb{Z}_M},
\end{equation}
where the $\mathbb{Z}_g$ embeds in the common diagonal (since $g = \mathrm{gcd}(N,M)$).

Let us now turn to some examples. Consider the special case $\Gamma = \mathbb{Z}_N \times \mathbb{Z}_N$. This generates the
5D $T_N$ theory \cite{Benini:2009gi} which has a manifest $\mathfrak{su}(N)^3$ flavor symmetry algebra. Our general considerations indicate that the global form of the flavor symmetry is $SU(N)^3 / \mathbb{Z}_N \times \mathbb{Z}_N$. As a piece of corroborating evidence for our proposal, we note that upon compactification on a circle, we obtain the 4D $T_n$ theories introduced in \cite{Gaiotto:2009we}, as can be obtained from compactification of $n$ M5-branes on the trinion (thrice-punctured sphere). The global 0-form symmetry for these 4D theories was recently investigated in reference \cite{Bhardwaj:2021ojs}, where it was argued on different grounds that the global form of this flavor symmetry is (again we are neglecting possible effects from mixing with R-symmetry) $SU(N)^3 / \mathbb{Z}_N \times \mathbb{Z}_N$. Let us also note that for $N = 3$ there is an additional enhancement in the flavor symmetry algebra to $\mathfrak{e}_6$, and the expectation from \cite{Bhardwaj:2021ojs} is that the non-abelian flavor group in this case is $E_6 / \mathbb{Z}_3$. While our geometric analysis does not directly detect such an enhancement, we can indeed see that this additional quotient by $\mathbb{Z}_3$ should be in operation since $E_6 / \mathbb{Z}_3$ contains the subgroup $SU(3)^3 / \mathbb{Z}_3 \times \mathbb{Z}_3$. As a final comment on this example, we note that the $T_3$ theory also has a gauge theory phase given by $SU(2)$ gauge theory coupled to five flavors in the fundamental representation \cite{Seiberg:1996bd}.

We now turn to discuss the general case for $\Gamma = \mathbb{Z}_n \times \mathbb{Z}_m$ with $n=mm'$ and $g_i'=m\;\! \textnormal{gcd}(m',k_i)$. Homology computations as in the previous subsections are more involved here and we instead make use of the prescription \eqref{eq:Solutions} which we argued for on general grounds via orbifold homology.

First we consider the subgroup $H\subset \Gamma$ generated by elements with fixed points. This is given by (see Appendix \ref{app:AppendixSU3} for details):
\be
H=\Z_{m \;\! \textnormal{gcd}(m',k_1)\;\! \textnormal{gcd}(m',k_2)\;\! \textnormal{gcd}(m',k_3)}\times \Z_m \subset \Z_n\times\,. \Z_m
\ee
The 1-form symmetry is therefore isomorphic to\footnote{The integers $\textnormal{gcd}(m',k_i)$ are pairwise coprime for if any pair were to share a factor larger than one then it would follow from the relation $k_1+k_2+k_3=n=mm'$ that all $k_i$ share a common factor. The group action would then not be faithful violating the assumption that we are describing an action by an abelian group of order $nm$.}
\be
\mathcal{A}\cong \Z_{m'/\textnormal{gcd}(m',k_1)\;\! \textnormal{gcd}(m',k_2)\;\! \textnormal{gcd}(m',k_3)}
\ee  by Armstrong's theorem. In our pair of short exact sequences:
\begin{align}
0 & \rightarrow \mathcal{C} \rightarrow Z_{\widetilde{G}} \rightarrow Z_{G} \rightarrow 0 \\
0 & \rightarrow \mathcal{C}^{\vee} \rightarrow \widetilde{\mathcal{A}}^{\vee} \rightarrow \mathcal{A}^{\vee} \rightarrow 0 ,
\end{align}
we now have:
\begin{align}\label{eq:Vgeneral}
0 & \rightarrow \mathbb{Z}_{m g''}\times \Z_m \rightarrow \mathbb{Z}_{g_1'} \times  \mathbb{Z}_{g_2'}  \times  \mathbb{Z}_{g_3'}  \rightarrow \lb \mathbb{Z}_{g_1'} \times  \mathbb{Z}_{g_2'}  \times  \mathbb{Z}_{g_3'} \rb /\lb \mathbb{Z}_{m g''} \times \Z_m\rb \rightarrow 0 \\
0 & \rightarrow \Z_{ m \;\!g''} \times \Z_m \rightarrow \mathbb{Z}_n\times \Z_m \rightarrow \Z_{m'/g''}\rightarrow 0. \label{secondo}
\end{align}
Here we introduced $g''=\textnormal{gcd}(m',k_1)\;\! \textnormal{gcd}(m',k_2)\;\! \textnormal{gcd}(m',k_3)$. Expanding out,
the flavor symmetry takes the form:
\be \ba\label{eq:FlavorGroupQuotient}
G&=\frac{SU(m\;\! \textnormal{gcd}(m',k_1))\times SU(m\;\! \textnormal{gcd}(m',k_2)) \times SU(m\;\! \textnormal{gcd}(m',k_3))}{\Z_{m\;\! \textnormal{gcd}(m',k_1)\;\! \textnormal{gcd}(m',k_2)\;\! \textnormal{gcd}(m',k_3)}\times \Z_m}\,.
\ea \ee
The embedding of $\mathcal{C}\cong \Z_{m\;\! \textnormal{gcd}(m',k_1)\;\! \textnormal{gcd}(m',k_2)\;\! \textnormal{gcd}(m',k_3)}\times \Z_m$ into the center $Z_{\widetilde{G}}=Z_{\widetilde{G}_1}\times Z_{\widetilde{G}_2}\times Z_{\widetilde{G}_3}$ is characterized in terms of the generators $\omega, \eta$ of $\Z_n,\Z_m$ respectively. We have
\be
Z_{\widetilde{G}_i}\cong\Z_{m\;\! \textnormal{gcd}(m',k_i)}=\langle  \omega^{c_i}\eta^{c_i'} \rangle \,, \qquad \mathcal{C}=\langle \omega^{c_1}\eta^{c_1'} ,\omega^{c_2}\eta^{c_2'} , \omega^{c_3}\eta^{c_3'}  \rangle
\ee
with integers $c_i,c_i'$ computed in Appendix \ref{app:AppendixSU3}. Owing to our specific parametrization of the group action we have $c_1=n/\textnormal{gcd}(n,k_1)$ and $c_1'=1$. Therefore $\Z_{m\;\! \textnormal{gcd}(m',k_1)}= \Z_{ \textnormal{gcd}(m',k_1)} \times\Z_m \subset \Z_n \times \Z_m$ generated by $\omega^{c_1},\eta$. We can therefore redefine generators as
\be \label{eq:CenterCommon}
\mathcal{C}\cong \langle \omega^{c_1}\eta^{c_1'} ,\omega^{c_2}\eta^{c_2'} , \omega^{c_3}\eta^{c_3'}  \rangle=\langle \omega^{c_1} , \omega^{c_2}, \omega^{c_3}   \rangle \times \langle  \eta \rangle = \langle \omega^{m'/g''} \rangle  \times \langle  \eta \rangle
\ee
where the final step follows from \eqref{eq:gcdc1c2c3}
\be
\textnormal{gcd}(c_1,c_2,c_3)=m'/\textnormal{gcd}(n,k_1)\;\! \textnormal{gcd}(n,k_2) \;\!  \textnormal{gcd}(n,k_3)=m'/g''\,.
\ee
This parametrization explicitly gives the embedding, fixed by mapping generators as
\be \ba\label{eq:embedding}
\mathcal{C} ~&\rightarrow ~ Z_{\widetilde{G}}=  Z_{\widetilde{G}_1} \times Z_{\widetilde{G}_2} \times Z_{\widetilde{G}_3}\,, \qquad
( \omega^{m'/g''},\eta)~&\mapsto~( \omega^{c_1}\eta^{c_1'} ,\omega^{c_2}\eta^{c_2'} , \omega^{c_3}\eta^{c_3'} ) \,.
\ea \ee

For this class of examples we are unaware of a known gauge theory phase not already occurring within the previously analyzed subclasses. Nevertheless, we can check for consistency with previous expressions. Consider for example the case $\Gamma=\Z_N\times \Z_M$ where $M$ divides $N$ belonging to (2)'. We compute $g''=N/M$ and we find trivial one-form symmetry $\mathcal{A}$, further $\mathcal{C}\cong \Z_N\times \Z_M$ matching \eqref{eq:FlavorGroup}.

As an explicit example not contained in the subclass of cases (2)' consider $\Z_9\times \Z_3$ with generators $\frac{1}{9}(1,1,7), \frac{1}{3}(0,1,2)$. We compute $H\cong \Z_3\times \Z_3$ and $\mathcal{A}\cong \Z_3$ and the sequences \eqref{eq:Vgeneral} take the form
\begin{align}
0 & \rightarrow \mathbb{Z}_{3}\times \Z_3 \rightarrow \mathbb{Z}_{3} \times  \mathbb{Z}_{3}  \times  \mathbb{Z}_{3}  \rightarrow \lb \mathbb{Z}_{3} \times  \mathbb{Z}_{3}  \times  \mathbb{Z}_{3} \rb /\lb \mathbb{Z}_{3} \times \Z_3\rb \rightarrow 0 \\
0 & \rightarrow \Z_{ 3} \times \Z_3 \rightarrow \mathbb{Z}_9\times \Z_3 \rightarrow \Z_{3}\rightarrow 0. \label{secondo}
\end{align}
Here $\mathcal{C}\cong \Z_3\times \Z_3=\langle \omega^3,\eta \rangle$ and $\Z_{m\textnormal{gcd}(m',g_i)}\cong \Z_3=\langle\eta \rangle,\langle \omega^3\eta^2 \rangle,\langle \omega^3\eta \rangle$. In \eqref{eq:embedding} we have explicitly $c_i=0,3,3$ and $c_i'=1,2,1$. For this example we also have a non-trivial 2-group.

\section{Elliptically Fibered Calabi-Yau Threefolds} \label{sec:ELLIPTIC}

In the previous section we focused on the special case where the M-theory background $X$ is defined by a global orbifold. Since the prescription of section \ref{sec:PRESCRIPTION} involves cutting and gluing the data of localized orbifold singularities, we expect it to apply to more general backgrounds. In this section we consider the special case of SQFTs obtained from M-theory on $X \rightarrow B$ an elliptically fibered Calabi-Yau threefold with section. In the closely related context of F-theory on an elliptically fibered Calabi-Yau threefold \cite{Vafa:1996xn, Morrison:1996na, Morrison:1996pp}, we get a 6D theory. Degenerations in the elliptic fiber provide a method for engineering gauge theories coupled to matter in different representations \cite{Bershadsky:1996nh, Katz:1996xe}. Moreover, compactification of F-theory on an elliptic $X \rightarrow B$ with a canonical singularity provides a general template for engineering 6D SCFTs \cite{Heckman:2013pva, Heckman:2015bfa}. Starting from such a  6D theory, compactification on a circle leads, in the limit of small circle size to a corresponding M-theory background on the same Calabi-Yau at large volume for the elliptic fiber. In this limit, we get a 5D SQFT when $X$ is non-compact. Moreover, further decoupling limits in the moduli space provide a general way to realize 5D SCFTs from compactification of 6D SCFTs \cite{DelZotto:2017pti}. More generally, a fruitful way to analyze 6D F-theory backgrounds is to instead treat their M-theory avatars since in this limit the blowup modes of the singular fiber are part of the 5D physical moduli space.

Now, the singular elliptic fibers occur at components of the discriminant locus of a Weierstrass model, which in affine coordinates can be written as:
\begin{equation}\label{Weierstrass}
y^2 = x^3 + fx + g.
\end{equation}
Over codimension one subspaces of the base $B$, there is a Kodaira classification of possible degenerations in the elliptic curve, as specified by the order of vanishing of $f$, $g$ and the discriminant locus $\Delta$ (see e.g. \cite{Morrison:1996na, Morrison:1996pp, Bershadsky:1996nh}), and in F-theory terms these specify a 7-brane. In the geometry of the singular fiber, this can be seen in terms of an affine Dynkin diagram of ADE type, the additional node indicating that we are dealing with a singular elliptic curve. Upon reduction on a circle, these flavor 7-branes descend to flavor 6-branes of the M-theory background.

Precisely because this is so close to the case of an orbifold singularity, we expect that our prescription of section \ref{sec:PRESCRIPTION} carries over to this case as well, where here, the flavor branes originate from non-compact singular Kodaira fibers. The main technical complication is how to properly treat the additional contribution from the elliptic fiber class. An additional benefit of treating this case in detail is that it will illustrate how we can also incorporate additional structures in flavor symmetries such as non-simply laced flavor groups. In the elliptically fibered model, this arises through the rearrangement of cycles in the singular fiber due to monodromy around some components in the base $B$ (see e.g. \cite{Bershadsky:1996nh}). We do not treat the case of ``frozen'' singularities \cite{Witten:1997bs, Tachikawa:2015wka, Bhardwaj:2018jgp} but expect that a suitable notion of gluing in singular homology and / or orbifold homology can also be extended to this case as well.

In the remainder of this section we show how our general prescription from section \ref{sec:PRESCRIPTION} applies to the case of $X \rightarrow B$ an elliptically fibered Calabi-Yau. We begin by showing how to generalize the prescription of section \ref{sec:PRESCRIPTION} to the case with singular elliptic fibers. We then turn to some examples of 5D SQFTs as obtained from the dimensional reduction on a circle of certain 6D SCFTs where the $B$ consists of a single linear chain of collapsing curves. The special case of the 5D SQFT obtained from reduction of 6D $(G,G)$ conformal matter is treated next. As a final example, we consider a case where the flavor symmetry algebra is not simply laced.

\subsection{Cutting and Gluing Elliptic Singularities}

We now show how to extend the prescription of section \ref{sec:PRESCRIPTION} to the case where we have singular elliptic fibers. In an elliptically fibered threefold $X \rightarrow B$, the corresponding discriminant locus decomposes into a collection of complex codimension one subspaces in $B$ which can possibly intersect further. To begin, then, we focus on the case of complex codimension one, which we can essentially treat by working with a twofold, and then we turn to how these building blocks fit together in a threefold.

As a warmup, we first treat the case of a single smooth component of the discriminant locus in an elliptically fibered non-compact twofold $\pi: Y \rightarrow \mathbb{C}$ and a marked point $0$ at which the elliptic fiber degenerates.
Our interest will be in the fiber $\mathbb{E}_{0} = \pi^{-1}(0)$. The fiber $ \mathbb{E}_0 = \pi^{-1}(0)$
is a degenerate elliptic curve with singular point $p\in \mathbb{E}_0$. Even though $\mathbb{E}_0$ is not smooth, we can still speak of a singular homology group $H_{1}(\mathbb{E}_0)$. Our main condition for counting such cycles is simply to require that in passing it around the geometry, a candidate 1-cycle does not shrink to zero size. Based on this, the Kodaira classification of singular elliptic fibers tells us that only fibers of type $I_n$ contain a non-trivial 1-cycle (the circle of the affine $\widehat{A}_{n-1}$ Dynkin diagram). Labelling the Kodaira
fiber type as $\Phi$, we have:
\be\label{eq:Singularfiber}
H_1(\mathbb{E}_0) = \begin{cases} \, 0\,,  \quad \,\Phi \neq I_n  \\ \,\Z \,, \quad  \Phi = I_n  \end{cases}
\ee
In more detail, this follows from crepant resolution of $Y$ where the singular fiber is blown up to a collection of rational curves, which contain a 1-cycle only in the case of $I_n$ singularities. When $\Phi\neq I_{n}$ the fiber $\mathbb{E}_{0}$ is topologically a sphere, and when $\Phi= I_{n}$ it is a pinched torus. This also implies that if we now delete the singular point $p$ from $\mathbb{E}_0$, then we have:
\be\label{eq:SingularfiberMinusPoint}
H_1(\mathbb{E}_0 \setminus p)=\begin{cases} \, 0\,,  \quad \,\Phi \neq I_n  \\ \,\Z \,, \quad  \Phi = I_n  \end{cases}
\ee
where, if the second case in \eqref{eq:Singularfiber} is generated by the $b$-cycle of $\mathbb{E}_0$, then the second case in \eqref{eq:SingularfiberMinusPoint} is generated by the conjugate $a$-cycle.\footnote{This is simply because in deleting $p$, we have destroyed the original $b$-cycle, but we can now consider a new non-contractible 1-cycle which encircles $p$.}

Next we compare $Y\setminus \mathbb{E}_0$ and $Y\setminus p$. The former deformation retracts onto a smooth elliptic fibration over a circle linking the origin of $\mathbb{C}$. Note that the homology groups of a manifold $X\rightarrow S^1$ fibered over a circle with fiber $Z$ are determined by the short exact sequence
\be \label{eq:SES}
0~\rightarrow ~ \textnormal{coker}\lb M_n-1\rb ~\rightarrow ~ H_n(X) ~\rightarrow ~ \textnormal{ker}\lb M_{n-1}-1\rb  ~\rightarrow ~ 0
\ee
where $M_n: H_n(Z)\rightarrow H_n(Z)$ is the monodromy map about the base circle lifted to $n$-cycles. We use this sequence repeatedly throughout this section and applied to the configuration at hand we have
\be
0  ~\rightarrow ~ \textnormal{coker}\lb M_n-1\rb~\rightarrow ~ H_n\lb Y\setminus \mathbb{E}_0 \rb~\rightarrow ~\textnormal{ker}\lb M_{n-1}-1\rb ~\rightarrow ~ 0
\ee
where $M_n: H_n(\mathbb{E})\rightarrow H_n(\mathbb{E})$ are the monodromy mappings on smooth fibers, of which only $M_1$ is non-trivial.
The first homology group of $Y\setminus \mathbb{E}_0$ is thus given by:
\be\label{eq:smoothfibers}
H_1(Y\setminus \mathbb{E}_0)=\Z\oplus \textnormal{coker}\lb M_1-1\rb=\begin{cases} \, \Z \oplus \mathrm{Ab}[\Gamma_{\Phi}] \,,  \quad  \Phi \neq I_n  \\ \,\Z^2\oplus \Z_{n+1} \,, \quad ~\: \Phi = I_n.  \end{cases}
\ee
where $\Gamma_{\Phi} \subset SU(2)$ a finite subgroup of ADE type associated with the ADE singularity $\mathbb{C}^2 / \Gamma_{\Phi}$ supported at $p$. The torsion cycles appearing in \eqref{eq:smoothfibers} are the same as those in the link of the ADE singularity. So as anticipated, for the case of flavor branes generated by singular elliptic fibers we can read off the torsional 1-cycles determined by the ADE type of the singularity from $Y\setminus \mathbb{E}_0$.

To explain this point in more detail, let $T$ denote a tubular neighbourhood of $\mathbb{E}_0\setminus p$ in $Y\setminus p$. Then, $ (Y\setminus \mathbb{E}_0)\cap T$ is a fibration over a circle with fibers homologous to $\mathbb{E}_0\setminus p$. Therefore
\be\label{eq:Intersection}
H_1((Y\setminus \mathbb{E}_0)\cap T)= H_1(S^1)\oplus H_1(\mathbb{E}_0\setminus p)=\begin{cases} \, \Z \,,  \quad \,\,\Phi \neq I_n  \\ \,\Z^2 \,, \quad  \Phi = I_n,  \end{cases}
\ee
where we have made use of the fact that \eqref{eq:SingularfiberMinusPoint} is generated by the monodromy invariant $a$-cycle when the fiber $\Phi = I_n $. Note that all these groups fit into the Mayer-Vietoris sequence for $Y\setminus p=(Y\setminus \mathbb{E}_0)\cup (\mathbb{E}_0\setminus p)$ which reads
\be
0 ~\rightarrow ~H_1((Y\setminus \mathbb{E}_0)\cap T) ~\rightarrow ~ H_1((Y\setminus \mathbb{E}_0))\oplus H_1(T)~\rightarrow ~H_1(Y\setminus p)~ \rightarrow ~0
\ee
where $H_1(T)=H_1(\mathbb{E}_0\setminus p)$ and $H_1(Y\setminus p)= \mathrm{Ab}[\Gamma_{\Phi}]$, as follows from the contractibility of $T$ to $\mathbb{E}_0\setminus p$ and the point $p$ supporting an ADE singularity. We have further used the fact that both components of the covering are connected and that $H_2(Y\setminus p)=0$ for ADE singularities.

With this building block in place, we now turn to the case of $X$ a non-compact elliptically fibered Calabi-Yau threefold. Consider the elliptic Calabi-Yau threefold $\pi: X\rightarrow B$ with discriminant locus $\Delta$ and singular fibers $F=\pi^{-1}(\Delta)$. We leave the compactly supported components implicit and denote the intersection of the non-compact components with the boundary by $\partial \Delta_i,K_i,\partial F_i$ with $i=1,\dots, N$. Here, we recall that $K_i$ denotes the locus of the flavor brane in the boundary geometry. Next note the nested inclusion
\be
K_i \subset \partial F_i \subset \partial X\,,
\ee
which gives three complements on the boundary
\be
\partial X_{F} =\partial X \setminus \cup_{i\;\!}  \partial F_i\,, \qquad \partial X^\circ=\partial X \setminus \cup_{i\;\!}   K_i\,,\qquad \partial F^\circ =\cup_{i\;\!}( \partial F_i\setminus K_i)\,.
\ee
Denoting by $T(K)$ a tubular neighborhood of $K$ in $\partial X$, we have the following coverings
\be
\partial X=\partial X^\circ \cup T(K)\,, \qquad \partial X^\circ = \partial X_{F} \cup T(\partial F^\circ)\,.
\ee
We first consider the Mayer-Vietoris sequence for the latter covering in degree one. It takes the form
\be\label{eq:MVElliptic1}
\dots~\xrightarrow[]{\,\partial_{2} \,} ~H_{1}\big(\partial X_{F} \cap T(\partial  F^\circ ) \big)  ~\xrightarrow[]{\,\iota_{1} \,} H_1\big(\partial X_{F}\big)\oplus H_1\big(T(\partial  F^\circ)\big) ~\xrightarrow[]{j_1-\ell_1 }  ~  H_1\big(\partial X^\circ \big) ~\xrightarrow[]{\,\partial_1\, }  ~0
\ee
The intersection $\partial X_{F} \cap T(\partial  F^\circ)$ projects to the base, with fibers homologous to $\partial F^\circ$. In the base we can split the geometry into parts tangential and normal to the discriminant. Then, restricting $\partial X_{F} \cap T(\partial  F^\circ)$ to the normal component, we observe that the local geometry near each component of the discriminant locus is precisely of the form already discussed in the special case of a twofold $Y \rightarrow \mathbb{C}$, but in which we fibered over a (boundary) circle. From \eqref{eq:Intersection} it now follows that for the different fiber types $\Phi$, we have:
\be
H_{1}\big(\partial X_{F} \cap T(\partial  F_i^\circ ) \big) =\begin{cases} \, \Z^2 \,,  \quad \qquad~ \Phi \neq I_{m}^{\textnormal{s}},I_{m}^{\textnormal{ns}}  \\ \,\Z^3 \,, \quad\qquad ~\Phi = I_{m}^{\textnormal{s}} \\\, \Z^2 \oplus \Z_2 \,,  \quad \Phi=I_{m}^{\textnormal{ns} }
 \end{cases}
\ee
where there is a universal factor of $\Z^2$ generated by a torus enclosing $\pi(\partial F_i)=\partial\Delta_i$ in the base. The remaining factor of $\Z$ or $\Z_2$ is generated by an $a$-cycle in the local geometry. Now, observe that $T(\partial  F_i^\circ ) $ deformation retracts to $\partial  F_i^\circ$.
We therefore have:
\be
H_{1}\big(\partial X_{F} \cap T(\partial  F_i^\circ ) \big)\cong H_1(T(\partial  F_i^\circ)) \oplus \Z \cong H_1(\partial  F_i^\circ) \oplus \Z
\ee
with the factor of $\Z$ generated by a circle linking the boundary discriminant component $\pi(\partial F_i)$ in the base. We can therefore remove a copy of $H_1(\partial  F_i^\circ)$ from the Mayer-Vietoris sequence by exactness and find
\be\label{eq:MVElliptic1}
\dots~\xrightarrow[]{} ~ \Z^N  ~\xrightarrow[]{\,\iota_{1} \,} H_1\big(\partial X_{F}\big) ~\xrightarrow[]{}  ~  H_1\big(\partial X^\circ \big) ~\xrightarrow[]{ }  ~0.
\ee
This is a useful simplification and allows us to compute $H_1\big(\partial X^\circ \big)$ from $H_1\big(\partial X_{F}\big)$. The latter is more easily computed from the elliptic fibration. Now \eqref{eq:MVElliptic1} immediately implies
\be\label{eq:Split}
\textnormal{Tor}\, H_1(\partial X^\circ)= \textnormal{Tor}\,H_1 (\partial X_F) \oplus \textnormal{Tor}\,H_1(\partial  B)
\ee
where the cycles $ \textnormal{Tor}\,H_1(\partial  B)$ are understood as torsional cycles in the total space by lifting them via the section.

At this point it should be clear that the prescription of section \ref{sec:PRESCRIPTION} does indeed extend to the case of elliptically fibered Calabi-Yau spaces with suitable non-compact components of the discriminant locus serving as flavor brane loci. We now apply this to some specific examples

\subsection{Generalized A-Type Bases}

We now apply this formalism in a large class of examples where the base of the elliptically fibered threefold $X \rightarrow B$ consists of a single spine of curves, intersecting according to a generalized A-type Dynkin diagram, but where we do not necessarily require all curves to have self-intersection $-2$. This situation occurs in the vast majority of 6D SCFTs engineered via F-theory \cite{Heckman:2013pva, Heckman:2015bfa}, but can also include more general 6D theories SQFTs and their reduction to 5D SQFTs \cite{Morrison:2012np} (see also \cite{Bhardwaj:2015oru}).

In what follows, we focus on the case where the geometry of the base is taken to be of generalized A-type:
\be\label{eq:Atypebase}
\begin{tikzpicture}
	\begin{pgfonlayer}{nodelayer}
		\node [style=none] (3) at (-5.75, 1.25) {$\big[\mathfrak{g}_1\big]$};
		\node [style=none] (4) at (1.75, 1.25) {$(k_{N})$};
		\node [style=none] (5) at (-5.25, 1.25) {};
		\node [style=none] (7) at (-4.75, 1.25) {};
		\node [style=none] (8) at (-3.75, 1.25) {};
		\node [style=none] (9) at (-3.25, 1.25) {};
		\node [style=none] (11) at (-4.25, 1.25) {$(k_1)$};
		\node [style=none] (12) at (3.25, 1.25) {$\big[\mathfrak{g}_N\big]$};
		\node [style=none] (13) at (2.75, 1.25) {};
		\node [style=none] (14) at (2.25, 1.25) {};
		\node [style=none] (15) at (-1.25, 1.25) {$(k_i)$};
		\node [style=none] (16) at (-1.25, 2.25) {$\big[\mathfrak{g}_i\big]$};
		\node [style=none] (17) at (-1.25, 1.9) {};
		\node [style=none] (18) at (-1.25, 1.6) {};
		\node [style=none] (21) at (-3, 1.25) {};
		\node [style=none] (22) at (-2.5, 1.25) {};
		\node [style=none] (23) at (-2.25, 1.25) {};
		\node [style=none] (24) at (-1.75, 1.25) {};
		\node [style=none] (25) at (-0.75, 1.25) {};
		\node [style=none] (26) at (-0.25, 1.25) {};
		\node [style=none] (27) at (0, 1.25) {};
		\node [style=none] (28) at (0.5, 1.25) {};
		\node [style=none] (29) at (0.75, 1.25) {};
		\node [style=none] (30) at (1.25, 1.25) {};
	\end{pgfonlayer}
	\begin{pgfonlayer}{edgelayer}
		\draw [style=ThickLine] (5.center) to (7.center);
		\draw [style=ThickLine] (8.center) to (9.center);
		\draw [style=ThickLine] (14.center) to (13.center);
		\draw [style=ThickLine] (17.center) to (18.center);
		\draw [style=DottedLine] (21.center) to (22.center);
		\draw [style=ThickLine] (23.center) to (24.center);
		\draw [style=ThickLine] (25.center) to (26.center);
		\draw [style=DottedLine] (27.center) to (28.center);
		\draw [style=ThickLine] (29.center) to (30.center);
	\end{pgfonlayer}
\end{tikzpicture}
\ee
Here the dots denote a linear chain of $N$ rational curves of self-intersection $-k_i$. We have also indicated the flavor symmetry algebra associated with a non-compact components of the discriminant locus by their corresponding Lie algebra $\mathfrak{g}_i$.
We further allow for non-compact discriminant loci $\Delta_i$ which we assume to intersect the boundary along Hopf circles
\be
\label{eq:assumption2}
\partial B \cap \Delta_i=\partial \Delta_i = S^1_i\,.
\ee
Part of our task will be to extract the global form of the flavor symmetry group directly from the boundary geometry.

Now, for generalized A-type bases, the linking boundary geometry is always of the form:
\begin{equation}
\partial B = S^3 / \mathbb{Z}_p,
\end{equation}
where the specific value of $p$ as well as the choice of group action $\frac{1}{p}(1,q)$ is obtained from the Hirzebruch-Jung continued fraction \cite{jung,MR0062842,Riemen:dvq}:
\begin{equation}
\frac{p}{q} = k_1 - \frac{1}{k_2 - ... \frac{1}{k_N}}.
\end{equation}
The base is permitted to contain any number of compact curves, supporting
arbitrary gauge algebras consistent with anomaly cancellation, i.e. the existence
of an elliptic fibration. The possible Calabi-Yau geometries of this
sort were classified in \cite{Heckman:2013pva, Heckman:2015bfa}.

Let us now turn to the 1-form symmetry of these systems. Geometrically, we are interested in non-compact 2-cycles which can extend out to the boundary. There is of course the contribution from non-compact 2-cycles supported purely in the base, and this will always contribute a factor of $\mathbb{Z}_p$ to the 1-form symmetry. The main challenge is to properly track the profile of the flavor components of the discriminant locus. To this end, we divide our discussion into two separate cases. First, we consider the case where we have no non-compact $I_n$ type fibers. We then turn to the case where there are possible $I_n$ fibers. We shall refer to a non-trivial identification in the basis of resolution cycles as a ``non-split'' fiber and the case of no identification as a ``split'' fiber, as in \cite{Bershadsky:1996nh}.

The first general comment is that the existence of any fiber which is not of $I_n$ type means that its corresponding $H_1(\mathbb{E}_{\mathrm{fiber}})$ is trivial. Consequently, we can use this fact to trivialize additional candidate 1-cycles in the boundary geometry. On the other hand, if all the fibers are of $I_n$-type, then there is the possibility that there is an additional contribution to the 1-form symmetry. In this case, there is again at most one non-trivial representative, so we conclude that the flavor symmetry could potentially contribute an additional $\mathbb{Z}_d$ factor, where $d$ depends on the details of the geometry in question.

We now explain this general point in more detail. Suppose first that the flavor locus has no $I_n$ type fibers.
We claim that in this case,
the 1-form symmetry of theories of type \eqref{eq:Atypebase} is \cite{Hubner:2022kxr}:
\be\label{eq:No2Group}
\mathcal{A}= \Z_p\,, \qquad \exists\:  \Phi_{i}\neq I_{n_{i}}^{\textnormal{s}},I_{n_{i}}^{\textnormal{ns}}
\ee
where the superscript $\mathrm{s}$ and $\mathrm{ns}$ refers to a split or non-split Kodaira fiber. Here, the $\Z_p$ is generated by the non-compact cycle of the base generating $H_1(\partial B)=\Z_p$ lifted to the total space by the section of the elliptic fibration. The absence of any other fibral contributions to the 1-form symmetry follows directly from equation \eqref{eq:Singularfiber}. Indeed, consider a non-compact 2-cycle of the bulk intersecting the boundary in a fibral 1-cycle. This cycle projects a semi-infinite line in the base and intersects the base boundary in a single point. Whenever there exists a non-compact discriminant component $S^1_i$ supporting a component of the discriminant with singular fiber $\Phi_{i}\neq I_{n_i}^{\textnormal{s}},I_{n_i}^{\textnormal{ns}}$, we can continuously deform this intersection point to that locus. The 1-cycle fibering the non-compact 2-cycle then necessarily collapses as the singular fibers have no 1-cycles of their own. Consequently, the non-compact 2-cycle is trivial in relative homology to begin with and does not contribute to $\mathcal{A}$. Note that all that is required to perform this analysis is that we have at least one fiber which is not of $I_n$ type. For a more detailed discussion on such structures see \cite{Hubner:2022kxr}.

Suppose now that there are only flavor loci supporting $I_n$ type singular fibers.
Whenever the $S^1_i$ exclusively support singularities of the type $ I_{n_i}^{\textnormal{s}},I_{n_i}^{\textnormal{ns}}$ we find \cite{Hubner:2022kxr}
\be\label{eq:Maybe2Group}
\mathcal{A}= \Z_p\oplus \Z_d\,, \qquad \forall \,\Phi_{i}=I_{n_i}^{\textnormal{s}},I_{n_i}^{\textnormal{ns}}
\ee
by the same arguments used above, since at least one 1-cycle of the elliptic fiber still collapses somewhere on the boundary. Next note that in both cases the base contribution $\Z_p$ does not arise in the geometry from an ADE locus. It cannot be detected restricting to their tubular neighbourhoods. Following our characterization \eqref{eq:chatgeo} we see that it does not participate in 2-group structures. Theories of line \eqref{eq:No2Group} therefore have no 2-group, but there is a chance that a 2-group will appear in theories of line (\ref{eq:Maybe2Group}).

As an additional comment, we note that for more general bases of the form $\mathbb{C}^2 / \Gamma_{U(2)}$ with $\Gamma_{U(2)}$ a finite subgroup of $U(2)$, we can also extract the contribution to the 1-form symmetry from the base geometry \cite{DelZotto:2015isa}.
Indeed, this contribution will simply be $\mathrm{Ab}[\Gamma_{U(2)}]$. The subtlety here is that whereas the boundary geometry for the
generalized A-type bases retain a simple characterization in terms of a Hopf fibration (which we used to analyze the flavor discriminant), in the more general setting there are some additional technical complications. Nevertheless, it is quite natural to expect that in this case as well, the existence of any fiber which is not of $I_l$-type would immediately trivialize any additional fibral contributions to the 1-form symmetry.

Having illustrated some general properties of models with a generalized A-type base, we now turn to the explicit computation of the various higher symmetries in some specific examples. To this end, it will be helpful to note that for generalized A-type bases, we can write equation \eqref{eq:Split} as:
\be\label{eq:CoolFormula}
\textnormal{Tor}\,H_1\big(\partial X^\circ \big) =\textnormal{Tor}\,H_1\big(\partial X_{F}\big)\oplus  \Z_p.
\ee
As far as characterizing the global form of the flavor symmetry and possible 2-group structures, the base is largely a spectator. Instead, all of this structure is dictated by the geometry $\partial X_{F}$.

We now proceed to some examples.

\subsection{5D Conformal Matter}

Let us now turn to some examples involving 5D conformal matter \cite{DelZotto:2017pti},
as obtained from the circle reduction of the partial tensor
branch deformation of 6D conformal matter \cite{DelZotto:2014hpa, Heckman:2014qba}.
This is realized by an elliptically fibered Calabi-Yau threefold with partial resolution given by:
\begin{equation}
[\mathfrak{g}_L] \,\, \overset{\mathfrak{g}}{2 }\,\,... \overset{\mathfrak{g}}{2 } \,\, [\mathfrak{g}_{R}],
\end{equation}
namely we have a collection of $-2$ curves intersecting according to an A-type Dynkin diagram. Over each curve we have a singular Kodaira fiber which yields a corresponding Lie algebra of type $\mathfrak{g}$ of ADE type. Further blowups in the base are needed to
get all fibers into Kodaira-Tate form, but this will not be needed in the discussion to follow.

As a general comment, it is well-known in the context of 6D SCFTs that there can be various enhancements in the flavor symmetry, and this often occurs when we have a low number of $-2$ curves. For our purposes here, however, we are primarily interested in these systems as 5D SQFTs, so we expect that various irrelevant operators generated by the explicit string compactification will lead to agreement between the answer we get from geometry, and what we expect from bottom up considerations. That being said, one can expect that in the limit where these irrelevant operators decouple from the physics, that there could be additional enhancements. It is also well-known that such ``accidents'' do not occur when the number of $-2$ curves is sufficiently large, but they certainly do arise at low rank gauge groups, and low numbers of $-2$ curves.

Indeed, it will prove simplest to first treat the case where we have no interior $-2$ curves, i.e. the base $B$ blows down to just $\mathbb{C}^2$. Using this building block, we can then quickly extend this analysis to the more general situation where $B = \mathbb{C}^2 / \mathbb{Z}_p$, i.e. the case of $N = p - 1$ curves of self-intersection $-2$.

\subsubsection{$\partial B = S^3$}

With this in mind, we first treat the case of $(G,G)$ conformal matter where $G$ is of ADE type, and where the base is just $B = \mathbb{C}^2 \equiv \mathbb{C}_1 \times \mathbb{C}_2$ so that $\partial B = S^3$. The flavor locus arises from singularities tuned on $\mathbb{C}_1\times \lbbb0\rbbb$ and $ \lbbb0\rbbb\times \mathbb{C}_2$. These intersect transversely and give a Hopf link in the boundary three-sphere. We view the boundary three-sphere as torus-fibered over the interval
\be
T^2 ~\hookrightarrow~ S^3 ~\rightarrow~ I\,.
\ee
The singularities $\Phi_{L,R}$ supported on Hopf fibers project to the ends of the interval $I$. Deleting these, we obtain a torus fibration over an open interval.  This space deformation retracts onto the torus fiber $T^2=S^1_L\times S^1_R$. Here $S^1_{L,R}$ links the circles supporting $\Phi_{L,R}$ respectively. The deformation retraction of $\partial X_F$ is therefore fibered as
\be
\mathbb{E} ~\hookrightarrow~ \partial X_F^{(r)} ~\rightarrow~ S^1_L\times S^1_R\,,
\ee
We now repeatedly apply \eqref{eq:SES} by first flipping either of the base circles into the fiber to define fibrations by three-manifolds
\be\ba
Y_3^{L} ~\hookrightarrow~ &\partial X_F^{(r)} ~\rightarrow~ S^1_R\\
Y_3^{R} ~\hookrightarrow~ &\partial X_F^{(r)} ~\rightarrow~ S^1_L\,.
\ea\ee
where $Y_3^{L,R} $ are themselves fibered over $S^1_{L,R}$. Now to compute the homology groups of $\partial X_F^{(r)}$ we first compute the homology groups of $Y_3^{L,R}$ using the short exact sequence
\be
0  ~\rightarrow ~ \textnormal{coker}\lb M_n^{L,R}-1\rb~\rightarrow ~ H_n\lb Y_3^{L,R}\rb~\rightarrow ~\textnormal{ker}\lb M_{n-1}^{L,R}-1\rb ~\rightarrow ~ 0
\ee
which follows from the Mayer-Vietoris sequence for spaces fibered over circles. Here
\be
M_n^{L,R}\, :\quad  H_n(\mathbb{E}) ~\rightarrow ~H_n(\mathbb{E})
\ee
are the monodromy mappings about $S^1_{L,R}$. We have
\be
H_1\lb Y_3^{L,R}\rb=\Z \oplus \textnormal{coker}\lb M_1^{L,R}-1\rb
\ee
Now we repeat for the remaining circle. The sequence reads
\be
0  ~\rightarrow ~ \textnormal{coker}\lb M_n^{R}-1\rb~\rightarrow ~ H_n\lb Y_3^{L}\rb~\rightarrow ~\textnormal{ker}\lb M_{n-1}^{R}-1\rb ~\rightarrow ~ 0
\ee
and we derive the key formula
\be\label{eq:DoubleQuotient}
H_1(\partial X_F^{})=H_1(\partial X_F^{(r)})=\Z^2\oplus \frac{\Z^2}{\textnormal{Im}(M_1^{L}-1)\cup \textnormal{Im}(M_1^{R}-1)}
\ee
where we have written out the cokernels and which is symmetric upon interchanging $L \leftrightarrow R$. Note analogous considerations determine the homology goup in \eqref{eq:chatgeo} to
\be\label{eq:LocalNhoods}
\textnormal{Tor}\,H_{1}\big(\partial X^\circ\cap T(K) \big) = \textnormal{Tor}\lb \textnormal{coker}\lb M_1^{L}-1\rb\rb \oplus \textnormal{Tor}\lb  \textnormal{coker}\lb M_1^{R}-1\rb \rb
\ee
and consequently the torsion subgroup in \eqref{eq:DoubleQuotient} sits diagonally in \eqref{eq:LocalNhoods}. That is we have
\be
\iota_1\,:\quad \textnormal{Tor}\,H_{1}\big(\partial X^\circ\cap T(K) \big) ~\rightarrow~  \textnormal{Tor}\,H_1(\partial X_F^{})
\ee
acts on the factors labelled by $L,R$ via quotienting by the images of $M_1^{R,L}-1$. We now make these maps explicit in a number of examples.

\medskip

\ni {\bf Transversely Intersecting $I_n,I_m$:} This setup engineers a hypermultiplet in the bifundamental representation of $\mathfrak{su}(n)\times \mathfrak{su}(m)$. The theory of the hypermultiplet consists of its kinetic terms as well as additional
irrelevant operator interactions which explicitly break the ``accidental'' enhancement back to $\mathfrak{su}(n+ m)$.
In this case we have:
\be
\textnormal{Tor}\,H_1(\partial X_F^{})=\Z_{\textnormal{gcd}(n,m)}\,,
\ee
and the map
\be
\iota_1:\Z_n\oplus \Z_m\rightarrow \Z_{\textnormal{gcd}(n,m)}
\ee acts on the first (resp. second) factor by modding out by $m$ (resp. $n$).
We have
\be
\textnormal{Tor}\,H_1(\partial X^\circ)=\Z_{\textnormal{gcd}(n,m)}
\ee
and
\be
\mathcal{C}^\vee= \Z_{\textnormal{gcd}(n,m)}\,.
\ee
We find the non-abelian flavor symmetry
\be
G_{\mathrm{non-ab}} = \frac{SU(n) \times SU(m)}{\mathbb{Z}_{\mathrm{gcd}(n,m)}},
\ee
where $\Z_{\textnormal{gcd}(n,m)}$ embeds diagonally in the two factors.
This is expected because bifundamental matter fields do not transform under this common diagonal sector.

The short exact sequences
\begin{align}
0 &~ \rightarrow~ \mathcal{C} ~\rightarrow~ Z_{\widetilde{G}} ~\rightarrow~ Z_{G} \rightarrow 0 \\
0 & ~\rightarrow~ \mathcal{C}^{\vee} ~\rightarrow~ \widetilde{\mathcal{A}}^{\vee} ~\rightarrow~ \mathcal{A}^{\vee} \rightarrow 0,
\end{align}
now respectively take the form
\begin{align}
0 &~ \rightarrow~  \Z_{\textnormal{gcd}(n,m)} ~\rightarrow~ \Z_{n}\oplus \Z_m ~\rightarrow~  \frac{\Z_{n}\times \Z_m}{\Z_{\textnormal{gcd}(n,m)}} \rightarrow 0 \\
0 & ~\rightarrow~  \Z_{\textnormal{gcd}(n,m)} ~\rightarrow~ \Z_{\textnormal{gcd}(n,m)} ~\rightarrow~ 0~ \rightarrow 0.
\end{align}
In particular, we note that $\mathcal{A}$ is trivial, and so there is no 1-form symmetry, or possible 2-group.

As a final comment on this case, we note that we have omitted the contribution from the $\mathfrak{u}(1)$ flavor symmetry factor (see \cite{Apruzzi:2020eqi} for further discussion). On the other hand, since we have now determined the non-abelian flavor symmetry to be:
\begin{equation}
G_{\mathrm{non-ab}} = \frac{SU(n) \times SU(m)}{\mathbb{Z}_{\mathrm{gcd}(n,m)}},
\end{equation}
we can piece together that the full flavor group is compatible with:
\begin{equation}
G = S [U(n) \times U(m)].
\end{equation}
Note also that the same considerations will clearly apply in the case of conformal matter with A-type flavor symmetries.

\ni {\bf $(\mathfrak{g},\mathfrak{g})$ Conformal Matter:} Consider conformal matter arising in the collision of two identical singularities $\Phi=\Phi_L=\Phi_R$ individually associated with the Lie algebra $\mathfrak{g}$. Again, this is really to be viewed as the theory of 5D conformal matter deformed by a collection of irrelevant operators which in the $\mathfrak{g} = \mathfrak{su},\mathfrak{so}$ cases explicitly breaks any low rank ``accidental'' enhancements.

By considerations analogous to those in the previous example we have
\be
\textnormal{Tor}\,H_1(\partial X^\circ)=\textnormal{Tor}\,H_1(\partial X_F^{})= \mathrm{Ab}[\Gamma_{\Phi}]\,,
\ee
where we model the flavor brane as a $\mathbb{C}^2 / \Gamma_{\Phi}$ singularity. In this case, the short exact sequences
\begin{align}
0 &~ \rightarrow~ \mathcal{C} ~\rightarrow~ Z_{\widetilde{G}} ~\rightarrow~ Z_{G} \rightarrow 0 \\
0 & ~\rightarrow~ \mathcal{C}^{\vee} ~\rightarrow~ \widetilde{\mathcal{A}}^{\vee} ~\rightarrow~ \mathcal{A}^{\vee} \rightarrow 0,
\end{align}
now respectively take the form
\begin{align}
0 &~ \rightarrow~   \mathrm{Ab}[\Gamma_{\Phi}] ~\rightarrow~  \mathrm{Ab}[\Gamma_{\Phi}] \times  \mathrm{Ab}[\Gamma_{\Phi}] ~\rightarrow~   \frac{\mathrm{Ab}[\Gamma_{\Phi}] \times  \mathrm{Ab}[\Gamma_{\Phi}]}{\mathrm{Ab}[\Gamma_{\Phi}]_{\mathrm{diag}}} \rightarrow 0 \\
0 & ~\rightarrow~   \mathrm{Ab}[\Gamma_{\Phi}] ~\rightarrow~  \mathrm{Ab}[\Gamma_{\Phi}] ~\rightarrow~ 0~ \rightarrow 0.
\end{align}
where in the first line, $ \mathrm{Ab}[\Gamma_{\Phi}]_{\mathrm{diag}} $ embeds diagonally in $ \mathrm{Ab}[\Gamma_{\Phi}]\times  \mathrm{Ab}[\Gamma_{\Phi}]$. Summarizing, the non-abelian flavor group extracted from geometry is just:
\begin{equation}
\mathrm{Non-Abelian\,Flavor} = \frac{G_L \times G_R}{Z_{\mathrm{diag}}},
\end{equation}
where $Z_{\mathrm{diag}}$ is just the diagonal flavor symmetry in the two factors of $G_L = G_R$.

\subsubsection{$\partial B = S^3 / \mathbb{Z}_p$}

Let us now turn to the more general case of higher rank 5D conformal matter, as obtained by taking multiple $-2$ curves. In this case,
the base $B$ has boundary $S^3 / \mathbb{Z}_p$ as dictated by the group action $(z_1, z_2) \mapsto (\omega z_1 , \omega^{-1} z_2)$,
with $\omega$ a primitive $p^{th}$ root of unity. From our general considerations presented earlier, we know that there is now a contribution to the 1-form symmetry, and in all cases it is just $\mathcal{A} = \mathbb{Z}_p$. Indeed, even in the case of $(SU,SU)$ conformal matter, we just saw that the fibers did not generate any contributions. Moreover, we also know from our previous discussion that this contribution to the 1-form symmetry from the base is essentially a spectator, so again, we know that the 2-group structure is trivial.

Turning next to the center flavor symmetry, we observe that in a configuration with a collection of $-2$ curves,
we can consider a limit in which the leftmost curve in the configuration:
\begin{equation}
[\mathfrak{g}_L] \,\, \overset{\mathfrak{g}}{2 }\,\,... \overset{\mathfrak{g}}{2 } \,\, [\mathfrak{g}_{R}],
\end{equation}
expands to large volume. In this case, we get two 5D conformal matter systems, and we can pass back to the original configuration by gauging a diagonal subgroup. By induction, we conclude that since the geometrically determined $\mathcal{C}$ is just the common center of $[\mathfrak{g}] - [\mathfrak{g}]$ conformal matter (with no $-2$ curves), then the process of gluing back together (i.e. by gauging a common diagonal) must retain this factor. Putting this together, we see that the analysis presented in the special case of no $-2$ curves carries through unchanged, and we can again read off the center flavor symmetry:
\begin{equation}
\mathrm{Non-Abelian\,Flavor} = \frac{G_L \times G_R}{\mathcal{C}_{\mathrm{diag}}}.
\end{equation}
where $\mathcal{C}_{\mathrm{diag}}$ is just the diagonal center flavor symmetry in the two factors of $G_L = G_R$.

\subsubsection{$\mathfrak{so}_{8+2m}$ on a $-4$ Curve}

As a final example, we also consider the special case of a single $-4$ curve supporting an $\mathfrak{so}_{8+2m}$ gauge algebra, with matter in the fundamental representation. The flavor symmetry algebra is of $\mathfrak{sp}$-type which is of interest precisely because it is non-simply laced. Our formalism captures such situations as well, as we now demonstrate.

To engineer this case, we consider the elliptic threefold $X$ given by a base with a single curve of self-intersection $-4$. Over the $-4$ curve we take an $I^{\ast,\mathrm{s}}_{m}$ fiber, realizing an $\mathfrak{so}_{8+2m}$ gauge theory. To obtain an anomaly free spectrum, we couple this to $2m$ hypermultiplets in the vector representation of the gauge group. We can arrange for these to be collected into a single $I_{4m}^{\mathrm{ns}}$ fiber, realizing a manifest $\mathfrak{sp}_{2m}$ flavor symmetry.\footnote{In our conventions $\mathfrak{sp}_1 \simeq \mathfrak{su}_2$, and the matter fields transform as half hypermultiplets under the $\mathfrak{sp}_{2m}$ symmetry.}
The local geometry takes the form:
\medskip
\be\label{eq:Quivers}\scalebox{1}{
\begin{tikzpicture}
	\begin{pgfonlayer}{nodelayer}
		\node [style=none] (0) at (0, 0) {$4$};
		\node [style=none] (1) at (0, 0.6) {$\mathfrak{so}_{8+2m}$};
		\node [style=none] (2) at (0.5, 0) {};
		\node [style=none] (3) at (1, 0) {};
		\node [style=none] (4) at (1.8, 0) {$\big[\mathfrak{sp}_{2m}\big]$};
	\end{pgfonlayer}
	\begin{pgfonlayer}{edgelayer}
		\draw [style=ThickLine] (2.center) to (3.center);
	\end{pgfonlayer}
\end{tikzpicture}}
\ee

Let us now turn to the boundary geometry. To begin, we note that the
base $B= {O}_{\P^1}(-4)$, with boundary $B = S^3 / \mathbb{Z}_4$, as induced by the group
action on $\mathbb{C}^2$ given by $(z_1 , z_2) \mapsto (\omega z_1 , \omega z_2)$ with $\omega$ a primitive $4^{th}$
root of unity. The boundary of $X$ fibers as:
\be
\partial X \rightarrow  S^3/\Z_4=\partial B\,,
\ee
The flavor brane is supported on the fiber class of ${O}_{\P^1}(-4)$ and intersects the base boundary on
a single circle $S^1_K$ of the $\partial B$, as discussed in \cite{Hubner:2022kxr}.
The base boundary is a smooth lens space and is fibered  as
\be
S^1 ~\hookrightarrow~ S^3/\Z_4 ~\rightarrow~ S^2\,,
\ee
and therefore deleting the singular fibers from $\partial X$ deletes a copy of the Hopf fiber $S^1_K$ from $\partial B =S^3/\Z_4 $. The base is now fibered over a punctured two-sphere which deformation retracts to a point. The deformation retract of $\partial X_F$ is therefore fibered as
\be
\mathbb{E} ~\hookrightarrow~ \partial X_F^{(r)} ~\rightarrow~ S^1_H\,.
\ee
where $S^1_H$ is a Hopf circle linking the orbifold locus $S^1_K$ in $\partial B$. Now note that $S^1_H$ also links the $\P^1$ in the bulk $B$ as it too is the boundary of a fiber class of ${O}_{\P^1}(-4)$. The elliptic monodromy action along $S^1_H$ is therefore that of an $I^{*}_{m}$ and $I_{4m}$ which read\footnote{Recall that the condition of split versus non-split does not impact the $SL(2,\mathbb{Z})$ monodromy (see e.g. \cite{Bershadsky:1996nh}).}
\be
M_{I_{m}^*}= \lb \begin{array}{cc} -1 & -m \\ 0 & -1 \end{array} \rb\,, \qquad M_{I_{4m}}=\lb \begin{array}{cc} 1 & 4m\\ 0 & 1 \end{array} \rb
\ee
and the total monodromy is
\be
M= \lb \begin{array}{cc} -1 & -5m \\ 0 & -1 \end{array} \rb\,.
\ee
We therefore have
\be
H_1(\partial X_F)=\Z\oplus \begin{cases} \Z_2\oplus \Z_2\,, \quad m \in 2\Z  \\ \Z_4\,, ~\quad\qquad m\in 2\Z+1 \end{cases}
\ee
and we conclude by \eqref{eq:CoolFormula}
\be
\textnormal{Tor}\,H_1(\partial X^\circ)=\begin{cases} \Z_2\oplus \Z_2 \oplus \Z_4 \,, \quad m \in 2\Z  \\ \Z_4\oplus \Z_4\,, ~\quad\qquad m\in 2\Z+1 \end{cases}
\ee
Now note for the boundary we have \cite{Hubner:2022kxr}
\be
\textnormal{Tor}\, H_1(\partial X)=\Z_2\oplus \Z_4
\ee
where $\Z_2, \Z_4$ are base and fiber contributions respectively. The factor of $\Z_2$ follows from the non-split $I_m^{\textnormal{ns}}$ locus. It is generated by the $b$-cycle which does not collapse at the discriminant locus \eqref{eq:Singularfiber}. However, due to the fiber being non-split we have $b\rightarrow -b$ upon traversing $S^1_K$. Therefore $b$ is a $\Z_2$ 1-cycle and we have \eqref{eq:Maybe2Group} with $p=2$. With this the sequence
\be
0 \rightarrow \mathcal{C}^{\vee} \rightarrow \widetilde{\mathcal{A}}^{\vee} \rightarrow \mathcal{A}^{\vee} \rightarrow 0
\ee
takes the form
\be
0 ~\rightarrow~ \Z_2 ~\rightarrow~ \Z_4\oplus \Z_4 ~\rightarrow~ \Z_2 \oplus \Z_4 ~\rightarrow~ 0
\ee
for even $m$, which we can split into base and fiber contributions respectively
\be\ba
0 ~\rightarrow~ \:0~\:\! ~\rightarrow~ &\Z_4 ~\rightarrow~ \Z_4 ~\rightarrow~ 0\\
0 ~\rightarrow~ \Z_2 ~\rightarrow~  &\Z_4 ~\rightarrow~ \Z_2 ~\rightarrow~ 0\,.
\ea\ee
For odd $m$ the analogous sequences take the form
\be\ba
0 ~\rightarrow~ \:0~\:\! ~\rightarrow~ &~~\,~\Z_4~\:\! \:\!  ~~ ~\rightarrow~ \Z_4 ~\rightarrow~ 0\\
0 ~\rightarrow~ \Z_2 ~\rightarrow~  &\Z_2\times \Z_2 ~\rightarrow~ \Z_2 ~\rightarrow~ 0\,.
\ea\ee
In the first case we have a 2-group and in the second case we do not.

From the discussion we see further see that in the map $\iota_1: H_{1}(\partial X^{\circ} \cap T(K)) \rightarrow H_1(\partial X^{\circ}) \oplus H_{1}(T(K))$, we have $\textnormal{ker}\,\iota_1=0$ and therefore the sequence
\be
0 ~ \rightarrow~ \mathcal{C} ~\rightarrow~ Z_{\widetilde{G}} ~\rightarrow~ Z_{G} \rightarrow 0
\ee
takes the form
\be
0 ~ \rightarrow~ \Z_2~\rightarrow~\Z_2 ~\rightarrow~ 0 \rightarrow 0
\ee
and we find the global symmetry group
\be
G=Sp(2m)/\Z_2\,.
\ee
The quiver \eqref{eq:Quivers} engineers 5D $\mathfrak{so}(8+2m)$ gauge theory with half hypermultiplets in the bifundamental representation of $\mathfrak{so}_{8+2m} \times \mathfrak{sp}_{2m}$. Choosing a purely electric polarization, the gauge group is $Spin(8+2m)$ and we match the results presented in \cite{Apruzzi:2021nmk}. In section \ref{sec:G2} we turn to a closely related example of this sort in the context of 4D $\mathcal{N} = 1$ theories engineered on local $G_2$ spaces.

\section{SQCD-Like Theories from $G_2$ Spaces}\label{sec:G2}

Having studied some 5D examples, we now turn to 4D $\mathcal{N}=1$ SQFTs engineered from M-theory on a non-compact $G_2$ space. In this setting as well, the field theory content of the system is dictated by the geometry of local orbifold singularities. As such, this ``geometrized'' version of type IIA realizations of SQFTs provides another arena to apply the techniques of section \ref{sec:PRESCRIPTION}.

Now, a well known difficulty in this regard is that the explicit construction of $G_2$ spaces remains a challenging problem, in part because we do not have the analog of Yau's theorem in the Calabi-Yau case. Nevertheless, physical considerations provide strong evidence that various IIA backgrounds with branes and orientifolds all have lifts to the $G_2$ setting.\footnote{See e.g. \cite{Cvetic:2001nr, Cvetic:2001kk, Cvetic:2001tj, Blumenhagen:2005mu} for some examples of such lifts in the context of string-based particle physics constructions.} These $G_2$ spaces, whose $G_2$-holonomy metric is conjectured to exist by IIA/M-theory duality, are circle fibrations over a non-compact Calabi-Yau threefold base with field strength\footnote{Mathematicians also refer to the cohomology class of $F$ as the Euler class of the fibration, denoted by $e$.} $F=dC_1$. Importantly, we can determine the topology of these circle bundles, meaning that we can then apply our Mayer-Vietoris procedure of the previous sections in this case as well.

Our main focus will be on the case of SQCD-like theories with gauge group given by either $SU(N_c)$ or $Spin(2N_c)$, with matter in the fundamental representation.\footnote{As in \cite{Lee:2021crt}, one can in principle take other choices for the global form of the gauge group associated with the $\mathfrak{spin}(2N_c)$ Lie algebra, but we defer the analysis of such possibilities to future work.} There are well-known type IIA constructions of SQCD-like theories, including their realizations in terms of D6-branes wrapping special Lagrangian manifolds in a non-compact Calabi-Yau threefold. Since SQCD has a vector-like matter spectrum, there can in principle be different ways to engineer the relevant matter content, and these lead to different boundary geometries. One possibility is to directly engineer a vector-like pair of matter fields, as in the IIA construction of \cite{Ooguri:1997ih}, as well as possible orientifolds of that construction. The $G_2$ lift of this case corresponds to gauge and flavor groups localized on codimension $4$ subspaces (i.e. $3$-cycles) and matter localized on codimension $6$ subspaces (i.e. $1$-cycles). The other possibility is to directly engineer chiral matter, as in the IIA constructions of \cite{Cachazo:2001sg,Feng:2005gw}, for example. In this case, the $G_2$ lift has matter localized on codimension $7$ subspaces (i.e. points). In both cases, we can make some of the resulting flavor symmetry manifest by coalescing all of the flavor branes at the same location. For example, in the case of $SU(N_c)$ gauge theory with $N_f$ flavors, the two possibilities result in the geometrized flavor symmetries:
\begin{align}\label{eq:twoflavsyms}
  &\textnormal{1)}\;  \; \; \; \mathfrak{su}(N_f)_{\mathrm{vec}}\;  \; \; \; \textnormal{(codimension 6 singularities)}\\
  &\textnormal{2)} \;\; \; \; \mathfrak{su}(N_f)_{L} \times \mathfrak{su}(N_f)_{R}\; \; \; \; \textnormal{(codimension 7 singularities)},
\end{align}
and similar considerations apply in the case of $Spin(2N_c)$ gauge theory, where in the IIA construction we include suitable orientifold planes. The $G_2$ lift amounts to including a suitable quotient by a geometric automorphism and / or monodromy action on the orbifold loci.
The different ways of engineering SQCD-like theories are associated with distinct manifest symmetries realized in geometry. This can occur
because different compactification effects can explicitly break some symmetries which may only emerge in flowing deep into the IR. Turning the discussion around, our analysis of higher symmetries provides a diagnostic in detecting the presence of such breaking terms in the first place.
Indeed, this also suggests that in a limit where we have not yet reached the deep infrared, the two theories are distinguishable.

Our aim in this section will be to compute the 0-form, 1-form and 2-group symmetries for these 4D SQFTs directly from geometry. As in earlier sections, we primarily focus on the contributions from non-abelian flavor symmetries, so we neglect for example the baryonic $U(1)$ global symmetry and $U(1)$ R-symmetry present $SU(N_c)$ SQCD-like theories. Additionally, in what follows, we focus on the electric polarization of the 1-form defect group, as captured by M2-branes wrapping non-compact 2-cycles in the geometry. There can also be M5-branes wrapped on non-compact 4-cycles which contribute to the magnetic 1-form symmetry. Our techniques extend to this case as well, but we leave a full analysis of magnetic symmetries to future work.

Having extracted the 0-form, 1-form and possible 2-group symmetries, we can match this data across Seiberg dual pairs of theories for $\mathfrak{su}$ and $\mathfrak{so}$ groups with fundamental flavors.\footnote{See \cite{Lee:2021crt} for a field theory analysis of higher symmetries in the $\mathfrak{so}$ case.} In some sense, this just follows from the general realization of such dualities as deformations which do not alter the boundary topology, and as such are irrelevant deformations to the IR physics.

The rest of this section is organized as follows. We begin by discussing the $G_2$ lift of type IIA realizations of 4D SQFTs. In particular, our aim will be to track the topology of the boundary geometry after performing this lift. Next, we consider the case of SQCD-like theories with $SU(N_c)$ gauge theory and matter engineered on a codimension $6$ singularity. We then turn to the related case with matter engineered on codimension $7$ singularities. With these results in place, we introduce the $G_2$ lifts of orientifolds to analyze the symmetries of
$Spin(2N_c)$ gauge theories. In Appendix \ref{app:G2LIFT} we provide some additional details on realizing SQCD-like geometries via $G_2$ spaces.

\subsection{Lifting IIA $D6/O6^-$ Configurations to M-Theory}
We now state our procedure for determining the topology of the IIA dilaton circle fibration on a non-compact Calabi-Yau threefold, $X_6$, via a Gysin sequence modified to the case where the circle fibers can become singular. The characterizing field strength of this circle bundle is the RR-flux $F=dC_1$ which is sourced by 6-branes on some loci $K$, and the important data for us will be the topological class $[F]\in H^2(X_6\backslash K)$ as well as $[F]|_{\partial X_6}\in H^2(\partial X_6\backslash \partial K)$.\footnote{It might perhaps be more useful to characterize $F$ as a class in differential cohomology which essentially extends this data of a cohomology class by a flat 1-form connection. See e.g. \cite{Freed:2000ta, Hopkins:2002rd, Freed:2006yc, Freed:2006ya, SimonsSullivan:2007, Davighi:2020vcm, Apruzzi:2021nmk} for more details.} Recall that we denote $X^\circ_6=X_6\backslash K$ and $\partial X^\circ_6=\partial X_6\backslash \partial K$. We will denote the M-theory geometry as $X_7$, which is expected to be a $G_2$ orbifold, while its boundary, $\partial X_7$, is expected to be a nearly K\"ahler manifold (possibly with singularities from flavor branes).
For backgrounds with $D6$-branes but no $O6$-planes, our procedure is as follows:
\begin{enumerate}
\item \textbf{Excision of Flavor Branes}: We compute the homology groups $H_2(\partial X_6^\circ)$, and denote its generators by $\beta_j$. Then the field strength $F=dC_1$ determines the Euler class of the M-theory circle fibration. It is helpful to convert this 2-form field strength into a cycle in the geometry. To accomplish this, we consider a deformation retract of $\partial X_{6}^{\circ}$ to a compact 4-manifold. On this 4-manifold, we can take the Poincar\'{e} dual of $F$, resulting in 2-cycles. We write this as:
\be
F_{\textnormal{PD}}=\sum_{j}n_j\beta_j,
\ee
where ``PD'' refers to Poincar\'{e} dual in the sense just defined, and the integers $n_j$ give the D6-brane flux through the 2-cycle $\beta_j$.
\item \textbf{Boundary Homology}: We fix a circle fibration
\be
\partial X^\circ_7 \rightarrow \partial X^\circ_6
\ee
with Euler class $e=F$. The homology groups of $\partial X^\circ_7$ then follow from the Gysin sequence
\be
\dots \rightarrow H^k(\partial X^\circ_7) \rightarrow H^{k-1}(\partial X^\circ_6) \xrightarrow[]{e{} \wedge} H^{k+1}(\partial X^\circ_6)\rightarrow  H^{k+1}(\partial X^\circ_7) \rightarrow \dots \\,.
\ee
\item \textbf{Gluing Flavor Branes}: We complete $\partial X^\circ_7$ to $\partial X_7$ by gluing back the flavor branes. This can be done using the Mayer-Vietoris sequence and the local model for each flavor brane.
\item \textbf{Bulk Homology}: The seven-manifold $X_7$ is fibered by a circle $X_7\rightarrow X_6$ and its homology groups follow from deformation retraction on the non-compact $X_6$. See Appendix \ref{app:G2LIFT} for an example.
\end{enumerate}
\noindent This procedure determines the homology groups of the $G_2$ space $X_7$ and its boundary $\partial X_7$. We can then add $O6$-planes since the orientifold action, $\sigma$, lifts to a $\Z_2$-action on $X_7$ which is locally of the form $(S^1,X_6)\mapsto (-S^1,\sigma(X_6))$.

With these geometric preliminaries in place, we now turn to the study of explicit SQCD-like theories.

\subsection{$SU(N_c)$ Gauge Theories with $N_f$ Flavors}

In this section we consider SQCD-like theories with an $SU(N_c)$ gauge group. Our starting point will be a IIA configuration of branes, which we then lift to a $G_2$ space. Again, since we only need the boundary topology of the construction, the physical expectation that a $G_2$ structure
exists will largely suffice for many purposes.

As already mentioned, we shall be interested in two related constructions of SQCD-like theories, one with vector-like matter localized on the same cycle, and one with chiral matter localized at distinct points in the geometry. In the IIA setting, our construction follows the one
given in reference \cite{Ooguri:1997ih}.
The main idea is to start with a deformed conifold, $X_6$, presented as
\begin{align}\label{eq:defcon5}
 X_6: \; & x^2+y^2=W \\
  & v^2+t^2=b-W
\end{align}
which is a $\mathbb{C}^{*}_{xy}\times \mathbb{C}^*_{vt}$-fibration over the complex plane $\mathbb{C}_W$. Here $b$ is a constant which sets the volume of the $S^3$ zero-section in $X_6\simeq T^*S^3$. The boundary is $\partial X_6=S^3\times S^2$, the simplest example of a Brieskorn space. This zero-section projects to the line segment $[0,b]$ in the W-plane and can engineer an $\mathfrak{su}(N_c)$ gauge theory by wrapping $N_c$ $D6$-branes on it.\footnote{The overall $U(1)$ in $U(N_c)$ realized in perturbative string theory decouples due to the generalized Green-Schwarz / Stuckelberg mechanism \cite{Blumenhagen:2005mu}.}

The zero-section is a special Lagrangian (sLag) cycle, as can be seen from the presentation $X_6\simeq T^*S^3$, but we review now the condition for a 3-manifold to be a sLag in this geometry more generally. The holomorphic 3-form on $X_6$ in the given coordinates is
\begin{equation}\label{eq:threeform}
  \Omega^{3,0}=\frac{dW\wedge dx\wedge dv}{4yt}
\end{equation}
where we have used the fact that $\frac{dx}{2y}$ is the global invariant 1-form on $\mathbb{C}^*_{xy}$.\footnote{One derives this by evaluating the holomorphic form of the ambient $\mathbb{C}^2$, $dx\wedge dy$, on the normal vector field $\frac{\partial }{\partial f}$,  where $f=x^2+y^2-W$.} The initial color stack wraps the interval $[0,b]\in \mathrm{Re}W$ times $\{x^2_R+y^2_R=W\}\times \{v^2_R+t^2_R=b-W\}$ where the subscript $R$ denotes ``real part". The color stack is then special Lagrangian with respect to $\mathrm{Re}\big({e^{i\theta}\Omega^{3,0}}\big)$ for $\theta=0$, so additional flavor branes must also be calibrated with respect to $\mathrm{Re}\big(\Omega^{3,0}\big)$ to preserve 4D $\mathcal{N}=1$ supersymmetry. Note that for, say, the $\mathbb{C}^*_{xy}$-fiber over a point $W$, the solution
\begin{equation}\label{eq:sol}
  x=e^{\textnormal{Arg}(W)/2}|x|, \; \; \; \; y=e^{\textnormal{Arg}(W)/2}|y|
\end{equation}
is calibrated with respect to $\textnormal{Re}(\frac{dx}{2y})$. More generally, we can define non-compact lines $L_\phi\subset \mathbb{C}^*_{xy}$ which are calibrated with respect to $e^{i\phi}\frac{dx}{2y}$. If we parameterize such a path as $x(s)$, then we are left to solve the differential equation (see e.g. \cite{Ooguri:1997ih, Shapere:1999xr}):
\begin{equation}\label{eq:xpath1}
  \frac{dx}{ds}=\mp 2  e^{i\phi}{\sqrt{W-x^2}}.
\end{equation}
which leads to a solution
\begin{equation}
  x(s)=\sqrt{W}\sin\bigg(\mp e^{i\phi}(2s+A_0)\bigg)
\end{equation}
for each choice of $\phi$, where $A_0$ is an integration constant. We see that if $e^{i\phi}\neq \pm 1$, then $L_\phi$ is non-compact inside $\mathbb{C}^*_{xy}$.

\subsubsection{Matter Engineered from Codimension 6 Singularities}

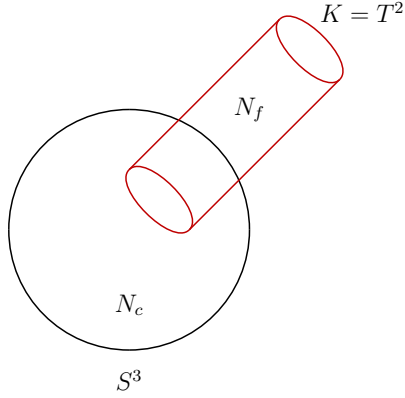
\begin{figure}
\centering
\scalebox{0.8}{
\begin{tikzpicture}
	\begin{pgfonlayer}{nodelayer}
		\node [style=none] (0) at (0, 2) {};
		\node [style=none] (1) at (0, -2) {};
		\node [style=none] (2) at (-2, 0) {};
		\node [style=none] (3) at (2, 0) {};
		\node [style=none] (4) at (0, 1) {};
		\node [style=none] (5) at (1, 0) {};
		\node [style=none] (6) at (2.5, 3.5) {};
		\node [style=none] (7) at (3.5, 2.5) {};
		\node [style=none] (8) at (3.875, 3.625) {$K=T^2$};
		\node [style=none] (9) at (0, -1.25) {$N_c$};
		\node [style=none] (10) at (0, -2.5) {$S^3$};
		\node [style=none] (11) at (2, 2) {$N_f$};
	\end{pgfonlayer}
	\begin{pgfonlayer}{edgelayer}
		\draw [style=ThickLine, in=-180, out=90] (2.center) to (0.center);
		\draw [style=ThickLine, in=90, out=0] (0.center) to (3.center);
		\draw [style=ThickLine, in=0, out=-90] (3.center) to (1.center);
		\draw [style=ThickLine, in=-90, out=-180] (1.center) to (2.center);
		\draw [style=RedLine, bend right=90, looseness=0.75] (4.center) to (5.center);
		\draw [style=RedLine, bend left=90, looseness=0.75] (4.center) to (5.center);
		\draw [style=RedLine] (4.center) to (6.center);
		\draw [style=RedLine] (5.center) to (7.center);
		\draw [style=RedLine, bend right=90, looseness=0.75] (6.center) to (7.center);
		\draw [style=RedLine, bend left=90, looseness=0.75] (6.center) to (7.center);
	\end{pgfonlayer}
\end{tikzpicture}}
\caption{Sketch of the IIA setup of type 1) with ``vector-like'' flavor symmetry. The total geometry is $T^*S^3$ with $N_c$ color D6-branes (black) wrapped on the compact $S^3$ and $N_f$ flavor D6-branes (red) wrapped on $T^2\times \mathbb{R}_+$ intersecting the $S^3$ in a circle and the boundary in $T^2$.}
\label{fig:vectorlike}

\end{figure}

We now turn to an SQCD-like theory with matter localized on codimension 6 singularities of the corresponding $G_2$ space. In this case, we expect that the geometry can capture an $\mathfrak{su}(N_f)_{\mathrm{vec}}$ ``vector-like'' flavor symmetry. Our task will be to extract directly from the geometry the global form of this 0-form symmetry, as well as possible 1-form symmetries and 2-group structures.

Returning to the IIA background, we add $N_f$ D6-branes along the locus
\begin{equation}\label{eq:codim6flav}
  [b,+\infty]\times \{x^2_R+y^2_R=W\}\times \{v^2_I+t^2_I=b-W\}
\end{equation}
where $I$ denotes ``imaginary" part. This is special Lagrangian since $\frac{dv}{2t}=\frac{idv_I}{i2t_I}=\frac{dv_I}{2t_I}$ is real. This intersects the color stack along a circle which leads to a vector-like pair of bifundamental chiral multiplets, which together transform as
\begin{equation}
  (\overline{\mathbf{N}}_c, \mathbf{N}_f)\oplus (\mathbf{N}_c, \overline{\mathbf{N}}_f).
\end{equation}
under $\mathfrak{su}(N_c)\times \mathfrak{su}(N_f)$.

The boundary of the deformed conifold, $\partial X_6=S^2\times S^3$, has a color brane along $S^2$ along with a flavor brane along a null-homologous $T^2$. Then $\partial X^\circ_6=S^3\times S^2 \backslash T^2$ and we can determine its homology using a Mayer-Vietoris sequence to calculate
\begin{equation}\label{eq:cohx6su}
  H_*(\partial X^\circ_6)= \{ \mathbb{Z},  0, \mathbb{Z}^2, \mathbb{Z}^3, 0,0  \}.
\end{equation}
$H_2$ is generated by the bulk $S_c^2$ of the original $S^2\times S^3$, and by another $S_{f}^2$ which links the flavor $T^2$ locus. Meanwhile, $H_3$ is generated by the bulk $S^3$, as well as two 3-cycles of the form $S_{f}^2\times S_{A,B}^1$ where the two choices are over the $a$- and $b$-cycle of the $T^2$. $H_4$ is empty because the $S^2\times T^2$ which surrounds the flavor locus is contractible in the same way the fundamental group of a sphere with one puncture is trivial.

We are now ready to apply the Gysin sequence to calculate $H_*(\partial X^\circ_7)$. The key piece of the long exact sequence is
\be
0 \rightarrow H^1(\partial X_7^\circ)\rightarrow H^0(\partial X_6^\circ) \xrightarrow[]{\,e_{N_f,N_c}\,}  H^2(\partial X_6^\circ) \rightarrow H^2(\partial X_7^\circ) \rightarrow 0
\ee
where the Gysin map is
\be
e_{N_f,N_c}\,:~ H^0(\partial X_6^\circ) ~\rightarrow~  H^2(\partial X_6^\circ)  \,, \qquad ~\:\! \alpha \; \;  ~\!\!\!\!\!\mapsto~ (\alpha N_f , \alpha N_c).
\ee
The Gysin map follows from the fact that the flux is $F=N_c\textnormal{vol}_{S_{N_f}^2}+N_f\textnormal{vol}_{S^2_{N_c}}$. The resulting cohomology groups are
\begin{equation}\label{eq:sug2co}
  H^*(\partial X_7^\circ)\cong\lbbb \Z, 0, \Z\times \Z_g,\Z^5,\Z^3,0,0\rbbb
\end{equation}
where $g \equiv \textnormal{gcd}(N_c,N_f)$, and after applying Poincar\'e duality and the universal coefficient theorem\footnote{Technically speaking, we are performing this step on a deformation retraction of $\partial X_7^\circ$ to a compact 4-manifold.} we have the homology groups
\begin{equation}\label{eq:sug2co1}
  H_*(\partial X_7^\circ)\cong\lbbb \Z, \Z_g, \Z,\Z^5,\Z^3,0,0\rbbb.
\end{equation}
We can also apply the Gysin sequence on the boundary of the tubular neighborhood of the flavor branes to obtain $\partial X_7^\circ \cap T(K)=T^2\times S^3/\Z_{N_f}$, so we now know the following groups
\begin{equation}
  \widetilde{\mathcal{A}}=\Z_g, \; \ ;\; \; \; Z_{\widetilde{G}}=\Z_{N_f}.
\end{equation}
We can solve for $\mathcal{A}$ by the following piece of the Mayer-Vietoris sequence
\be
 \rightarrow H_1(\partial X_7^\circ\cap T(K)) \xrightarrow[]{\,\textnormal{im}=\Z_g\,}  H_1(\partial X_7^\circ) \xrightarrow[]{\,\textnormal{ker}=\Z_g\,}  H_1(\partial X_7)\rightarrow 0
\ee
which implies $\mathcal{A}=0$. One can then read off from our two main short exact sequences:
\begin{align}
0 & \rightarrow \mathcal{C} \rightarrow Z_{\widetilde{G}} \rightarrow Z_{G} \rightarrow 0\\
0 & \rightarrow \mathcal{C}^{\vee} \rightarrow \widetilde{\mathcal{A}}^{\vee} \rightarrow \mathcal{A} \rightarrow 0.
\end{align}
that
\begin{equation}
  \mathcal{C}=\Z_g, \; \; \; \; \; Z_G=\Z_{N_f/g}.
\end{equation}
So in other words, the global form of the non-abelian flavor symmetry is:
\begin{equation}
G_{\mathrm{non-ab}} = SU(N_f) / \mathbb{Z}_g.
\end{equation}

Summarizing, we see that for our SQCD-like theory engineered with codimension 6 matter, there is no 1-form symmetry and $Z_G = \Z_{N_f / g}$.
We further note that the naive 1-form symmetry is indeed $\widetilde{\mathcal{A}} = \mathbb{Z}_g$, but that this is fully screened by the quotienting group $\mathcal{C}$. This is to be expected since the matter fields transform in the bifundamental representation. As a final comment, we note that the non-abelian part of the global 0-form symmetry as well as the 1-form symmetry both agree with the results expected for $N_f$ massive Dirac fermions in reference \cite{Hsin:2020nts}. This mass term explicitly breaks the $\mathfrak{su}(N_f)_L \times \mathfrak{su}(N_f)_R$ to a diagonal flavor symmetry.\footnote{As already stated, we have also neglected various $\mathfrak{u}(1)$ symmetry factors, including baryonic and R-symmetry factors. In reference \cite{Hsin:2020nts} the flavor symmetry for the massive Dirac fermion case is presented as
$U(N_f) / \mathbb{Z}_{N_c}$. Decomposing $N_c = \mathrm{gcd}(N_c,N_f) \times N^{\prime}_{c}$, the match to geometry follows.} In this sense, one can also view our model with codimension 6 singularities as obtained from a limit where all diagonal mass terms have been switched off.

\subsubsection{Matter Engineered from Codimension 7 Singularities}\label{sssec:sucodim7}

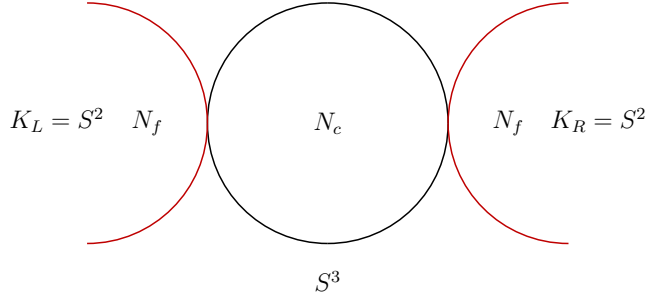
\begin{figure}
\centering
\scalebox{0.8}{
\begin{tikzpicture}
	\begin{pgfonlayer}{nodelayer}
		\node [style=none] (0) at (0, 2) {};
		\node [style=none] (1) at (0, -2) {};
		\node [style=none] (2) at (-2, 0) {};
		\node [style=none] (3) at (2, 0) {};
		\node [style=none] (9) at (0, 0) {$N_c$};
		\node [style=none] (10) at (0, -2.625) {$S^3$};
		\node [style=none] (11) at (-3, 0) {$N_f$};
		\node [style=none] (12) at (4, 2) {};
		\node [style=none] (13) at (4, -2) {};
		\node [style=none] (14) at (-4, -2) {};
		\node [style=none] (15) at (-4, 2) {};
		\node [style=none] (16) at (3, 0) {$N_f$};
		\node [style=none] (17) at (4.5, 0.05) {$K_R=S^2$};
		\node [style=none] (18) at (-4.5, 0.05) {$K_L=S^2$};
	\end{pgfonlayer}
	\begin{pgfonlayer}{edgelayer}
		\draw [style=ThickLine, in=-180, out=90] (2.center) to (0.center);
		\draw [style=ThickLine, in=90, out=0] (0.center) to (3.center);
		\draw [style=ThickLine, in=0, out=-90] (3.center) to (1.center);
		\draw [style=ThickLine, in=-90, out=-180] (1.center) to (2.center);
		\draw [style=RedLine, in=90, out=0] (15.center) to (2.center);
		\draw [style=RedLine, in=0, out=-90] (2.center) to (14.center);
		\draw [style=RedLine, in=90, out=-180] (12.center) to (3.center);
		\draw [style=RedLine, in=-180, out=-90] (3.center) to (13.center);
	\end{pgfonlayer}
\end{tikzpicture}}
\caption{Sketch of the IIA setup of type 2) with chiral pairs. The total geometry is again $T^*S^3$ with $N_c$ color D6-branes wrapped on the compact $S^3$ (black) and two stack of $N_f$ flavor D6-branes (red) wrapped on two copies of $\mathbb{R}^3$ intersecting the $S^3$ in the north and south pole. These intersect the boundary in two two-spheres $K_{L,R}$.}
\label{fig:chiral}
\end{figure}

We now turn to a different construction of an SQCD-like theory, with matter fields localized at codimension $7$ singularities of the local $G_2$ space. We again describe the geometry by prescribing a circle fibration over a type IIA background. For a top-down construction which represents the resulting geometry as glued from two Acharya-Witten cones see Appendix \ref{app:G2LIFT}.

The IIA background we consider here is once more the deformed conifold $T^*S^3$ with a stack of $N_c$ color D6-branes wrapped on the compact $S^3$. However, we now consider two flavor stacks of $N_f$ D6-branes each intersecting the color stack in points rather than circles. We take these points to be the north and south pole of the three-sphere. In this case the supersymmetric cycles supporting the flavor branes are
\begin{equation}\label{eq:codim7flav1}
  [b,+i\infty]\times \{L_{-\pi/2}\}\times \{|v|^2+|t|^2=|b-W|\}\,,
\end{equation}
and
\begin{equation}\label{eq:codim7flav1}
  [0,-i\infty]\times \{L_{-\pi/2}\}\times \{|v|^2+|t|^2=|W|\}\,,
\end{equation}
each topologically a copy of $\R^3$. Here the intervals denote vertical rays in the $\mathbb{C}_W$ plane and $L_\phi$ is a non-compact line in $\mathbb{C}^*$ discussed in the beginning of this section.
These loci are special Lagrangian since the angles of the first two factors cancel as so $\pi/2-\pi/2=0$ while they are identical for the last. Each intersection leads to chiral matter, but in conjugate representations:
\begin{equation}
  (\overline{\mathbf{N}}_c, \mathbf{N}_f, \mathbf{1})\oplus (\mathbf{N}_c, \mathbf{1}, \overline{\mathbf{N}}_f).
\end{equation}
under the symmetry algebra $\mathfrak{su}(N_c)\times\mathfrak{su}(N_f)_L\times \mathfrak{su}(N_f)_R$.

Geometrically, the flavor loci at the boundary consists of two two-spheres $K_{L,R}\cong S^2$, see figure \ref{fig:chiral}. The boundary is simply $S^2\times S^3$ and excising $K_{L,R}$ we delete two copies of $S^2$ at fixed points of $S^3$. This twice punctured $S^3$ deformation retracts to a two-sphere $S^2_{N_c}$ while full geometry $\partial X^\circ_6$ retracts to
\be
\partial X^\circ_{6,\textnormal{Retract}}=S^2_{N_c} \times S^2_{N_f}\,
\ee
where the $S^2_{N_f}$ are boundary 2-cycles. The two-sphere $S^2_{N_c}$ links both flavor loci. The two-sphere $S^2_{N_f}$ lives in the fiber of the deformed conifold and therefore links the three-sphere supporting the color stack. Consequently these two two-spheres are threaded by D6-brane flux and we have
\be
F=N_c \textnormal{vol}_{S_{N_f}^2}+N_f\textnormal{vol}_{S^2_{N_c}}
\ee
which determines a circle bundle over $X^\circ_{6,\textnormal{Retract}}$ via fixing the Euler class $e=F$. Again, using the Gysin sequence we compute the homology groups of this circle bundle. The sequence gives
\be
H^0(\partial X_7^\circ )\cong H^5(\partial X_7^\circ)\cong \Z
\ee
and further splits as
\be\ba
0 &\rightarrow H^1(\partial X_7^\circ)\rightarrow H^0(S^2_{N_c}\times S^2_{N_f}) \rightarrow H^2(S^2_{N_c}\times S^2_{N_f}) \rightarrow H^2(\partial X_7^\circ) \rightarrow 0 \\
0 &\rightarrow H^3(\partial X_7^\circ)\rightarrow H^2(S^2_{N_c}\times S^2_{N_f}) \rightarrow H^4(S^2_{N_c}\times S^2_{N_f}) \rightarrow H^4(\partial X_7^\circ) \rightarrow 0
\ea\ee
where the central maps are wedging with the Euler class of the fibration. These are only non-trivial in even degree and there we have
\be\ba
e_0\,\wedge\, &: \Z ~\rightarrow ~ \Z^2\,, \qquad k ~\mapsto ~ (kN_f,kN_c)\,, \\
e_2\,\wedge\, &: \Z^2 ~\rightarrow ~ \Z\,, \qquad (n,m) ~\mapsto ~ nN_f+mN_c\,.
\ea\ee
This gives the cohomology groups
\be
H^*(\partial X_7^\circ)\cong \lbbb \Z ,0, \Z \oplus \Z_{\tn{gcd}(N_f,N_c)}, \Z, \Z_{\tn{gcd}(N_f,N_c)}, \Z \rbbb\,.
\ee
and dualizing to homology\footnote{Again, we are performing this step on the contraction of $\partial X_7^\circ$ to a compact 5-manifold.}  we find
\be\label{eq:HomoSmooth}
H_*(\partial X_7^\circ)\cong \lbbb \Z , \Z_{\tn{gcd}(N_f,N_c)},  \Z, \Z\oplus \Z_{\tn{gcd}(N_f,N_c)},0,\Z \rbbb\,.
\ee
Now we glue the two flavor loci, topologically two two-spheres, back into $\partial X_7^\circ$, completing it to $\partial X_7$, via an application of the Mayer-Vietoris sequence. This follows because the intersection of the tubular neighborhoods of the flavor loci in $\partial X_7$ with $\partial X_7^\circ$ are simply copies of $\partial X_7^\circ$. We therefore find
\be\label{eq:SmoothBdryG22}
H_*(\partial X_7)\cong \lbbb \Z, 0, \Z \oplus \Z_{\tn{gcd}(N_f,N_c)}, 0 ,\Z\oplus  \Z_{\tn{gcd}(N_f,N_c)},0 , \Z \rbbb\,.
\ee
We remark further on the tubular neighbourhoods. The local geometry of the orbifold loci is $\C^2/\Z_{N_f} \rightarrow S^2$ and therefore the tubular neighborhoods are fibered as $S^3/\Z_{N_f} \rightarrow S^2$ where the Hopf circle of the Lens space is the M-theory circle. Now $S^2$ links the color stack and is threaded by $N_c$ units of D6 flux, twisting the M-theory circle over $S^2$. The overall geometry has a torsional 1-cycle of order $g=\textnormal{gcd}(N_c,N_f)$ as correctly computed in \eqref{eq:HomoSmooth}. This implies:
\begin{equation}
  \widetilde{\mathcal{A}}=\Z_g, \; \;\; \; \; Z_{\widetilde{G}} =\Z_{g}\times\Z_{g}.
\end{equation}
with $g = \mathrm{gcd}(N_c,N_f)$. The 1-form symmetry follows from the Mayer-Vietoris sequence, which gives
\be
 \rightarrow H_1(\partial X_7^\circ\cap T(K)) \xrightarrow[]{\,\textnormal{im}=\Z_g\,}   H_1(\partial X_7^\circ) \xrightarrow[]{\,\textnormal{ker}=\Z_g\,}  H_1(\partial X_7)\rightarrow 0\,.
\ee
This implies a trivial 1-form symmetry $\mathcal{A}=H_1(\partial X_7)^\vee=0$. It further follows that leftmost map takes the form $\iota_1:\Z_g^2\rightarrow \Z_g$ mapping $(k,k)\mapsto k$. By our general formalism we now find
\begin{equation}
  \mathcal{C}=\mathbb{Z}_g, \; \; \; \; \; Z_G=\mathbb{Z}_g.
\end{equation}
The pair of short exact sequences
\begin{align}
0 & \rightarrow \mathcal{C} \rightarrow Z_{\widetilde{G}} \rightarrow Z_{G} \rightarrow 0 \\
0 & \rightarrow \mathcal{C}^{\vee} \rightarrow \widetilde{\mathcal{A}}^{\vee} \rightarrow \mathcal{A}^{\vee} \rightarrow 0 ,
\end{align}
therefore takes the form
\begin{align}
0 & \rightarrow  \Z_{g} \rightarrow \Z_{g}\times\Z_{g} \rightarrow  \Z_{g} \rightarrow 0 \\
0 & \rightarrow  \Z_{g} \rightarrow  \Z_{g} \rightarrow 0\rightarrow 0 ,
\end{align}
The 1-form symmetry is trivial, consequently there is no 2-group. The naive global symmetry derived from geometry has center $ Z_{\widetilde{G}}$. We therefore find
\be
\widetilde{G}= SU(N_f) /\Z_{N_f/g} \times SU(N_f) /\Z_{N_f/g}
\ee
which is corrected to the global symmetry:
\be
G = \frac{SU(N_f) /\Z_{N_f/g} \times SU(N_f) /\Z_{N_f/g}}{\Z_g}.
\ee
As a comment, when comparing with the related field theory analysis in \cite{Hsin:2020nts} one should keep in mind that we are dealing with SQCD as opposed to QCD. Indeed, we also have a global $\mathfrak{u}(1)$ R-symmetry. For SQCD in the conformal window, the R-charge of all the squarks is $R_{\mathrm{squark}} = (N_f - N_c) / N_f$, and so we explicitly see that $g = \mathrm{gcd}(N_c,N_f)$ cancels out of this ratio. This appears to match with expectations from geometry, though we leave a more complete analysis for future work.

\subsection{$Spin(2N_c)$ Gauge Theories with $2N_f$-Flavors}

We now consider an example where we engineer an $\mathfrak{so}(2N_c)$ gauge algebra with $2N_f$ flavors in the fundamental representation.
As noted in reference \cite{Lee:2021crt}, an interesting feature of this class of examples is that it can also exhibit a 2-group structure, which can in principle be matched across Intriligator-Seiberg duality. As before, we begin with a IIA construction, which we then lift to a $G_2$ space. Again, we emphasize that for our present purposes, we primarily only need to know the topology of the boundary space rather than the explicit form of the $G_2$ metric in the interior.

We now study placing a $O6^-$-plane along the zero-section of $X_6=T^*S^3$ to engineer $Spin(2N_c)$ SQCD with $2N_f$ chiral mutliplets in the vector representation. From the $G_2$ point of view, this will appear as a suitable $\Z_2$ quotient, so we can simply adapt our results from the previous section to extract the 0-, 1-form symmetry, and will now see a 2-group structure emerge.

The difference in flavor algebras in the codimension 6 and codimension 7 cases comes from the IIA perspective by how a $D6$-brane intersects an $O6$. Explicitly, we study the following action on $X_6$ (see e.g. \cite{Kachru:2001je,Cvetic:2001kk}):
\begin{equation}
  \sigma: (x,y,t,v,z)\mapsto (x^*,y^*,t^*,v^*,z^*)
\end{equation}
which maps the flavor $D6$ stack to itself in the codimension 6 case, and to an image brane stack in the codimension 7 case. One can then use string perturbation theory to show that the former case yields an $\mathfrak{sp}$ algebra while the latter retains an $\mathfrak{su}$ algebra (see e.g. \cite{Blumenhagen:2005mu} for a review).

In the left of the geometry with orientifolds, we expect $Spin$-type gauge groups to originate from a D-type singularity, while the flavor stack will originate from A-type singularities. Much as in our discussion from section \ref{sec:ELLIPTIC}, the $\mathfrak{sp}$ factor originates from an $A_{\mathrm{even}}$-type singularity with monodromy (i.e. quotienting by an outer automorphism). This implies that the action of $\sigma$ on the $\partial X_7$ geometry for the $SU$-case either does ($\mathfrak{sp}$ flavor) or does not ($\mathfrak{su}$ flavor) act on the torsional 1-cycles. In what follows, we refer to this new $G_2$ space as $Y_7$, and its boundary as $\partial Y_7$. Even though we cannot obtain $Y_7$ as a $\mathbb{Z}_2$ quotient of $X_7$, the topology of the boundary can be viewed in this way, and this greatly simplifies the analysis to come.

\subsubsection{Matter Engineered from Codimension 6 Singularities}
We now extract the flavor and 2-group structure for the $Spin(2N_c)$ gauge theory with $\mathfrak{sp}(N_f)$ flavor algebra from the geometry. Our approach will be to compute $\pi_1$ for the various manifolds of interest, which come from the $SU$ case after quotienting by $\sigma$. We define $\partial Y_7 \equiv \partial X_7/\sigma$, then because $\partial X_7$ is simply-connected and the action of $\sigma$ is free (it only has fixed points in the bulk), then we know that
\begin{equation}
  \pi_1(\partial Y_7)=\mathbb{Z}_2
\end{equation}
which implies that the 1-form symmetry is $\mathcal{A}=\Z_2$. This is expected from the field theory because chiral multiplets in the vector representation screen some of the gauge Wilson lines but not the ones in a spinor representation.\footnote{Wilson lines with either spinor chirality are equivalent if there are dynamical fields in the vector representation. } This opens up the possibility to have a non-trivial 2-group. Moving on to $\pi_1(\partial Y^\circ_7)$, this is of the general form $\Z_g \rtimes \Z_2$ where the precise $\Z_2$ action can be motivated as follows. Our setup has $\sigma$ turning the color locus in the bulk from an A-type singularity to a D-type singularity, while keeping the flavor locus to be of A-type. We can rewrite $\pi_1(\partial X^\circ_7)=\Z_g$ as $\Z_{N_c}/(y^{N_f}\sim 0)$, with the geometric interpretation being that torsional 1-cycles coming from the Hopf fibers of the color stack can trivialize in $N_f$-bunches along the flavor loci. Then quotienting by $\sigma$ to produce a D-type singularity for the color locus\footnote{Note that our charge conventions imply a re-scaling $N_c\rightarrow 2N_c$ and $N_f\rightarrow 2N_f$.} implies that we introduce an element $\delta$ into $\pi_1(\partial Y^\circ_7)$ such that it has the presentation
\begin{equation}\label{eq:pi1Y7}
  x^{2N_c}=1, \; \; \; \delta^2=x^{N_c}, \; \; \; \delta x \delta = x^{-1}, \; \; \; x^{2N_f}=1.
\end{equation}
The first three relations of (\ref{eq:pi1Y7}) are precisely the defining relations of a D-type finite subgroup of $SU(2)$, while the last one imposes the trivialization condition we had before quotienting by $\sigma$. Note that because $2N_f$ is even, the abelianization of the group \ref{eq:pi1Y7} will be invariant under $2N_f$. Additionally, this same reasoning can be repeated for $\pi_1(\partial Y^\circ_7\cap T(K))$ to see that it is $\Z_g \rtimes \Z_2$ with the same semi-direct product structure as $\pi_1(\partial Y^\circ_7)$. After taking abelianizations, we arrive at the homology groups of interest. This produces
\begin{equation}
  Z_{\widetilde{G}}=\widetilde{\mathcal{A}}=\begin{cases}
                                            \Z_2\times \Z_2, & \mbox{if}\;  2N_c\equiv 0\; \mathrm{mod} \; 4 \\
                                              \Z_4, & \mbox{if}\; 2N_c \equiv 2\; \mathrm{mod} \; 4
                                            \end{cases}
\end{equation}
which further implies that $Z_G=\mathcal{C}=\Z_2$ for both cases. In this case, we have a flavor group:
\begin{equation}
G = Sp(N_f) / \mathbb{Z}_2,
\end{equation}
and we observe that there is a non-trivial 2-group structure when $N_c$ is odd, which agrees with the pattern found for $Spin(2N_c)$ SQCD in reference \cite{Lee:2021crt}.

\subsubsection{Matter Engineered from Codimension 7 Singularities}
We next consider $X_7$ with codimension 7 singularities, and its cousin $Y_7$ obtained from the $G_2$ lift of the corresponding
IIA model. Since again $\partial X_7$ is simply connected and the action of $\sigma$ is free, $\partial Y_7=\partial X_7/\sigma$ has a fundamental group of $\Z_2$. Thus again we have $\mathcal{A}=\Z_2$. We also again know the general form $\pi_1(\partial Y^\circ_7\cap T(K))=\Z_g \rtimes \Z_2$ since $\sigma$ turns the color loci into a D-type singularity, and keeps the flavor loci as an A-type singularity. It has the same semidirect product structure as the codimension-7 case. So we now regain the result
\begin{equation}
  Z_{\widetilde{G}}=\widetilde{\mathcal{A}}=\begin{cases}
                                              \Z_2\times \Z_2, & \mbox{if}\;  2N_c\equiv 0\; \mathrm{mod} \; 4 \\
                                              \Z_4, & \mbox{if}\; 2N_c \equiv 2\; \mathrm{mod} \; 4
                                            \end{cases}
\end{equation}
along with $Z_G=\mathcal{C}=\Z_2$. In this case we get a flavor group:
\begin{equation}
G = SU(2N_f) / \mathbb{Z}_2.
\end{equation}
The 2-group structure dependence on $N_c$ again agree with what was
found in \cite{Lee:2021crt} for SQCD with gauge group $Spin(2N_c)$
and $2N_f$ flavors in the fundamental representation.

\newpage

\section{Conclusions} \label{sec:CONC}

In this paper we have studied the structure of 0-form, 1-form and 2-group symmetries in SQFTs engineered via glued orbifold singularities.
We have shown that all of these can be extracted purely from the boundary geometry. This is encoded in a category of boundaries, and is also captured by orbifold homology. We exhibited this general structure in the case of 5D orbifold SCFTs, 5D gauge theories obtained from elliptically fibered Calabi-Yau threefolds, as well as local $G_2$ spaces engineering 4D SQCD-like theories. In the remainder of this section we discuss some further avenues for investigation.

Our analysis has primarily focused on the global form of the continuous non-abelian flavor symmetry. There can also be additional $U(1)$ factors, as well as possible discrete factors which would all be interesting to investigate further. For some recent discussion of higher symmetries and their mixing with $U(1)$ factors in the context of string compactifications, see e.g. \cite{Cvetic:2021sxm}.

We have mainly studied global structures with singularity type dictated by orbifold singularities. This covers a broad class of examples, but there are well-known cases where this is not the case. It would be interesting to see whether the same structure of cutting and gluing could be extended to such situations.

An important subtlety in our analysis involves the issue of possible ``accidental'' flavor enhancements / reductions as one passes to a regime of strong coupling. Such effects are known to occur in a variety of SCFTs, so it is natural to ask whether geometry can provide a guide and / or a constraint on such phenomena. Turning the question around, the appearance of such accidental symmetries may suggest a quantum generalization of cutting and gluing of M-theory background geometries.

Clearly, it would also be of interest to study further examples involving elliptically fibered Calabi-Yau threefolds.
A rather natural class of examples in this regard are the F-theory backgrounds which realize certain ``small instanton'' 6D SCFTs, which in M-theory terms are obtained from the worldvolume theory of M5-branes probing an ADE singularity wrapped by an $E_8$ nine-brane \cite{Aspinwall:1997ye, DelZotto:2014hpa, Heckman:2015bfa}. Upon compactification on a circle, this also gives rise to a rich class of 5D SQFTs.
A related class of questions concerns the behavior of the global flavor symmetry under Higgs branch flows. In many cases, these can be characterized by group theoretic data associated with a nilpotent orbit of the flavor symmetry algebra (as in the case of conformal matter) \cite{Heckman:2015bfa, Heckman:2016ssk, Heckman:2018pqx}, or a finite group homomorphism (in the case of orbi-instanton theories) \cite{Heckman:2015bfa, Frey:2018vpw}. Since this process often involves the decoupling of various flavor branes, it is natural to suspect that this can be isolated via a procedure of cutting and gluing along the lines used in this work.

We expect that higher structures such as 3-groups will arise when excision of the flavor branes still results in a singular space. It would be interesting to investigate this possibility further.

The main emphasis in this paper has been on the development of a set of computational techniques for extracting the higher symmetries directly from geometry. Given this, it would seem important to extract further details, as captured by topologically robust quantities such as anomalies. Perhaps this can be calculated along the lines of reference \cite{Apruzzi:2021nmk} (see also \cite{Heckman:2017uxe}).

Finally, the main thrust of our analysis has been in the context of SQFTs engineered on a non-compact geometry $X$. For $X$ compact, gravity is again dynamical and we expect that these symmetries are either explicitly broken by compactification effects, or are instead gauged, with anomaly inflow from the rest of the bulk geometry.\footnote{For a recent example of this sort of analysis for 8D and 7D vacua, see e.g. \cite{Cvetic:2021sxm,Cvetic:2022uuu}.} Since we now have an explicit way to cut and glue local contributions to such symmetries, it is natural to apply this same method of analysis in this broader setting as well.


\section*{Acknowledgments}

We thank L. Bhardwaj, M. Del Zotto, M. Dierigl, I. Garcia-Etxebarria, C. Lawrie, L. Lin, S. Meynet, R. Moscrop,
S. Schafer-Nameki, Y. Wang, H.Y. Zhang, and G. Zoccarato for helpful discussions.
Part of this work was performed at the conference ``Geometrization of (S)QFTs in $D \leq 6$'' held at the
Aspen Center for Physics, which is supported by National Science Foundation grant PHY-1607611.
The work of MC and JJH is supported by the DOE (HEP)
Award DE-SC0013528. The work of MC and MH is supported by
the Simons Foundation Collaboration grant 724069 on ``Special
Holonomy in Geometry, Analysis and Physics''.
MC also acknowledges support from the Fay R. and Eugene L. Langberg Endowed
Chair, and the Slovenian Research Agency (ARRS No. P1-0306).

\newpage

\appendix

\section{$G_2$ Spaces for SQCD-Like Theories} \label{app:G2LIFT}

The $G_2$ spaces in section \ref{sec:G2} are presented as circle fibrations with Calabi-Yau threefold bases.
In this Appendix we give a top-down construction for the model described in section \ref{sssec:sucodim7}
as the gluing of two Acharya-Witten cones. Related constructions were recently considered in \cite{DelZotto:2021ydd} where the resulting space was described as Taub-NUT spaces fibered over a collection of intersecting associative submanifolds.

\subsection{Gluing Acharya-Witten Cones}
\label{sec:GluingCones}

\begin{figure}
\centering
\scalebox{0.8}{
\begin{tikzpicture}
	\begin{pgfonlayer}{nodelayer}
		\node [style=none] (0) at (1.25, -2) {};
		\node [style=none] (1) at (1.25, 2) {};
		\node [style=none] (2) at (5.25, -2) {};
		\node [style=none] (3) at (1.25, 1) {};
		\node [style=none] (4) at (2.75, -0.5) {};
		\node [style=none] (5) at (4.25, -2) {};
		\node [style=none] (6) at (5.25, -2.5) {$R_{N_c}$};
		\node [style=none] (7) at (0.625, 1.875) {$R_{N_f}$};
		\node [style=NodeCross] (8) at (2.75, -2) {};
		\node [style=NodeCross] (9) at (1.25, -0.5) {};
		\node [style=Circle] (10) at (2.75, -0.5) {};
		\node [style=none] (11) at (0.625, -0.5) {$S^2_{N_f}$};
		\node [style=none] (12) at (2.75, -2.5) {$S^2_{N_c}$};
		\node [style=none] (13) at (5.375, 0) {$(S^3/\Z_{N_c}\times S^3/\Z_{N_f})/U(1)$};
		\node [style=none] (14) at (0.75, 1) {$R$};
		\node [style=none] (15) at (4.25, -2.425) {$R$};
	\end{pgfonlayer}
	\begin{pgfonlayer}{edgelayer}
		\draw [style=ArrowLineRight] (0.center) to (1.center);
		\draw [style=ArrowLineRight] (0.center) to (2.center);
		\draw [style=DashedLine] (3.center) to (4.center);
		\draw [style=DashedLine] (4.center) to (5.center);
	\end{pgfonlayer}
\end{tikzpicture}
}
\caption{Sketch of an Acharya-Witten cone as a fibration over the quadrant $R_{N_c},R_{N_f}\geq 0$. The fibers are copies of $(S^3/\Z_{N_c}\times S^3/\Z_{N_f})/U(1)$. Whenever $R_{N_c},R_{N_f}=0$ fibers degenerate to two-spheres. The link of the cone at radius $R$ projects to the dashed line.}
\label{fig:AWCone}
\end{figure}
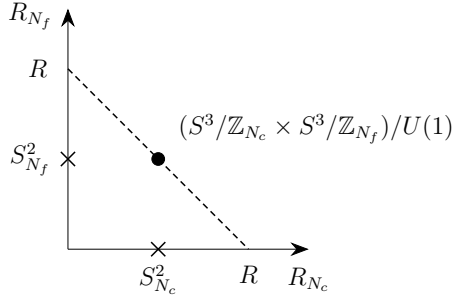

The cones of Acharya and Witten \cite{Acharya:2001gy} describe the M-theory lift of two stacks of D6-branes wrapping distinct supersymmetric three-planes in $\R^6$ and filling space-time. We distinguish the color and flavor stack which support $N_c$ and $N_f$ branes respectively. The M-theory lift is the purely geometric background
\be\label{eq:AWCone}
X_{N_c,N_f}=\frac{ \mathbb{C}^2/\Z_{N_c}\times \mathbb{C}^2 /\Z_{N_f} }{U(1)}
\ee
where the $U(1)$ acts with charges $\pm 1$ on the two ADE singularities respectively. The seven-manifold $X_{N_c,N_f}$ is conjectured to admit a $G_2$ holonomy metric. We now describe their basic features and describe parameterizations favorable for gluing of two such cones.

\begin{figure}
\centering
\scalebox{0.8}{
\begin{tikzpicture}
	\begin{pgfonlayer}{nodelayer}
		\node [style=none] (0) at (-2, -2) {};
		\node [style=none] (1) at (-2, 2) {};
		\node [style=none] (2) at (2, -2) {};
		\node [style=none] (3) at (-2, 1) {};
		\node [style=none] (4) at (1, 1) {};
		\node [style=none] (5) at (1, -2) {};
		\node [style=none] (6) at (2, -2.5) {$R_{N_c}$};
		\node [style=none] (7) at (-2.625, 2) {$R_{N_f}$};
		\node [style=NodeCross] (8) at (-0.5, -2) {};
		\node [style=NodeCross] (9) at (-2, -0.5) {};
		\node [style=none] (11) at (-2.625, -0.5) {$S^2_{N_c}$};
		\node [style=none] (12) at (-0.5, -2.5) {$S^2_{N_f}$};

	\end{pgfonlayer}
	\begin{pgfonlayer}{edgelayer}
		\draw [style=ArrowLineRight] (0.center) to (1.center);
		\draw [style=ArrowLineRight] (0.center) to (2.center);
		\draw [style=DashedLine] (3.center) to (4.center);
		\draw [style=DashedLine] (4.center) to (5.center);
	\end{pgfonlayer}
\end{tikzpicture}
}
\caption{Alternative parametrization for the boundary of an Acharya-Witten cone. }
\label{fig:AWCone2}
\end{figure}
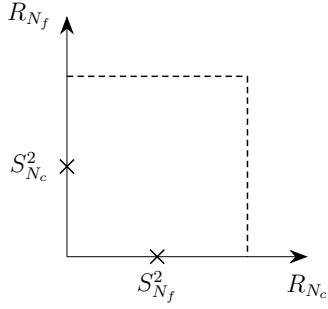

Let us parametrize the first and second factor of $\C^2$ by complex coordinates $u_1,u_2$ and $v_1,v_2$ respectively. These then parametrize $X_{N_c,N_f}$ up to the equivalence \cite{Acharya:2001gy}
\be
(u_1,u_2,v_1,v_2) \sim (\omega^{N_f} u_1,\omega^{-N_f} u_2, \omega^{N_c}v_1, \omega^{-N_c}v_2)
\ee
with phases $\omega=\exp(i\psi/N_cN_f)\in U(1)$. Next we introduce radii on the ADE singularities \eqref{eq:AWCone}
\be
u_1\bar u_1 + u_2\bar u_2={R_{N_f}} \,, \qquad v_1\bar v_1 + v_2\bar v_2={R_{N_c}}
\ee
and denote their sum by $R=R_{N_f}+R_{N_c}$. The link of the cone \eqref{eq:AWCone} is a slice of constant $R$ and was argued in \cite{Acharya:2001gy} to be the quotient of a weighted projective space
\be
\partial X_{N_c,N_f}=\C\P^3_{n_c,n_c,n_f,n_f}/\Z_g
\ee
with $g=\textnormal{gcd}(N_c,N_f)$ and $g(n_c,n_f)=(N_c,N_f)$.  Note that $X_{N_c,N_f}$ is fibered over the quadrant parametrized by $R_{N_c}$ and $R_{N_f}$ and consequently $\partial X_{N_c,N_f}$ is fibered over an interval. See figure \ref{fig:AWCone}. The topology of this boundary remains unaltered if we instead place it at constant values of $R_{N_c},R_{N_f}$ respectively as depicted in figure \ref{fig:AWCone2}.

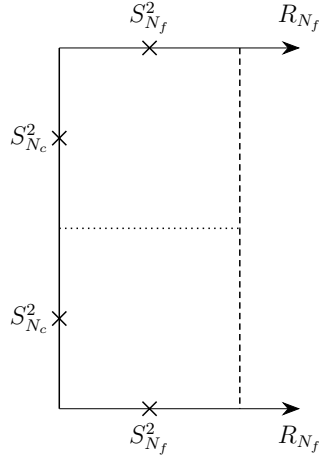
\begin{figure}
\centering
\scalebox{0.8}{
\begin{tikzpicture}
	\begin{pgfonlayer}{nodelayer}
		\node [style=none] (0) at (-1.5, -1.5) {};
		\node [style=none] (1) at (-1.5, 4.5) {};
		\node [style=none] (2) at (2.5, -1.5) {};
		\node [style=none] (3) at (-1.5, 1.5) {};
		\node [style=none] (4) at (1.5, 1.5) {};
		\node [style=none] (5) at (1.5, -1.5) {};
		\node [style=none] (6) at (2.5, -2) {$R_{N_f}$};
		\node [style=NodeCross] (8) at (0, -1.5) {};
		\node [style=NodeCross] (9) at (-1.5, 0) {};
		\node [style=none] (11) at (-2, 0) {$S^2_{N_c}$};
		\node [style=none] (12) at (0, -2) {$S^2_{N_f}$};
		\node [style=none] (13) at (-1.5, 4.5) {};
		\node [style=none] (14) at (2.5, 4.5) {};
		\node [style=none] (15) at (1.5, 4.5) {};
		\node [style=none] (16) at (2.5, 5) {$R_{N_f}$};
		\node [style=NodeCross] (17) at (0, 4.5) {};
		\node [style=none] (18) at (0, 5) {$S^2_{N_f}$};
		\node [style=none] (19) at (-2, 3) {$S^2_{N_c}$};
		\node [style=NodeCross] (20) at (-1.5, 3) {};
	\end{pgfonlayer}
	\begin{pgfonlayer}{edgelayer}
		\draw [style=ArrowLineRight] (0.center) to (2.center);
		\draw [style=DashedLine] (4.center) to (5.center);
		\draw [style=ArrowLineRight] (13.center) to (14.center);
		\draw [style=DashedLine] (15.center) to (4.center);
		\draw [style=ThickLine] (13.center) to (0.center);
		\draw [style=DottedLine] (3.center) to (4.center);
	\end{pgfonlayer}
\end{tikzpicture}
}
\caption{Gluing of two Acharya-Witten cones along a common subset of their boundary (dotted). The boundary of the new space is fibered over the dashed line.}
\label{fig:AcharyaWittenCone3}
\end{figure}

We now take two Acharya-Witten cones and glue these along their boundaries with constant $R_{N_c}$ as depicted in figure \ref{fig:AcharyaWittenCone3} and denote the resulting space by $X_7$. The gluing identifies the fibers
\be
\Pi_{N_c,N_f}=(S^3/\Z_{N_c}\times S^3/\Z_{N_f})/U(1)
\ee
over this locus and is therefore well-defined only when the values of $N_c,N_f$ match for both of the glued cones. This is equivalent to anomaly cancellation. Let us discuss the geometry of the resulting space. Note first that the locus $R_{N_f}=0$, which is a copy of $\R^3$ in each cone, is compactified to a three-sphere where the northern/southern hemispheres are thought of as belonging to either of the gluing blocks. The space $X_7$ therefore contains an $S^3$ worths of $A_{N_c-1}$ singularities. The loci $R_{N_c}=0$ in each building block remain separated and we therefore find two loci, topologically copies of $\R^3$, supporting $A_{N_f-1}$ singularities. The singularities intersect in two points thought of as north/south pole of the $S^3$ cycle.

We now study the homology groups of $X_7$ and its boundary and match the result \eqref{eq:SmoothBdryG22}. First note that $X_7$ deformation retracts to a three-sphere which follows immediately from figure \ref{fig:AcharyaWittenCone3} and determines its homology groups. The homology groups of the boundary $\partial X_7$ are computed via an application of the Mayer-Vietoris sequence. We decompose
\be\label{eq:Covering}
\partial X_7=\partial X_7^{(n)}\cup \partial X_7^{(s)}
\ee where each factor is the set of fibers $\Pi_{N_c,N_f}$ fibered over constant $R_{N_c}$ in each gluing block. These project onto the top and bottom half of the dashed line in figure \ref{fig:AcharyaWittenCone3} respectively. These two sets intersect in a copy of the fiber
\be
\Pi_{N_c,N_f}=\partial X_7^{(n)}\cap  \partial X_7^{(s)}
\ee
and individually deformation retract onto a two-sphere. In order to evaluate the Mayer-Vietories sequence we therefore require the homology groups of $\Pi_{N_c,N_f}$.

The fiber itself is circle fibered
\be\label{eq:MtheoryCircle}
S^1~\hookrightarrow ~\Pi_{N_c,N_f}\rightarrow S^2_{N_c}\times S^2_{N_f}
\ee
where the $S^1$ is the diagonal of the Hopf circles of the Lens spaces. The Euler class of the fibration is therefore
\be
e= N_f \textnormal{vol}_{S^2_{N_c}}+N_c \textnormal{vol}_{S^2_{N_f}}\,.
\ee
The Gysin long exact sequence applied to this circle fibration gives
\be
H^0(\Pi_{N_f,N_c})\cong H^5(\Pi_{N_f,N_c})\cong \Z
\ee
and further splits as
\be\ba
0 &\rightarrow H^1(\Pi_{N_f,N_c})\rightarrow H^0(S^2_{N_c}\times S^2_{N_f}) \rightarrow H^2(S^2_{N_c}\times S^2_{N_f}) \rightarrow H^2(\Pi_{N_f,N_c}) \rightarrow 0 \\
0 &\rightarrow H^3(\Pi_{N_f,N_c})\rightarrow H^2(S^2_{N_c}\times S^2_{N_f}) \rightarrow H^4(S^2_{N_c}\times S^2_{N_f}) \rightarrow H^4(\Pi_{N_f,N_c}) \rightarrow 0
\ea\ee
where the central maps are wedging with the Euler class of the fibration
\be
e\,\wedge \,: H^*(S^2_{N_c}\times S^2_{N_f}) ~\rightarrow ~H^{*+2}(S^2_{N_c}\times S^2_{N_f})\,.
\ee
These are only non-trivial in even degree and there we have
\be\ba
e_0\,\wedge\, &: \Z ~\rightarrow ~ \Z^2\,, \qquad k ~\mapsto ~ (kN_f,kN_c)\,, \\
e_2\,\wedge\, &: \Z^2 ~\rightarrow ~ \Z\,, \qquad (n,m) ~\mapsto ~ nN_f+mN_c\,.
\ea\ee
This gives the cohomology groups
\be
H^*(\Pi_{N_f,N_c})\cong \lbbb \Z ,0, \Z \oplus \Z_{\tn{gcd}(N_f,N_c)}, \Z, \Z_{\tn{gcd}(N_f,N_c)}, \Z \rbbb\,.
\ee
and dualizing to homology we find
\be
H_*(\Pi_{N_f,N_c})\cong \lbbb \Z , \Z_{\tn{gcd}(N_f,N_c)},  \Z, \Z\oplus \Z_{\tn{gcd}(N_f,N_c)},0,\Z \rbbb\,.
\ee
which were already computed in \eqref{eq:HomoSmooth}.

With these results we return to computing the homology groups of $\partial X_7$. The Mayer-Vietoris sequence for the covering \eqref{eq:Covering} now takes the form
\be
\dots~\xrightarrow[]{\,\partial_{k+1} \,} ~H_{k}\big(\Pi_{N_f,N_c}\big)  ~\xrightarrow[]{\,\iota_{k} \,} H_k\big(S^2_{N_f}\big)\oplus H_k\big(S^2_{N_f}\big) ~\xrightarrow[]{j_k-\ell_k }  ~  H_k\big(\partial X_7\big) ~\xrightarrow[]{\,\partial_k\, }  ~\dots.
\ee
and it follows straightforwardly
\be
H_*(\partial X_{7})\cong \lbbb \Z, 0, \Z \oplus \Z_{\tn{gcd}(N_f,N_c)}, 0 ,\Z\oplus  \Z_{\tn{gcd}(N_f,N_c)},0 , \Z \rbbb
\ee
which are the same homology groups computed in \eqref{eq:SmoothBdryG22} from uplifting the D6-brane setup described in section \ref{sssec:sucodim7}. Here we make no claims regarding metric data and, as already for the case of the Acharya-Witten cones, rely on M-theory in conjecturing that a space of the above topology should admit a $G_2$ holonomy metric.

\subsection{Reduction to IIA}

We now substantiate our claim, that the seven-manifold $X_7$ constructed in Appendix \ref{sec:GluingCones} is topologically identical to the circle bundle constructed in section \ref{sssec:sucodim7}, by determining the IIA background $X_7/U(1)_M$. The M-theory circle $U(1)_M$ is taken to act on each $\mathbb{C}^2$ factor in the Acharya-Witten cone \eqref{eq:AWCone} by phase rotations with the same charge $+1$ on the two ADE factors. It is therefore contained in the fiber $\Pi_{N_c,N_f}$ as the diagonal Hopf circle, this is precisely the circle fiber of \eqref{eq:MtheoryCircle}. Consequently
\be
\Pi_{N_c,N_f}/U(1)_M=S^2_{N_c}\times S^2_{N_f}
\ee
which determines the boundary $\partial X_7/U(1)_M$ to be fibered over an interval as
\be
S^2_{N_c}\times S^2_{N_f} ~\hookrightarrow~\partial X_7/U(1)_M~\rightarrow ~I\,.
\ee
At the ends of the interval $S^2_{N_c}$ collapses - it traces out a three-sphere $S^3$. We therefore find
\be
\partial X_7/U(1)_M=S^3\times S^2_{N_f}\,.
\ee
Varying the radius of $S^2_{N_f}$ in $\partial X_7/U(1)_M$ sweeps out the six-manifold $X_7/U(1)_M$, therefore
\be
 X_7/U(1)_M=T^*S^3\,.
\ee
The fixed point locus of the $U(1)_M$ action restricts on the boundary to two copies of $S^2_{N_f}$ located at the north/south pole of the $S^3$. Here the orbits of $U(1)$ in \eqref{eq:AWCone} and $U(1)_M$ coincide and consequently the M-theory circle $U(1)_M$ must collapse. By identical arguments the central $S^3$ of the bulk is also a fixed point locus and overall we determine the D6-brane content to two stacks of non-compact $N_f$ D6-branes, topologically $\mathbb{R}^3$, intersecting one stack of $N_c$ D6-brane transversely, wrapped on $S^3$. This is topologically precisely the IIA background we took as starting point in section \ref{sssec:sucodim7}.

Finally we comment on the construction of the above IIA backgrounds. These appear naturally as the local models of the geometries discussed in \cite{Feng:2005gw} of which we now discuss a simple example. Consider the local Calabi-Yau threefold $X_6$ with hypersurface equation
\be\label{eq:HS}
P(x_1,x_2,x_3,x_4;\lbbb \mu_k \rbbb )=x_1^2+x_2^2+x_3^2+x_4^4+\mu_2 x_4^2+\mu_3 x_4+\mu_4=0 \,.
\ee
which describes a $T^*S^2$ fibration over the $x_4$-plane. The polynomial $F(x_4)=x_4^4+\mu_2 x_4^2+\mu_3 x_4+\mu_4$ vanishes for four values $x_4=\ell_i$ with $i=1,\dots 4$. The fibral two-sphere traces out a three-sphere over any path connecting two of these values. Of these three-spheres three are independent in homology with volumes controlled by the parameters $\mu_{k}$. There are no other compact cycles in the geometry. The holomorphic top-form of the Calabi-Yau threefold takes the standard form
\be\label{eq:topform}
\Omega=\frac{dx_1\wedge dx_2\wedge dx_3 }{\partial P/\partial x_4}\,.
\ee
We require the generators of the homology groups in degree three to be calibrated with respect to $\Omega$, i.e. $\textnormal{Im}\,\Omega$ vanishes restricted to the three three-spheres. This constrains the parameters $\mu_k$ \cite{Feng:2005gw}. We can study this constraint locally at intersection between the three-spheres. Locally near $x_4 = \ell_i$, we have the geometry
\be
x_1^2+x_2^2+x_3^2+\tilde x_4=0
\ee
where we have redefined $x_4$ via shifts and rotations to $\tilde x_4$. Two three-spheres intersecting at $\tilde x_4=0$ project to intervals in $\tilde x_4$-plane. We require these to intersect at an angle of $\theta=\pm 2\pi /3$ to ensure that both are calibrated with respect to $\textnormal{Im}\,\Omega$ locally near $\ell_i$. Indeed, in this case the two calibrated loci are mapped onto each other by a phase rotation by  $\mp 2\pi /3$ on $x_1,x_2,x_3$ respectively which leaves $\Omega$ in variant. Schematically we have the setup
\be\label{eq:FibrationPic}\scalebox{0.8}{
\begin{tikzpicture}
	\begin{pgfonlayer}{nodelayer}
		\node [style=none] (0) at (-3, 2) {};
		\node [style=none] (1) at (3, 2) {};
		\node [style=none] (2) at (3, -2) {};
		\node [style=none] (3) at (-3, -2) {};
		\node [style=none] (4) at (-0.5, -1) {};
		\node [style=none] (5) at (0.5, 1) {};
		\node [style=none] (6) at (2, 1) {};
		\node [style=none] (7) at (-2, -1) {};
		\node [style=none] (8) at (-2.5, 1.5) {$x_4$};
		\node [style=Circle] (9) at (0.5, 1) {};
		\node [style=Circle] (10) at (2, 1) {};
		\node [style=Circle] (11) at (-0.5, -1) {};
		\node [style=Circle] (12) at (-2, -1) {};
		\node [style=none] (13) at (0.5, 1.5) {$\ell_3$};
		\node [style=none] (14) at (2, 1.5) {$\ell_4$};
		\node [style=none] (15) at (-2, -1.5) {$\ell_1$};
		\node [style=none] (16) at (-0.5, -1.5) {$\ell_2$};
		\node [style=none] (17) at (-2, 1) {};
		\node [style=none] (18) at (-2, 2) {};
		\node [style=none] (19) at (-3, 1) {};
	\end{pgfonlayer}
	\begin{pgfonlayer}{edgelayer}
		\draw [style=ThickLine] (0.center) to (1.center);
		\draw [style=ThickLine] (2.center) to (1.center);
		\draw [style=ThickLine] (2.center) to (3.center);
		\draw [style=ThickLine] (3.center) to (0.center);
		\draw [style=ThickLine] (7.center) to (4.center);
		\draw [style=ThickLine] (4.center) to (5.center);
		\draw [style=ThickLine] (5.center) to (6.center);
		\draw [style=ThickLine] (19.center) to (17.center);
		\draw [style=ThickLine] (17.center) to (18.center);
	\end{pgfonlayer}
\end{tikzpicture}
}
\ee
where each interval connecting $\ell_{i}$ to $\ell_{i+1}$ lifts to a three-sphere. We denote the three-sphere connecting $\ell_i$ to $\ell_j$ by $S^3_{ij}$.

We now wrap $N_f$ D6-branes on $S^3_{12},S^3_{34}$ and $N_c$ D6-branes on the three-sphere $S^3_{23}$. The two stacks of $N_f$ D6-branes each intersect the stack of $N_c$ D6-branes transversely once.
We next take a local limit centered on the central three-sphere. The two stacks of $N_f$ D6-branes are demoted to flavor stacks. This limit sends $\ell_1,\ell_4$ in \eqref{eq:FibrationPic} to infinity. The resulting geometry $X_{6}^{\textnormal{loc}}$ is topologically the deformed conifold whose compact cycle is a single compact three-sphere $S^3_{23}$. The constrains on angles now simply amounts to the standard
constraint for D6-branes intersecting at angles to preserve supersymmetry.

In IIA string theory this setup engineers 4D $\mathcal{N}=1$ SQCD with gauge algebra $\mathfrak{g}=\mathfrak{su}(N_c)$ and $N_f$ flavors in the fundamental representation.

\section{Finite Abelian Subgroups of $SU(3)$} \label{app:AppendixSU3}

In this Appendix we characterize the finite abelian subgroups of the special unitary group $SU(3)$. We begin with following structure theorem\cite{Ludl:2011gn}:\smallskip

\noindent \textit{Every finite abelian subgroup $\Gamma\subset SU(3)$ is isomorphic to $\Z_n\times \Z_m$ with $m$ dividing $n$ and
\be \label{eq:Maximum}
n=\max_{\gamma \in \Gamma} \textnormal{ord}(\gamma)\,.
\ee
}\smallskip

\vspace{-15pt}
\noindent First, notice that for all $\gamma\in \Gamma$ we have $\gamma^n=1$. For if there were to exist a $\sigma \in \Gamma$ of order $l$ with $\sigma^n\neq 1$ then $\sigma\epsilon^g$ is of order $\textnormal{lcm}(n,l)>n$ which yields a contradiction. Here $\epsilon\in \Gamma$ is the order $n$ element guaranteed to exist by \eqref{eq:Maximum} and $g=\textnormal{gcd}(l,n)<l$ as $l$ does not divide $n$. We can therefore define $\omega=\exp(2\pi i/n)$ and every element in $\Gamma$ now has the form
\be
\gamma_{kl}=\diag\Big( \omega^k,\omega^l,\omega^{-k-l}\Big)\,.
\ee
Therefore $\Gamma$ is a subgroup of $\Z_n\times \Z_n$ generated by $\diag(\omega,1,\omega^{-1})$ and $\diag(1,\omega,\omega^{-1})$. Next consider the prime decomposition $n=p_1^{r_1}\cdots p_s^{r_s}$ and conclude by the structure theorem on abelian groups that
\be \label{eq:StructureThm1}
\Z_n\times \Z_n\cong \lb \Z_{p_1^{r_1}} \times \Z_{p_1^{r_1}} \rb \times \dots \times   \lb \Z_{p_s^{r_s}} \times \Z_{p_s^{r_s}}\rb\,.
\ee
The structure theorem on subgroups of abelian groups now implies
\be
\Gamma\cong  \lb \Z_{p_1^{u_1}} \times \Z_{p_1^{v_1}} \rb \times \dots \times   \lb \Z_{p_s^{u_s}} \times \Z_{p_s^{v_s}}\rb\cong \Z_u\times \Z_v\,.
\ee
with integers $0\leq u_i\leq v_i\leq r_i$ where transposition in relation to \eqref{eq:StructureThm1} are made in order to realize $u_i\leq v_i$. Here $u=p_1^{u_1}\cdots p_s^{u_s}$ and $v=p_1^{v_1}\cdots p_s^{v_s}$ and $u\leq v$ and $u$ divides $v$. We know of an order $n$ subgroup, therefore $v=n$, and we finish by setting $u=m$.

Next we derive canonical representations for generators. First, consider the case $\Gamma\cong\Z_n$. Then the generator $\omega$ of $\Gamma$ acts as follows on $\mathbb{C}^3$, parametrized by coordinates $(z_1,z_2,z_3)$,
\be \label{eq:GroupAction}
\omega\,: \quad (z_1,z_2,z_3)~\mapsto (\omega^{k_1}z_1,\omega^{k_2}z_2,\omega^{k_3}z_3)\,.
\ee
Here $\omega$ is primitive $n$-th root of unity and $0\leq k_i \leq n-1$. Now $k_1+k_2+k_3=0$ mod $n$ and therefore the $k_i$ sum to $n$ or $2n$. In the latter case we instead consider the generator $\omega^{-1}$ which implies the
redefinition $k_i\rightarrow n-k_i$. Without loss of generalization we can therefore assume $k_1+k_2+k_3=n$.

Next we note that the group action \eqref{eq:GroupAction} is necessarily faithful when assumed to be of order $n$. Define $q_i=n/\textnormal{gcd}(n,k_i)$. The subgroup generated by $\omega^{q_i}$ does not act on the coordinate $z_i$. Faithfulness then amounts to requiring $\omega^q$ to generate the trivial subgroup, that is we impose $q=n$, where
\be \label{eq:LCM}
q=\textnormal{lcm}(q_1,q_2,q_3)\,.
\ee
With this we arrive at the following result: the generators of subgroups of $SU(3)$ with $\Gamma\cong \Z_n$ are characterized by triples $(k_1,k_2,k_3)$ where $0\leq k_i \leq n-1$ and $k_1+k_2+k_3=n$ and $q=n$. We often denote the generator of $\Gamma$ characterized in this way as $\frac{1}{n}(k_1,k_2,k_3)$. These triples are of course not unique.

Now consider the case $\Gamma\cong \Z_n\times \Z_m$ with $m$ dividing $n$. We write $n=mm'$ and denote the generator of $\Z_n$,  $\Z_m$ by $\omega = \frac{1}{n}(k_i)$, $\eta=\frac{1}{m}(b_i)$ respectively. A subgroup $\Z_m\times \Z_m$ is generated by $\rho,\eta$ where $\rho=\omega^{m'}=\frac{1}{m}(a_i)$. Here $0\leq a_i\leq m-1$ and $a_i=k_i$ mod $m$ and $a_1+a_2+a_3=m$ (whenever the sum equates to $2m$ we redefine $\rho\rightarrow \rho^{-1}$). By construction $\frac{1}{m}(a_i)$ describe a group action of order $m$ and therefore \eqref{eq:LCM} evaluates to $m$. We therefore have two copies of abelian subgroups of order $m$ with generators $\rho,\eta$ canonically represented as in the previous paragraph.

We now argue that redefinitions allow us to put the generators of the $\Z_m\times \Z_m$ subgroup into the form
\be \label{eq:FinalForm}
\rho=\frac{1}{m}(1,0,m-1)\,, \qquad \eta= \frac{1}{m}(0,1,m-1)\,.
\ee
For this note that there exists index $i_0$ such that $a_{i_0}> b_{i_0}$. We now take successive differences. We begin by making the redefinition of generators
\be
(\rho,\eta)~\rightarrow~(\rho_1,\eta)\,, \qquad \rho_1=\rho \eta^{-t_1}=\frac{1}{m}(a_1^{(1)},a_2^{(1)},a_3^{(1)})
\ee
where $t_1$ is the largest integer such that $0\leq a_{i_0}^{(1)}<b_{i_0}$ with $a_{i_0}^{(1)}=a_{i_0}-t_1 b_{i_0}$. Then we redefine generators as
\be
(\rho_1,\eta)~\rightarrow~(\rho_1,\eta_1)\,, \qquad \eta_1=\eta \rho_1^{-s_1}=\frac{1}{m}(b_1^{(1)},b_2^{(1)},b_3^{(1)})
\ee
where $s_1$ is the largest integer such that $0\leq b_{i_0}^{(1)}<a_{i_0}^{(1)}$ with $b_{i_0}^{(1)}=b_{i_0}-s_1 a_{i_0}^{(1)}$. We iterate this alternating redefinition of generators either $2r-1$ or $2r$ times until either $a^{(r)}_{i_0}=0$ or $b^{(r)}_{i_0}=0$ respectively. These redefinitions are invertible and therefore each step yields a pair of generators of the full $\Z_m\times \Z_m$ subgroup.

Let us assume for concreteness $i_0=2$. If the process terminates after an odd number of iterations be have $\rho_r=\frac{1}{n}(1,0,m-1)$ (possibly after replacing the generator with its inverse). We can now make the final redefinition
\be
(\rho_r,\eta_{r-1})~\rightarrow~(\rho_r,\eta_{r-1}')\,, \qquad \eta_{r-1}'=\eta_{r-1}\rho_r^{-b^{(r-1)}_1}
\ee
to achieve the form \eqref{eq:FinalForm}. For $ \eta_{r-1}'$ the entries $1,m-1$ are also possibly transposed, in this case replace the generator with its inverse. Other values of $i_0$ and the case in which redefinition terminate after an even number of iterations are treated similarly.

We conclude, generators for $\Gamma=\Z_n\times \Z_m$ can be taken to be of the form
\be \label{eq:PresentationFavorable}
\omega=\frac{1}{n}(k_1,k_2,k_3)\,, \qquad  \eta=\frac{1}{m}(0,1,m-1)
\ee
where \eqref{eq:LCM} evaluates to $n$ and $0\leq k_i \leq n-1$ and $k_1+k_2+k_3=n$. Further we require $\textnormal{gcd}(n,k_1)$ and $m$ to be co-prime as otherwise $\Z_m$ and $\Z_n$ have nontrivial intersection when generated by  \eqref{eq:PresentationFavorable} which would violate our assumption of describing a subgroup of order $|\Gamma|=nm$. The result \eqref{eq:FinalForm} immediately imples that whenever $n=m$ we can improve our choice of generators to
\be \label{eq:vgoodgenerators}
\omega=\frac{1}{n}(1,0,n-1)\,, \qquad  \eta=\frac{1}{n}(0,1,n-1)\,.
\ee

Let us discuss the fixed point loci. An element $\gamma \in \Gamma$ fixes $z\in\mathbb{C}^3$ whenever it preserves all of its coordinates $\gamma\cdot z_i =z_i$ which is satisfied whenever $z_i=0$ or expanding $\gamma=\omega^{l_1}\eta^{l_2}=\frac{1}{n}(\gamma_1,\gamma_2,\gamma_3)$ we have $\gamma_i=0$ mod $n$. We have $\gamma_1+\gamma_2+\gamma_3=0$ mod $n$ so the latter condition can only be realized for a single coordinate and the other coordinates are necessarily set to vanish. Fixed point sets therefore consist of planes $F_{ij}$ characterized by $z_i=z_j=0$ where $(i,j)=(1,2),(2,3),(3,1)$ and are necessarily ADE singularities of type A for if a single $\gamma_i$ vanishes the remaining entries must be equal and opposite.

We consider the two cases in more detail and begin with $\Gamma\cong \Z_n$ generated by $\frac{1}{n}(k_1,k_2,k_3)$. Setting all but $z_i$ to zero be see that the subgroup generated by $\omega^{q_i}$ with $q_i=n/\textnormal{gcd}(n,k_i)$ acts trivially on this hyperplane. Therefore the fixed point locus consists of three planes of $A_{\textnormal{gcd}(n,k_i)-1}$ singularities intersecting at the origin of $\mathbb{C}^3$.

Now consider $\Gamma\cong \Z_n\times \Z_m$ with generators \eqref{eq:PresentationFavorable}. We immediately conclude that there is a $\Z_{\textnormal{gcd}(n,k_1)}\times \Z_m= \Z_{m \;\! \textnormal{gcd}(n,k_1)}= \Z_{m \;\! \textnormal{gcd}(m',k_1)}\subset \Z_n\times \Z_m$ acting trivially on the plane $F_{23}$, which follows from $\textnormal{gcd}(n,k_1)$ and $m$ being co-prime. Next we conclude $\Z_{\textnormal{gcd}(n,k_2)}\subset \Z_n$ acts trivially on $F_{13}$. However there exists a diagonal subgroup in $\Z_n\times \Z_m$ which also fixes $F_{13}$, it is determined by requiring phase rotations on $z_2$ to cancel
\be
c_2 k_2+c_2' m'=0 ~~(\textnormal{mod }n)
\ee
with integers $c_2,c_2'$. This is solved by $c_2=m'/\textnormal{gcd}(m',k_2)$ and $c'_2=-k_2/\textnormal{gcd}(m',k_2)$ mod $n$. We conclude that $\omega^{c_2}\eta^{c_2'}$ generate a subgroup of elements with fixed points and with order
\be
\frac{n}{m'/\textnormal{gcd}(m',k_2)}=m \;\! \textnormal{gcd}(m',k_2)\,.
\ee
This follows as $\eta^{c_2'}$ raised to that power is trivial. Now we argue that this subgroup contains $\Z_{\textnormal{gcd}(n,k_2)}$ described above as a subgroup. We raise the generator $\omega^{c}\eta^{c'}$ to the power $( n/\textnormal{gcd}(n,k_2))/(m'/\textnormal{gcd}(m',k_2))$ and find $\omega^{n/\textnormal{gcd}(n,k_2)}$ which is the generator of $\Z_{\textnormal{gcd}(n,k_2)}\subset \Z_n$. The fixed locus $F_{12}$ is analyzed similarly introducing the integers $c_3,c_3'$.

The subgroups of $\Gamma=\Z_n\times \Z_m$ and their fixed loci are therefore, where $n=mm'$,
\be\ba \label{eq:FixedPts}
&\Z_{m\;\! \textnormal{gcd}(m',k_1)}\,, && F_{23}=\lbbb z_2=z_3=0\rbbb\\
 &\Z_{ m \;\! \textnormal{gcd}(m', k_2)}\,, && F_{31}= \lbbb z_1=z_3=0\rbbb\\
  &\Z_{m \;\!  \textnormal{gcd}(m', k_3)}\,, && F_{12}=\lbbb z_1=z_2=0\rbbb\,.\\
\ea \ee
As a check of \eqref{eq:FixedPts}, consider the case $n=m$ with generators \eqref{eq:vgoodgenerators}. We find three copies of $A_{n-1}$ singularities.

Next we study the subgroup $H$ of $\Z_n\times \Z_m$ with fixed points. It is generated by
\be \ba
H&=\langle \omega^{n/\textnormal{gcd}(n,k_1)},\eta, \omega^{c_2}\eta^{c_2'},  \omega^{c_3}\eta^{c_3'}\rangle \\
&= \langle \omega^{n/\textnormal{gcd}(n,k_1)}, \omega^{c_2},  \omega^{c_3}\rangle \times \langle \eta \rangle
\ea \ee
and is therefore isomorphic to $\Z_{n/k}\times \Z_m$ where, recalling the definition of $c_i$,
\be\ba\label{eq:gcdc1c2c3}
k&=\textnormal{gcd}\lb  \frac{n}{\textnormal{gcd}(n,k_1)},\frac{m'}{\textnormal{gcd}(m',k_2)}, \frac{m'}{\textnormal{gcd}(m',k_3)}\rb\\
& = \textnormal{gcd}\lb  \frac{n}{\textnormal{gcd}(m',k_1)},\frac{m'}{\textnormal{gcd}(m',k_2)}, \frac{m'}{\textnormal{gcd}(m',k_3)} \rb \\
& = \textnormal{gcd}\lb  n ,\frac{m'}{\textnormal{gcd}(m',k_1)\textnormal{gcd}(m',k_2)\textnormal{gcd}(m',k_3)} \rb \\
&=\frac{m'}{\textnormal{gcd}(m',k_1)\textnormal{gcd}(m',k_2)\textnormal{gcd}(m',k_3)}
\ea\ee
where the final result is integral because the individual gcd's are pairwise co-prime. Overall we find
\be
H=\Z_{m\;\!\textnormal{gcd}(m',k_1)\textnormal{gcd}(m',k_2)\textnormal{gcd}(m',k_3) } \times \Z_{m} =\langle \omega^{m'/\textnormal{gcd}(m',k_1)\textnormal{gcd}(m',k_2)\textnormal{gcd}(m',k_3)},\eta \rangle
\ee
and therefore
\be
\Gamma/H=\Z_{m'/\textnormal{gcd}(m',k_1)\textnormal{gcd}(m',k_2)\textnormal{gcd}(m',k_3)}=\langle \omega \rangle\,.
\ee

\newpage

\bibliographystyle{utphys}
\bibliography{HigherGluing}

\end{document}